\documentclass{aa}  
\usepackage{rotating} 
\usepackage{afterpage}
\usepackage{graphicx}  
\usepackage{txfonts}  
\usepackage{hyperref}  
\usepackage{subcaption}  
\usepackage{siunitx}  

\usepackage{caption}

\bibpunct{(}{)}{;}{a}{}{,} 

\begin{document} 

   \title{Comparing the dynamics of Jupiter-family Comets and comet-like fireballs \thanks{The data used in this study is available at \url{https://doi.org/10.5281/zenodo.10671333}}}

   \author{P.M. Shober\inst{1}\fnmsep\thanks{E-mail: \href{mailto:patrick.shober@obspm.fr}{patrick.shober@obspm.fr}}
          \and G. Tancredi\inst{2}
          \and J. Vaubaillon\inst{1}
          \and H.A.R. Devillepoix\inst{3,4}
          \and S. Deam\inst{3}
          \and S. Anghel\inst{1,5}
          \and E.K. Sansom\inst{3,4}
          \and F. Colas\inst{1}
          \and S. Martino\inst{2}
          }

   \institute{Institut de Mécanique Céleste et de Calcul des Éphemerides, Observatoire de Paris, PSL, 75014, Paris, France
         \and Departmento de Astronomía, Facultad de Ciencias, Iguá 4225, 11400 Montevideo, Uruguay
         \and Space Science and Technology Centre, Curtin University, GPO Box U1987, Perth WA 6845, Australia
         \and International Centre for Radio Astronomy Research, Curtin University, GPO Box U1987, Perth WA 6845, Australia
         \and Astronomical Institute of the Romanian Academy, Cutitul de Argint 5, 040557 Bucharest, Romania
             }

   \date{Received February 17, 2024; accepted DATE}

\abstract
{Jupiter-family comets (JFCs), which originate from the Kuiper belt and scattered disk, exhibit low-inclination and chaotic trajectories due to close encounters with Jupiter. Despite their typically short incursions into the inner solar system, a notable number of them are on Earth-crossing orbits, with fireball networks detecting many objects on ``JFC-like'' ($2<T_{J}<3$) orbits.}
{This investigation aims to examine the orbital dynamics of JFCs and comet-like fireballs over $10^{4}$\,yr timescales, focusing on the trajectories and stability of these objects in the context of gravitational interactions within the solar system.}
{We employed an extensive fireball dataset from Desert Fireball Network (DFN),  European Fireball Network (EFN), Fireball Recovery and InterPlanetary Observation Network (FRIPON), and Meteorite Observation and Recovery Project (MORP), alongside telescopically observed cometary ephemeris from the NASA HORIZONS database. The study integrates 646 fireball orbits with 661 JFC orbits for a comparative analysis of their orbital stability and evolution.}
{The analysis confirms frequent Jupiter encounters among most JFCs, inducing chaotic orbital behavior with limited predictability and short Lyapunov lifetimes ($\sim$120 years), underscoring Jupiter's significant dynamical influence. In contrast, "JFC-like" meteoroids detected by fireball networks largely exhibit dynamics divergent from genuine JFCs, with 79-92\% on ``JFC-like'' orbits shown not to be prone to frequent Jupiter encounters; in particular, only 1-5\% of all fireballs detected by the four networks exhibit dynamics similar to that of actual JFCs. In addition, 22\% (16 of 72) of near-Earth JFCs are on highly stable orbits, suggesting a potential main belt origin for some of the bodies.}
{This extensive study delineates the stark dynamical contrast between JFCs and JFC-like meteoroids detected by global fireball networks. The majority of centimeter- and meter-scale meteoroids on JFC-like orbits exhibit remarkably stable trajectories, which starkly differ from the chaotic paths of their km-scale counterparts. Our findings suggest that the JFC-like objects observed by fireball networks predominantly originate from the outer main belt, with only a minor fraction being directly attributable to traditional JFCs.}


   \maketitle


\section{Introduction}

Jupiter-family comets (JFCs) are a distinct group of low-inclination ($\lesssim$30$^{\circ}$) comets predominantly originating from the Kuiper belt and scattered disk, a vast region of icy bodies extending beyond Neptune’s orbit, where volatile components are retained \citep{fernandez1980existence,boehnhardt2004split,duncan1997disk,duncan2004dynamical}. The transformation of these trans-Neptunian objects into the inner solar system is driven and controlled completely by a series of chaotic close encounters with Neptune and then Jupiter, as reflected in the name of the population \citep{tancredi1992evolution,levison1997kuiper,disisto2009population,nesvorny2017origin}. However, in the inner solar system, JFCs exhibit short orbital periods, typically less than 20 years. Previously defined by these shorter orbital periods, it has since become standard to classify these bodies by their Tisserand's parameter, which  stays predominantly within the range $2<T_{J}<3$ \citep{carusi1985long,levison1994long}. 

The incursions of JFCs into the inner solar system are infrequent and short-lived, with only $\sim30$\% of JFCs becoming active and part of the "visible" population (i.e., q$<$2.5\,au) during their lifetimes \citep{levison1997kuiper}. This specific distance is used as a threshold because it is at this point that the heating by the Sun becomes significant enough to cause sublimation of water ice, resulting in observable cometary activity such as the development of a coma or tail \citep{duncan1988origin,levison1994long}. Significantly fewer make it all the way to a near-Earth orbit, and those that do only spend a few thousand years normally in the region \citep{fernandez2002there}. This is because the frequent close encounters with Jupiter will quickly remove the JFCs. As a result, the Lyapunov lifetimes (the inverse of the Lyapunov characteristic exponent) for these bodies tend to concentrate between 50-150\,years, as the encounters with Jupiter make predicting their orbital evolution impossible on very short timescales \citep{tancredi1995dynamical,tancredi1998chaotic,tancredi2014criterion}. The dynamical lifetime of JFCs is found to be around \num{1.5e5}\,yrs, however, they only typically spend $\sim10^{3}$\,yrs in near-Earth space (q$<$1.3\,au). This implies that of the JFCs that serendipitously end up in near-Earth space at some point during their lifetime, this  phase only accounts for about 1\% of their total dynamical lifetime. 

Based on telescopic observations and dynamical modeling, there is a paucity of JFCs at sub-kilometer scales \citep{meech2004comet,fernandez2006population,nesvorny2017origin}. This lack of objects at smaller sizes is thought to be related to the inferred brief physical lifetimes ($10^{3}-10^{4}$\,yrs) of JFCs in the inner solar system \citep{levison1997kuiper,fernandez2002there,disisto2009population}. Due to their low bulk density and high volatile content, JFCs should fade away more quickly when they are within 2.5\,au of the Sun. Typically, the JFCs lose mass through volatile sublimation and dust entrainment, with rates of $\sim10^{1-3}$\,kg/s near perihelion, depending on the active surface area \citep{ahearn1995ensemble}. However, towards the end of their physical lifetime, these objects are believed to either go dormant as they have built up an insulating outer layer of devolatilized material or (much more likely) fragment and disintegrate \citep{rickman1990formation,disisto2009population}. This fragmentation process is likely why there is a paucity of JFCs at smaller sizes. However, this material does not simply disappear. Large amounts of debris are generated from these cometary fracturing and splitting and are left behind. For example, the Andromedid meteor shower, a currently weaker shower with outbursts of meteoroids of $\sim10^{-7}$\,kg, is linked to the short-period comet 3D/Biela. This parent comet is believed to have completely disintegrated due to a catastrophic splitting event after being observed to fragment in 1845/1846 and a final observation in 1852 \citep{wiegert2013return}. Several years later, on November 27, 1872, the debris from this catastrophic splitting event manifested in a spectacular meteor display labeled the "Bielids," with reports of several thousand meteors per hours \citep{hoffleit1988yale}. The prevalence of JFC fracturing and splitting has been demonstrated with numerous observations of cometary splitting and fracturing \citep{chen1994rate,boehnhardt2004split,fernandez2005albedos,fuse2007observations,fernandez2009s,dones2015origin,li2015disappearance,graykowski2019fragmented}. The debris from the destruction of these JFCs should produce ample numbers of micron to sub-millimeter meteors on Earth. 

The zodiacal cloud (ZC) is a faint, diffuse population of small (a few micrometers to a few millimeters) interplanetary dust particles, primarily consisting of material released by comets and asteroid collisions \citep{gustafson1994physics,nesvorny2010cometary}. The thermal emission and scattered light from the ZC, aptly called "zodiacal light," has been observed for centuries as a faint, diffuse glow extending around the ecliptic plane of the solar system. Cassini's 17th-century study was the first to attribute the zodiacal light to sunlight reflecting off a disk-shaped interplanetary dust cloud, challenging the notion of it as an atmospheric phenomenon \citep{cassini_1685}. The structure of the dust within the ZC was first mapped by the Infrared Astronomical Satellite (IRAS) \citep{hauser1984iras,sykes1988iras}, and has also been detected and characterized by other interplanetary space missions such as the Cosmic Background Explorer (COBE), Pioneer 10, AKARI, the Parker Solar Probe (PSP), and Juno missions \citep{kelsall_cobe,matsumoto_pioneer10,szalay_parker_solar_probe,jorgensen2021distribution_juno}. Dust within the ZC has been linked to both asteroids and comets based on concentrations in the ecliptic latitudes, which have been linked to asteroid families \citep{kehoe2015signatures} and from narrow trails of dust associated with comets \citep{sykes1992cometary}. However, research suggests that JFCs contribute significantly more to the ZC, with more than 90\% of the mid-infrared emission in the ZC attributed to dust grains released by JFCs \citep{nesvorny2010cometary}. This abundance of mass being contributed to the ZC from JFCs suggests that 85\% of the total mass flux hitting the Earth's atmosphere annually (\num{3e7}\,kg\,yr$^{-1}$) originates from JFCs \citep{nesvorny2010cometary,plane2012cosmic}. However, the mass input required to maintain the ZC is several times larger than the mass loss from JFCs due to regular activity due to volatile sublimation. This implies spontaneous disruptions and splittings of JFCs are contributing the bulk of the mass within the ZC \citep{nesvorny2010cometary,nesvorny2011dynamical,rigley2022comet}. This fragmentation and splitting process should occur relatively frequently (lower limit of 0.01\,yr$^{-1}$), producing enough dust to support the observed population \citep{chen1994rate}. 

The Rosetta mission's observations using its OSIRIS cameras at comet 67P/Churyumov-Gerasimenko revealed the ejection of boulders, alongside smaller dust particles, challenging our understanding of cometary ejecta \citep{agarwal2016acceleration,ott2017dust}. The observation and detection of cm to m particles or aggregates in the inner coma of 67P/Churyumov-Gerasimenko have confirmed the fact that comets can eject boulders in addition to small dust particles. However, there is limited information about the nature of these objects, leaving the question open as to whether they are fluffy aggregates or more cohesive bodies. This intermediate-size cometary debris being produced could have significantly short physical lifetimes \citep{beech2001endurance,boehnhardt2004split,fernandez2009s,jewitt2016fragmentation}. 

The steady-state debiased near-Earth object (NEO) model by \citet{granvik2018debiased} predicted that $\sim$10\% of the meter-scale NEO population below diameters of 100\,m originate from the JFC population; however, this estimate decreases to only a maximum of 1.7\% if cometary splitting is considered a more likely end-state \citep{bottke2002debiased,nesvorny2010cometary,2023AJ....166...55N}. This prediction would seem reasonable based on the proportion of fireballs with JFC-like pre-impact orbits ($2<T_{J}<3$) \citep{brown2000fall,borovivcka2013kovsice,spurny2013trajectory,madiedo2014trajectory,trigo2019jupiter,pena2022orbital,hughes2022analysis}. While cometary fragmentation is believed to contribute substantial amounts of dust to the ZC \citep{nesvorny2010cometary,rigley2022comet}, the contribution of more substantial debris, ranging from centimeters to meters (observable by fireball networks), remains an open question. In recent studies, alternative approaches to determining the origins of these meteoroids have also been explored. Notably, \citet{borovivcka2022_two} conducted a fragmentation analysis, concluding that the evidence supports an abundance of weaker cometary debris on JFC-like ($2<T_{J}<3$) orbits. However, the fragmentation characteristics and meteoroids' strengths are difficult to compare as many factors influence the observed atmospheric fragmentations. While cometary debris are statistically weaker and more friable, it has also been shown that asteroidal debris can be extremely weak due to macro-scale features of the sample (cracks, porosity, etc.; \citealp{popova2011very}). 

The dynamics of the fireball orbits offer a much clearer picture of the meteoroid source regions because the dominant forces acting on centimeter-to-meter debris and the kilometer-scale JFCs do not differ significantly on 10,000\,yr timescales. Over periods of thousands of years in the inner solar system, the orbits for both size ranges of objects are primarily perturbed through close encounters and mean-motion resonances (MMRs); whereas Poynting-Robertson drag needs to be considered for smaller sub-millimeter debris \citep{nesvorny2011dynamical}. Solar radiation pressure and Yarkovsky-induced drift are also considerable at these size ranges, however, they are insignificant on the concerned timescales of $10^{3}-10^{4}$\,years. The JFC population has many close encounters with Jupiter, causing the orbits to rapidly and chaotically jump in and out of the inner solar system every few thousand years \citep{tancredi1995dynamical,tancredi1998chaotic,tancredi2014criterion}. This is in stark contrast with the orbital evolution of nearly all near-Earth asteroids, which behave predictably on similar periods. However, there is a tiny fraction of the near-Earth asteroid population that does exist on similarly chaotic orbits. This group of bodies has been referred to as the "asteroid cometary orbit" (ACO) population previously, making up $<$0.05\% of the entire asteroid population and $<$0.1\% of the near-Earth asteroid population \citep{tancredi2014criterion}. The work of \citet{fernandez2002there} and \citet{fernandez2014assessing} suggests that these ACOs could originate from the main-belt asteroid population, which has diffused out due to perturbations from the terrestrial planets. However, further dynamical evolution analysis needs to be done on this population to discern the prevalence of dormant, inactive comets. 

Thus, to accurately gauge the source of fireballs on JFC-like orbits, whether it is the MB or the JFC population, an in-depth analysis of the orbital stability is necessary. \citet{shober2021main} studied the orbital stability of 50 sporadic fireballs detected by the Desert Fireball Network, originating from JFC-like orbits. The dynamical analysis showed that almost none ($<$6\,\%) of the sporadic JFC-like fireballs are dynamically linked to Jupiter. Entirely all of the sporadic fireballs were interlopers from the main belt. In this study, we aim to build upon the foundational work established in \citet{shober2021main} by examining the orbital stability of fireballs and the JFC comet population more extensively. This direct comparison of the orbital dynamics is pivotal for a conclusive source region analysis. Our expanded work includes the examination of 646 fireball orbits and 661 JFC orbits. This massive comparison of fireball data and the JFC population and their orbital stabilities on 10,000\,yr timescales will provide a clear picture of what is happening at the centimeter- and meter-size range in this region of the solar system.

\section{Materials and methods}
\subsection{Overview}
We employed a multifaceted approach to identify the sources of meteoroids and comets on orbits with Tisserand's parameters of \(2 < T_{J} < 3\) in near-Earth space. This research is characterized by its extensive collection of fireball data from four separate continental-scale fireball observation networks, combined with the ephemeris data of JFCs. This methodology is notable for its direct comparison of the dynamics of meteoroid and comet populations within a single study, which has not been explored in previous research.

Our methods encompass a range of techniques that focus on determining the source populations of these objects based on their orbits and dynamics. The scale of data analysis and the direct comparison of meteoroid and comet orbit dynamics are key aspects that distinguish this study. These elements contribute to a more comprehensive understanding of these objects in near-Earth space and provide valuable insights. The methodology comprises the following steps:
1. Collection and processing of data on fireballs and comets.
2. Identification of fireballs that qualify as asteroids in cometary orbits (ACOs), using the criterion defined by \citet{tancredi2014criterion}.
3. Monte Carlo integration of meteoroid and comet orbits over a span of 10,000 years.
4. Estimation of Lyapunov lifetimes for meteoroids and comets.
5. Calculation of debiased source regions for Near-Earth Objects (NEOs) of meteoroids and comets.
6. Identification of meteorite falls.
7. Identification of meteor showers.
These enumerated steps underscore our comprehensive approach and provide a clear summary of our methodologies, contributing to a deeper understanding of the dynamics of these objects in near-Earth space.

The dynamics of the comets and meteoroids in this study are characterized based on the 10\,kyr simulations, Lyapunov lifetimes, and source region likelihood based on a debiased NEO model. On the other hand, the meteorite fall and meteor shower analysis helps us better understand how the dynamics relate to the populations and what they can tell us about the meteorites we find on Earth.  

\subsection{Data collection and processing}
This study has collected the largest fireball dataset of objects originating from orbits with a Tisserand's parameter of $2.0 < T_{J} - 3 \sigma$ and $3.0 > T_{J} + 3 \sigma$, and has completed a detailed analysis of the stability and dynamics of this population. The data used here comes from four continental-scale fireball networks: Desert Fireball Network (DFN), European Fireball Network (EFN), Fireball Recovery and InterPlanetary Observation Network (FRIPON), and Meteorite Observation and Recovery Project (MORP). The data for the DFN and FRIPON were accessed as the authors of this study are collaborators on the projects. Given the reduced quality of data for less observed events, only FRIPON fireballs with at least four observations were considered. In contrast, the data used within this study from the EFN and MORP were taken from previously published datasets. Except for the MORP network, all the networks are still active and continually making fireball observations. 

Slight variations existing between the fireball datasets are noted and discussed within the study (orbits, velocity uncertainties, etc.), however, the source of these variations is beyond the scope of this individual study. Ongoing work is being conducted on comparing the data reduction pipelines and methodologies used by several of the largest meteor physics research groups in order to ensure accurate interpretations of meteor and fireball data \citep{shober2023comparison}.

\begin{figure*}[]
\centering
\includegraphics[width=\textwidth]{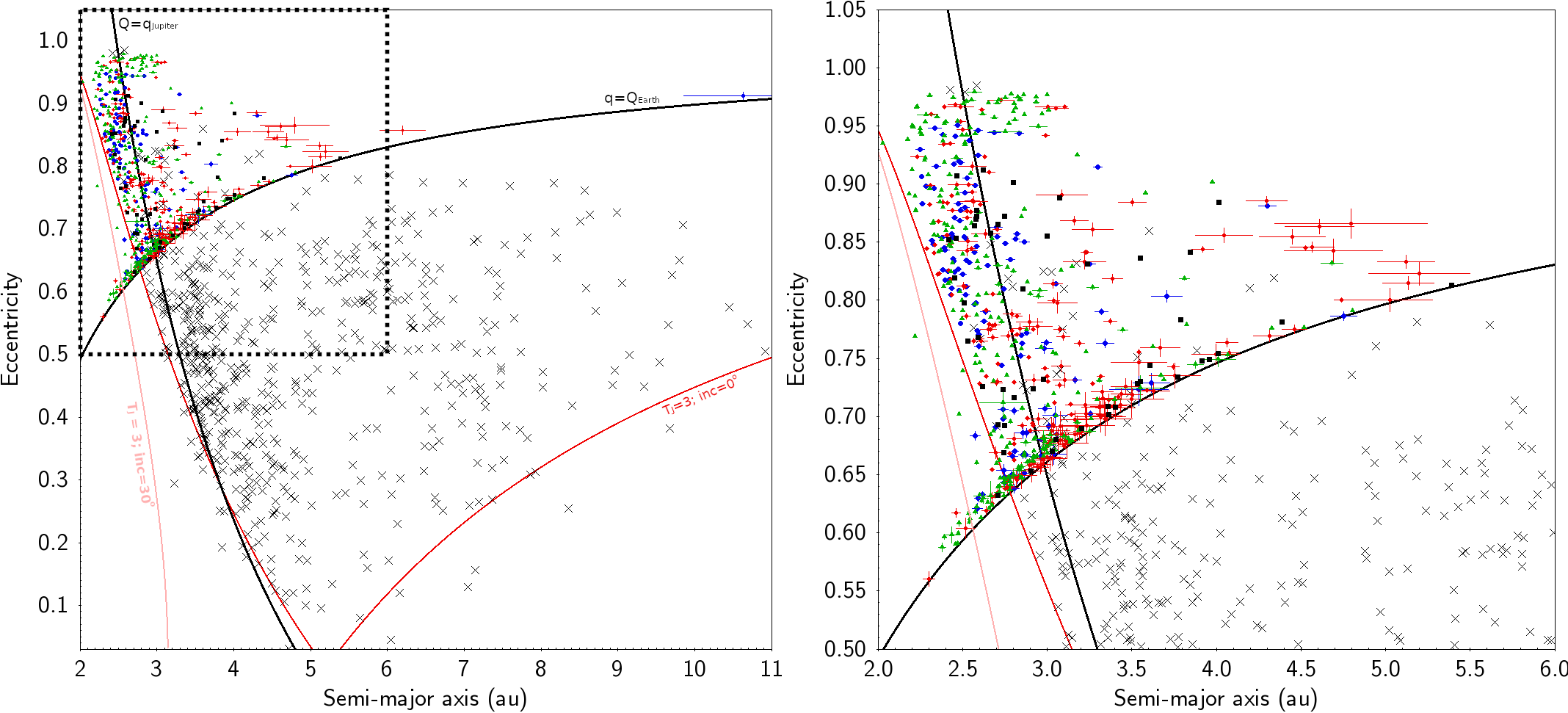} 
\caption{Semi-major axis (au) versus eccentricity distribution of fireballs on JFC-like orbits (i.e., $2<T_{J}<3$), where the plot on the right is the dashed-zone on the left plot. The fireball observations for the DFN (blue dots), EFN (red diamonds), FRIPON (green triangle), and MORP (black squares) are shown. The horizontal and vertical bars of the fireball data indicate the 1-$\sigma$ uncertainties in the semi-major axis and eccentricity. In addition, the 661 JFCs used in this study taken from NASA's HORIZONS database are also shown as black crosses.}
\label{fig:ae_all} 
\end{figure*}

\begin{figure}[]
\centering
\includegraphics[width=0.5\textwidth]{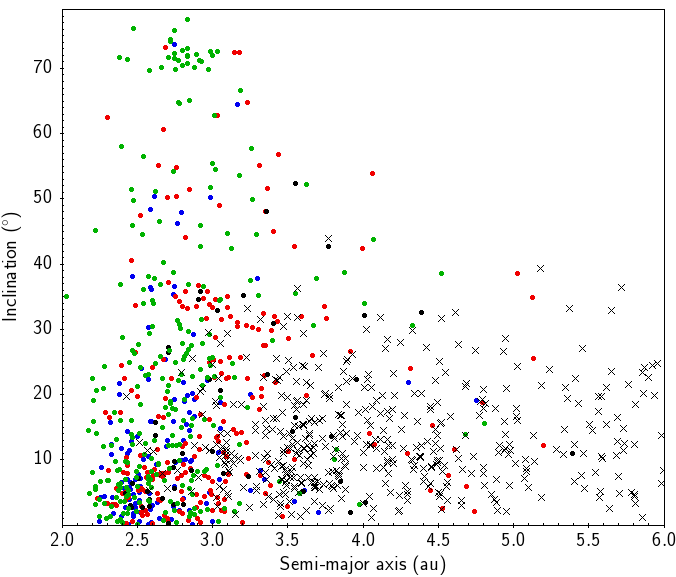} 
\caption{Semi-major axis (au) versus inclination distribution of fireballs on JFC-like orbits (i.e., $2<T_{J}<3$). The fireball observations for the DFN (blue), EFN (red), FRIPON (green), and MORP (black) are shown. In addition, the 661 JFCs used in this study taken from NASA's HORIZONS database are also shown as black crosses.}
\label{fig:ai_all} 
\end{figure}

\subsubsection{DFN}
 Desert Fireball Network (DFN) is a massive network of cameras covering around 2.5 million km$^{2}$ of the Australian outback \citep{bland2012australian,howie2017build}. This is over one-third of the Australian landmass, making the DFN the largest contiguous photographic fireball network globally. Initiated with a robust focus on reliability and autonomy, the DFN began assembling its first digital prototypes in 2013, leading to a final design rolled out in 2014–2015 \citep{howie2017build}. This expansion resulted in the establishment of approximately 50 digital fireball stations. Technologically advanced, the DFN utilizes high-resolution digital single-lens reflex (DSLR) cameras paired with all-sky fish-eye lenses. This combination allows for long exposure captures, enhanced by GNSS synchronized liquid crystal shutters, a method proven instrumental for accurate meteorite fall position determinations \citep{howie2017submillisecond,devillepoix2019observation}. The observatories exhibit a limiting magnitude of zero, allowing the capture of meteoroids as small as approximately 5 centimeters while they are still at higher altitudes before significant atmospheric deceleration occurs. This sensitivity also enables the observation of the brightest ablation phases of larger, half-meter-sized meteoroids despite some sensor saturation challenges.

Strategically situated in desert locales, the DFN optimizes the likelihood of meteorite recovery by being positioned in favorable regions for meteorite searching. Since its inception, the network and its team have helped lead to the recovery of 15\,meteorites ($\sim$30\% of total recovered with orbit information) \citep{King_Winchcombe2022SciA,shober2022arpu,Devillepoix_Madura_Cave,Anderson2022ApJ}. The DFN has since expanded globally, forming the Global Fireball Observatory (GFO) \citep{devillepoix2020global}. The GFO\footnote{\url{https://gfo.rocks/}} consists of ten partner networks and 18 collaborating institutions spread over nine countries globally, all using observatories originally developed by the DFN.

Atmospheric trajectories of fireball events captured by the DFN are determined by employing a refined straight-line least squares (SLLS) approach and a velocity profile is extracted using an extended Kalman smoother \citep{Borovicka_1990BAICz,sansom2015novel}. This methodology accounts for the uncertainties in both the trajectory and the velocity data, stemming from the observational errors and the fitting process. The Kalman filter plays a critical role in managing these uncertainties and helping the uncertainties be realistic. For calculating pre-entry orbits, the meteoroids' states are numerically integrated beyond Earth's gravitational influence, considering all significant perturbative effects \citep{jansen2019comparing}. A Monte Carlo technique is applied to estimate the associated orbital uncertainties, which integrates various initial state samples within the uncertainty bounds at the entry point of the Earth’s atmosphere \citep{shober2019identification,shober2020did}.

\subsubsection{FRIPON}
FRIPON\footnote{\url{https://www.fripon.org/}} is a ground-breaking international initiative aimed at the meticulous monitoring and recovery of meteoroids and meteorites, enhancing our understanding of interplanetary matter and the evolution of the solar system. The FRIPON consortium started in 2015 in France by deploying fully automated cameras and radio receivers \citep{colas2020fripon}. In the years that followed, many neighboring nations joined the consortium, combining the data from over 15 countries on 4 continents, covering an area of $\sim$2 million km$^{2}$ with over 250 cameras and 40 radio receivers. The cameras are wide-angle CCDs recording at 30 frames per second, producing high-time-resolution data. The detections and underlying data are openly available on the consortium webpage \footnote{\url{https://fireball.fripon.org}}. The limiting magnitude of any single frame is about zero, but longer exposures are taken every 10 minutes to acquire better astrometry and photometry, resulting in a good signal-to-noise ratio (S/N) up to a stellar magnitude of 6 for dark sites \citep{anghel2019photometric,jeanne2019calibration,jeanne2020methode,colas2020fripon}. FRIPON’s automation facilitates rapid response for meteorite recovery campaigns, aiming to recover meteorites with final masses estimated at 500 grams or more. Its international expansion and collaboration have further bolstered the network’s coverage and capabilities, heralding a new era of meteorite recovery and interplanetary observation. The network and team have assisted in the tracking and recovery of seven meteorites to date. 

The standard data reduction process of FRIPON observations is described in \citet{jeanne2019calibration}, \citet{jeanne2020methode}, and \citet{colas2020fripon}. However, within this study, we decided to process the FRIPON observations using the Monte Carlo triangulation approach described by \citet{vida2020estimating}. The method begins with an initial trajectory estimation using intersection points (IP) and lines of sight (LoS) methods to fit observed meteor paths, using the angular residuals as a direct measure of uncertainty \citep{Gural_2012M&PS,Weryk_Brown_2012P&SS}. These residuals inform the generation of Monte Carlo simulations, wherein Gaussian noise is added to the observations, and the trajectory is recalculated to produce a variety of geometrically plausible solutions within the bounds of measurement uncertainty. A key advancement of this method is the selection of the best solution based on the consistency of observed meteor dynamics across different stations, rather than relying solely on a geometric fit. The open-source Python code\footnote{\url{https://github.com/wmpg/WesternMeteorPyLib}} for the simulation and solver is available for public use, which increases the transparency and reproducibility of the method.

\subsubsection{EFN}
 Established in 1963, EFN\ has since evolved tremendously over the decades, adapting various technological advancements to improve its meteor detection capabilities \citep{borovivcka2022_one}. Over the several decades of operation, the network and team have aided in the recovery of at least 13 meteorites \citep{spurny2003photographic,spurny2013trajectory,spurny2017ejby,spurny2020vzvdar}. The network primarily utilizes digital autonomous fireball observers (DAFOs), which are weather-proof, fully autonomous systems, to capture images of the entire sky continuously during favorable weather conditions. Each DAFO employs DSLR cameras equipped with fisheye lenses, similar to the DFN, providing images with remarkable clarity and detail. Regarding sensitivity and detection capabilities, EFN's cameras are configured to collect digital images of meteors brighter than an absolute magnitude of about -2, and they can perform high-temporal-resolution radiometric measurements of those brighter than a magnitude of approximately -4 \citep{borovivcka2022_one}. The EFN is proficient in detecting meteoroids larger than 5 grams and can capture high-velocity meteoroids down to masses of about 0.1\,grams. The EFN comprises 26 stations equipped with DAFOs spread across central Europe, covering about one million km$^{2}$. 

The EFN data used within this study was published and described in \citet{borovivcka2022_one} and \citet{borovivcka2022_two}. The fireball's atmospheric trajectory is deduced using the SLLS method, assuming a straight-line trajectory in space \citep{Borovicka_1990BAICz}, similar to the DFN. However, the velocity method differs from the DFN. Time data are projected onto a computed SLLS trajectory and fitted with a physical four-parameter function that accounts for atmospheric drag and ablation \citep{pecina1983BAICz..34..102P}. This function includes the preatmospheric velocity, ablation coefficient, and mass-related parameters, and it is fine-tuned using atmospheric density models like CIRA72 or NRLMSISE-00. When deceleration is significant, manual adjustments or alternative models may be employed. Outliers are carefully weighed or excluded to ensure accuracy, and systematic discrepancies between cameras are resolved before finalizing the data. The heliocentric orbits are then calculated with a slightly modified version of the method described in \citet{Ceplecha_1987}, correcting for Earth's rotation and gravity. 

\subsubsection{MORP}
MORP was an operational fireball network in Western Canada from 1971 to 1985 \citep{halliday1996detailed}. It was a landmark initiative to observe bright meteors and recover meteorites to discern their pre-collision orbits. MORP utilized a network of 60 cameras distributed across 12 stations, enabling near-comprehensive sky coverage. MORP successfully recorded over 1000 fireballs throughout its operational years, contributing significantly to meteoroid orbit research. However, only a small fraction (285 in total) of these fireballs had orbits reduced and published \citep{campbell2004new}. A milestone achievement was the recovery and analysis of the Innisfree meteorite, marking a pivotal success in linking meteorites to their original orbits \citep{halliday1978innisfree}. However, the project faced several challenges, such as lost original measurements and uncertainties in time determinations, which somewhat affected the precision and comprehensiveness of the data. Despite these hurdles, MORP’s legacy remains important in the meteor science research landscape. The MORP orbital data used within this study was taken from \citet{halliday1996detailed}. 

\subsubsection{Case studies}
In addition to the fireballs observed by DFN, EFN, FRIPON, and MORP, we also completed a stability analysis of three individual fireballs claimed to have originated from the JFC population. The first two of these fireballs were recorded by another fireball network: Spanish Meteor Network (SPMN). The first SPMN event was described in \citep{madiedo2014trajectory}, where they observed a $-13.0\pm0.5$ fireball over the center of Spain. The second event, also observed by the SPMN in 2008, similarly reached an absolute magnitude of $-18\pm1$ and the authors claimed the fireball could have produced meteorites \citep{trigo2009observations}. The third event was described in the study of \citet{hughes2022analysis}, where a very bright fireball of peak magnitude of -19 was detected to impact over the eastern coast of Florida. 

\subsubsection{Jupiter-family comets}
In addition to the fireball data, we also performed the same exact analysis on 661 JFCs, extracted from the NASA HORIZONS\footnote{\url{https://ssd.jpl.nasa.gov/horizons/}} database, excluding only the cometary fragments from the study. Each object has been given a comet designation and fulfills the \citet{carusi1985long} Tisserand parameter JFC definition. If the JFC population is producing the debris we observe with fireball networks around the globe, the dynamics of their orbits should be extremely similar. All ephemeris information of JFCs available through the HORIZONS databases were used, excluding cometary fragments. \\

\noindent 
All fireball and JFC data used in this study is openly available\footnote{\url{https://doi.org/10.5281/zenodo.10671333}}. 

\subsection{Orbital analysis}

Asteroidal and cometary orbits are traditionally classified based on Tisserand's parameter (\(T_J\)), a widely used criterion introduced by \citet{kresak1979} and further supported by subsequent studies \citep{carusi1985long,levison1994long,levison1996taxonomy}. According to this classification, minor bodies are categorized based on their \(T_J\) values. Asteroids are defined as having \(T_J > 3\). On the other hand, comets are classified as having \(T_J \le 3\), with further subdivisions into Jupiter-family comets (JFCs) if \(2 < T_J \le 3\) and Halley-type comets if \(T_J \le 2\). This classification helps in understanding the dynamic properties and evolutionary trajectories of these minor celestial bodies.

However, there are problems with this simple method to classify objects, since there are some border cases: objects that behave dynamically as comets but have not shown any type of activity (asteroids in cometary orbits, ACOs), and objects in asteroidal orbits that present evidence of gas and/or dust ejections (main belt comets, MBCs; or active asteroids). \citet{tancredi2014criterion} (hereafter, T14) considered it necessary to develop a more adequate orbital criterion for the distinction between asteroids and comets, and to determine the border cases. The criterion is based on Tisserand's parameter and the minimum orbital intersection distance (MOID); it also considers information on perihelic and aphelic distances of the objects and the location of the resonances. A summarized version of the criterion is presented here. We recommend reading the original article to understand the classification scheme's details. According to this scheme, there are four classes of periodic comets: Halley-type comets, Jupiter-family comets, “Comets” in asteroidal orbits (ACOs), and centaur comets. On the other hand, ACOs are grouped into three categories: ACO-Jupiter family types (AJFs), Centaur asteroids (CAs), and ACO-Halley types (AHTs). Moreover, ACOs in the Jupiter Family have $2 < T_J < 3.05$, $q < Q_J$, are not in resonances, and have low MOID respect to the giant planets; Centaur asteroids have $T_J > 2$ and $Q_J < q < a_{Ura}$; and Halley-type ACOs have $T_J < 2$; where $q$ is the object perihelion, $Q_J$ is the Jupiter aphelion, and $a_{Ura}$ is the Uranus semi-major axis. We will use the traditional simple criterion and this other one and compare the results for the population of fireballs.

\subsection{Dynamic analysis}
The diagnostic feature of an object originating from the JFC population is its frequency of chaotic close encounters with the gas giant Jupiter, on timescales of centuries to millennia \citep{tancredi1992evolution,fernandez2015jupiter}. These encounters characterize the JFCs and lead to extremely short predictability of the orbits (typically less than 50-150\,yrs; \citealp{tancredi1995dynamical,tancredi1998chaotic,tancredi2014criterion}). Conversely, small bodies originating from the main asteroid belt on similar orbits rarely have close encounters with Jupiter, giving them fairly predictable orbits on the same 10,000\,yr timescales \citep{tancredi2014criterion}.

Within this study, we replicated the orbital analysis of \citet{shober2021main} and expanded upon the work of \citet{fernandez2015jupiter}. Using the IAS15 integrator implemented by the REBOUND python module \citep{rebound2012A&A...537A.128R,rein2015ias15}, each fireball orbit and JFC from the HORIZONS database had their trajectory integrated backward 10,000\,yrs. The IAS15, a high-order, adaptive timestep algorithm, ensures exceptional accuracy, a crucial attribute for modeling the intricate and sensitive dynamics of JFC-like debris. This high accuracy is imperative given that minor deviations in initial conditions or perturbative forces can yield substantial divergences in orbital evolution over time. Crucially, the IAS15's proficiency in handling close planetary encounters is of paramount importance. JFCs frequently undergo close approaches with Jupiter and the terrestrial planets, necessitating an integrator capable of accurately resolving these encounters without significant loss of precision. The adaptive timestep mechanism of IAS15 dynamically adjusts time steps in response to varying dynamical conditions, a feature that is particularly beneficial in the simulation of planet-encountering objects where rapid changes in gravitational forces are common. 

We generated 1000 clones  within the published orbital uncertainties of the EFN, FRIPON, and DFN fireball events. Covariance matrices were not used to generate the clones within the simulations as the DFN, EFN, and MORP did not output matrices. The diagonal elements were used to initialize the Monte Carlo simulations, giving an overall uncertainty volume slightly larger than and enclosing, the actual one. The same was done for the JFCs taken from the NASA HORIZONS service, with 1000 clones generated within the given observational uncertainties. The MORP fireballs dataset had no listed uncertainties; therefore, an assumption had to be made about the uncertainties of the orbital data of the historical dataset. The MORP orbital simulations were initiated with 1-sigma uncertainties corresponding to 0.01\,au, 0.001, 0.1$^{\circ}$, 0.01$^{\circ}$, 0.01$^{\circ}$ for the semi-major axis, eccentricity, inclination, longitude of ascending node, and argument of perihelion. Each simulation included the gravitational influence of the Sun, the planets (Mercury through Neptune), and the Moon. The planetary ephemeris was obtained from online solar system data and ephemeris computation service NASA HORIZONS, and the simulations were initiated at the beginning of the fireball observation. The results of the integrations can be accessed through the supplementary materials. 

The 10,000\,yr orbital simulations performed using REBOUND recorded the positions and osculating heliocentric orbital elements at each time step. In addition, the minimum orbital intersection distance (MOID) with Jupiter and Saturn were also calculated at each timestep using the technique described in \citet{baluev2019fast}. The MOID value indicates the minimum distance between the two Keplerian orbits, but it is important to remember that the bodies may never reach this distance in reality due to MMRs and perturbations. The methodology of \citet{baluev2019fast} produces accurate and fast results with its algebraic approach, solving a 16th-order polynomial to determine all the critical points of the distance function. The speed and accuracy of the MOID algorithm was necessary to process all 1000\,timesteps of the 646,000 particles integrated within the simulations. 

\subsubsection{Assessing orbital instability}

JFC orbits can be analyzed through a mean dynamical path derived from a set of outcomes for a specific comet and its multiple clones. To quantify the instability of these orbits, certain metrics are utilized:

Initially proposed by \citet{fernandez2014assessing}, the $f_{q}$ index quantifies the instability by measuring the duration, in the preceding 10,000 years, that the clones generated within the observational uncertainties of a JFC have a perihelion distance greater than 2.5\,au. This distance is significant because comets predominantly remain beyond this limit under Jupiter’s gravitational influence, rarely approaching the Sun, in agreement with many previous studies \citep{fernandez2006population,fernandez2014assessing,fernandez2015jupiter}. The metric is mathematically represented as:

\begin{equation}
    f_{q} = \frac{\sum_{j=1}^{N}\Delta t_{j}}{N \times 10^{4}}
,\end{equation}

where $\Delta t_{j}$ signifies the time duration (in years) with the specified orbital conditions among the $N$ clones (here, $N=1000$). 

The \textit{fa} index, also formulated by \citet{fernandez2014assessing}, reflects the fraction of the past 10,000 years during which a JFC or its clones maintain a semi-major axis exceeding 7.37\,au. Expressed mathematically:

\begin{equation}
    f_{a} = \frac{\sum_{j=1}^{N}\Delta t'_{j}}{N \times 10^{4}}
.\end{equation}

An observed correlation between $f_{q}$ and $f_{a}$ indices indicates that comets with higher values of these indices typically follow unstable orbits, spending significant time outside the defined thresholds. 

Another metric we introduce here is the percentage of the particle clones that reach at least a perihelion value of 2.5\,au. This criterion is useful and meaningful in addition to the $f_{q}$ statistic introduced by \citet{fernandez2014assessing}. Instead of being weighted by the proportion of time spent with $q>$\,2.5\,au, this metric is only concerned with the likelihood that any event reaches $q>$\,2.5\,au within the 10,000\, year integration.

In \citet{shober2021main}, the variation in the orbital elements and close encounters with Jupiter during the orbital integrations were manually assessed. This is an intensive process of evaluating the orbital histories of every particle and is only effective if the distinction between the dynamics is extremely drastic. For the 50 events analyzed, which were mostly confined to smaller semi-major axis values, this was true. However, for larger datasets, and those with more data well beyond 3\,au, the dynamic differences become less and less obvious. Therefore, in this study, a manual reduction of the data was performed, as described in \citet{shober2021main}, to demonstrate how the $f_{q}$, $f_{a}$, and the likelihood of $q>$\,2.5\,au metrics produce similar results but are more informative and scale-able in comparison. 

\subsubsection{Calculation of Lyapunov lifetimes} 
In addition to the 10,000\,yr orbital simulations to characterize the prevalence of close encounters with Jupiter, a separate set of simulations was done to estimate the Lyapunov characteristic exponents of the orbits. These simulations were also conducted using the REBOUND architecture. However, using the WHfast integrator with a 0.01\,yr timestep \citep{rein2015whfast} for 20,000\,yrs. This timescale is consistent with previous studies \cite{tancredi1995dynamical,tancredi1998chaotic}, and was also found here to be a good duration that produced reproducible results based on our testing of integrations between 1000 and 25,000\,yrs. 

The REBOUND integration module employs a symplectic integrator tailored for gravitational dynamics to calculate the Lyapunov Characteristic Exponent (LCE), a quantitative measure of chaos in dynamical systems. The methodology involves integrating the equations of motion for the system alongside a variational equation representing a small deviation from the initial conditions. REBOUND tracks the divergence of this deviation over time, providing a direct measure of the system's sensitivity to initial conditions. The LCE is then computed by averaging the exponential growth rate of this divergence. The accuracy and reliability of the LCE as computed by REBOUND have been validated against established benchmarks in the field, confirming its efficacy in characterizing the dynamical stability of celestial configurations \citep{rein2015ias15,rein2015whfast}.

\subsubsection{Debiased NEO model source region estimate} 
The fireball pre-encounter orbits were also compared to the debiased NEOMOD\footnote{\url{https://www.boulder.swri.edu/~davidn/NEOMOD_Simulator/}} model to compare the $10^{4}$\,yr dynamics of these objects to the long-term dynamics of NEOs on similar orbits. \citet{2023AJ....166...55N} developed a novel orbital model for NEOs, integrating asteroid orbits from the main belt and aligning these with NEO observations from the Catalina Sky Survey (CSS). This model represents NEOs as a transitional population originating from the main asteroid belt and trans-Neptunian scattered disk. Their lifecycle concludes either via planetary impact, solar disintegration, or ejection from the solar system. The study utilized the \textit{astorb.dat}\footnote{\url{https://asteroid.lowell.edu/main/astorb/}} catalog from the Lowell Observatory for initial orbit setup of main-belt asteroids, which, as of early 2022, contained nearly 1.2 million entries. The main-belt asteroids evolved into resonances by the Yarkovsky thermal effect, although the exact mechanism of their entry into these resonances was deemed less significant for the NEO orbital distribution. Orbital elements for the eight planets (Mercury to Neptune) were obtained from NASA/JPL Horizons database. Using the Swift rmvs4 N-body integrator, the orbital evolution of planets and 105 test bodies per source was followed. The integrations utilized 2000 Ivy Bridge cores of NASA's Pleiades Supercomputer, each core handling eight planets and 50 test bodies. The initial integrations ran for 100 Myr, with NEO orbits extended to 500 Myr. The study found the contribution of long-lived NEOs beyond 500 Myr to be insignificant. 

The CSS data, comprising nearly 22,000 NEO detections from 2005-2012, provided a basis for calibration. The survey's pointing history and detection efficiency were crucial for understanding the observational bias. The detection probability of an object in a CSS Field of View (FOV) was broken down into geometric probability, photometric probability, and trailing loss. The study employed MultiNest for model selection, parameter estimation, and error analysis. This Bayesian evidence-based approach helped in dealing with high-dimensional parameter spaces and multiple posterior modes. The log-likelihood definition in MultiNest was used to compute the joint probability over all bins, considering detection probability and the number of objects detected by CSS. The model considered bin sizes and ranges as follows: semi-major axis (\textit{a}) from 0 to 4.2\,au in 42 bins of 0.1\,au each, eccentricity (\textit{e}) from 0 to 1 in 20 bins of 0.05 each, inclination (\textit{i}) from 0 to 88$^{\circ}$ in 22 bins of 4$^{\circ}$ each, and absolute magnitude (\textit{H}) from 15 to 25 in 40 bins of 0.25 each \citep{2023AJ....166...55N}.

For our study, we found the corresponding bin within the NEOMOD model for each of the JFC-like fireballs considered here to estimate the source region likelihood. The \textit{H} value of 25.0 was used in the model for the fireball data as it was the highest absolute magnitude (smallest object) defined in the model. For an object with a geometric albedo of 0.25, this roughly translates to an object of $\sim$27\,m in diameter, i.e., at least one order of magnitude or more larger than any of the meteoroids analyzed in this study. This caveat may prove to be significant as the source region variations between $>$100\,m asteroids/comets used to calibrate the model and the cm-m meteoroid populations differ, a concern that can be tested here by comparing the JFC source region prediction of the model to ours based on the dynamics of the objects during a 10\,kyr simulation. 

\subsection{Meteorite fall determination}
In this study, the determination of meteorite falls is crucial for discerning the nature of meteoroids on comet-like orbits, particularly those with Tisserand's parameters between 2 and 3. This analysis helps ascertain whether meteoroids from stable or dynamically active orbits are more likely to contribute to meteorite falls on Earth. Given that our data appears to predominantly feature objects from stable orbits \citep{shober2021main}, finding meteorite falls among these provides insight into the source regions of the meteoroids and aids in understanding the compositional diversity of this population. 

The alpha-beta (\(\alpha-\beta\)) methodology is a pivotal approach in the analysis of fireball dynamics and their potential to result in meteorite falls. This methodology, which calculates the ballistic coefficient (\(\alpha\)) and the mass-loss parameter (\(\beta\)), was notably developed and refined in several studies \citep{gritsevich2006extra,gritsevich2007og,lyytinen2016implications,sansom2019determining}. \citet{sansom2019determining} then applied these theoretical constructs, demonstrating the practical utility of the \(\alpha-\beta\) parameters in quickly identifying potential meteorite-dropping events. 

While a theoretical construct, the \(\alpha-\beta\) methodology gains its significance in practical applications, as demonstrated by \citet{sansom2019determining}. They not only elucidated the theoretical underpinnings of these parameters but also applied them in real-world scenarios to identify potential meteorite-dropping events. This application involved analyzing the trajectory and physical properties of observed fireballs to calculate their \(\alpha\) (ballistic coefficient) and \(\beta\) (mass-loss parameter). By doing so, \citet{sansom2019determining} effectively bridged the gap between theoretical meteor dynamics and empirical observations, offering a methodological framework that enhances the predictive accuracy of meteorite fall determinations.

The limiting lines for a 50\,g final mass are based on the methodology of \citet{sansom2019determining} using the following equations for meteoroids of different densities \citep{consolmagno2008significance}. The boundary lines denote the regions below where, given a shape change coefficient ($\mu$), a meteorite of at least 50\,g would survive:

For meteoroids with an ordinary-chondrite density of 3500 kg/m$^{3}$:
\begin{equation}
\ln(\beta) = \ln(13.2 - 3 \ln(\alpha \sin\gamma)), \mu = 0
\label{eq:oc_mu0} 
,\end{equation}
\begin{equation}
\ln(\beta) = \ln(4.4 -  \ln(\alpha \sin\gamma)), \mu = 2/3
\label{eq:oc_mu23} 
.\end{equation}

For meteoroids with a carbonaceous density (2240 kg/m$^{3}$):
\begin{equation}
\ln(\beta) = \ln(14.09 - 3 \ln(\alpha \sin\gamma)), \mu = 0
\label{eq:cc_mu0} 
,\end{equation}
\begin{equation}
\ln(\beta) = \ln(4.7 -  \ln(\alpha \sin\gamma)), \mu = 2/3
\label{eq:cc_mu23} 
.\end{equation}

Here, \(\gamma\) is the entry angle of the meteoroid, and \(\mu\) is the shape change coefficient representing the rotation of a meteoroid body ($0\leq \mu \leq 2/3$). For more comprehensive insights into the development and applications of the \(\alpha-\beta\) methodology, we refer to the following works: \cite{gritsevich2006extra, gritsevich2007og, lyytinen2016implications, sansom2019determining, shober2021main}. The \(\alpha-\beta\) parameters can be calculated from any event with velocity and height data and determine if a meteorite is on the ground\footnote{\url{https://github.com/desertfireballnetwork/alpha_beta_modules}}.

\subsection{Meteor shower identification}
Identifying meteor showers in our dataset serves as a key method for tracing the origin of meteoroids observed entering Earth’s atmosphere, thus providing a window into the dynamic processes shaping their source regions. By correlating atmospheric trajectories and orbital characteristics with known meteor showers and their parent bodies, we can ascertain whether these meteoroids stem from the main asteroid belt, are part of an evolved comet population, or directly originate from the JFCs. This comparative analysis between identified showers and the sporadic background allows us to unravel these meteoroids' dynamical history and potential evolutionary paths, offering profound insights into the mechanisms that govern their delivery to near-Earth space.

Several established methods exist to identify meteor showers or orbital similarity. The conventional approach has been the use of an orbital similarity criterion. The first of these was introduced by \citet{southworth1963statistics}; however, other improved orbital criteria have since been proposed \citep{drummond1981test,jopek1993remarks,jopek2008meteoroid}. 

The criterion chosen to be used in this study is based on a distance function ($D_{N}$) involving four geocentric quantities directly linked to observations proposed by \citet{valsecchi1999meteoroid}. This method, defined in a space with as many dimensions as the number of independently measured physical quantities, differs from the conventional orbital similarity criteria. It is based on the components of the geocentric velocity at the encounter, essential for \"{O}pik's theory of close encounters, and two of the new variables are near-invariant with respect to the principal secular perturbation affecting meteoroid orbits \citep{valsecchi1999meteoroid}. The approach aims to overcome the limitations of previous methods by focusing on quantities that can be computed directly from observed data without relying solely on the derivation of conventional orbital elements.

\section{Results}

\subsection{Orbital distribution \label{orbitaldistribution}}
The incursion of a JFC onto an Earth-crossing orbit is a rare event. Only a small portion of the JFC population exists on Earth-crossing orbits (Fig. \ref{fig:ae_all}). In the NASA HORIZONS database, 72 of the 661 JFCs (excluding fragments) are also NEOs -- and we note that this is an overestimation as there is a significant bias towards the discovery of near-Earth JFCs. Also, the contribution from the main asteroid belt is much more significant, long-lived, and easier to produce. According to \citet{granvik2018debiased}, the JFC component of the near-Earth population is at most 10\% depending on the H-value and assuming that dormancy is a more likely outcome than disintegration. This implies that if the dormancy assumption is incorrect, the JFC component is even less than predicted.

Moreover, only $\sim$30\% of JFCs are predicted to become active bodies, with an orbit below 2.5\,au where the water begins to sublimate \citep{levison1997kuiper}. Additionally, of the bodies that encroach even closer to the sun to near-Earth space (q$<$1.3\,au), they typically only do so for a few thousand years \citep{fernandez2002there}. As seen in Fig. \ref{fig:ae_all}, only a small portion of the JFCs overlap in orbital space with the fireballs detected around the globe. Most of the known JFCs exist on orbits well beyond the orbit of the Earth, mostly concentrated with aphelion values crossing the orbit of Jupiter. The inclination distribution tends to stay $<$30$^{\circ}$, expected for short-period comets originating from the scattered disk (Fig. \ref{fig:ai_all}). A majority of the fireball orbits overlap with the inclination distribution of the JFC population, but this does not signify any link as the JFC distribution is low-inclination. The asteroid population also has low inclinations, typically more concentrated toward the ecliptic plane. Surprisingly, the EFN also seems to contain more events beyond 3\,au compared to the other networks in this study, with a concentration around 3.2-3.5\,au. This concentration could be indicative of the larger proportions of fireballs originating from the 2:1\,MMR within the EFN dataset or some observational bias. The 2:1\,MMR is a wide, powerful resonance centered at 3.28\,au \citep{tancredi2014criterion}.

\begin{figure}[]

\begin{minipage}{0.5\textwidth}
  \centering
  \begin{subfigure}[b]{0.41\textwidth}
    \includegraphics[width=\textwidth]{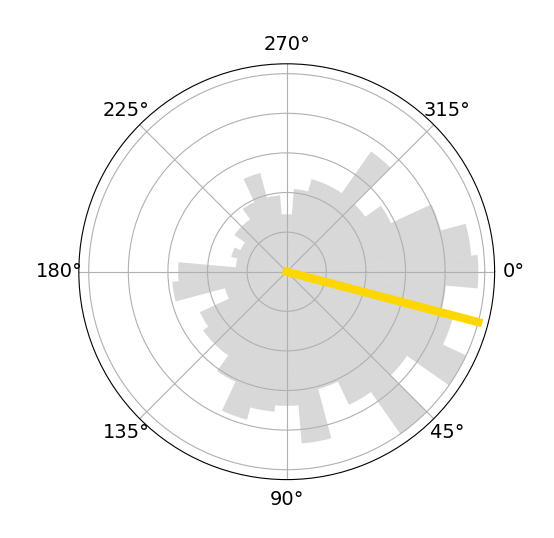}
    \caption{JFCs}
    \label{fig:polar_JFCs}
  \end{subfigure}
  \hspace{10mm}
  \begin{subfigure}[b]{0.41\textwidth}
    \includegraphics[width=\textwidth]{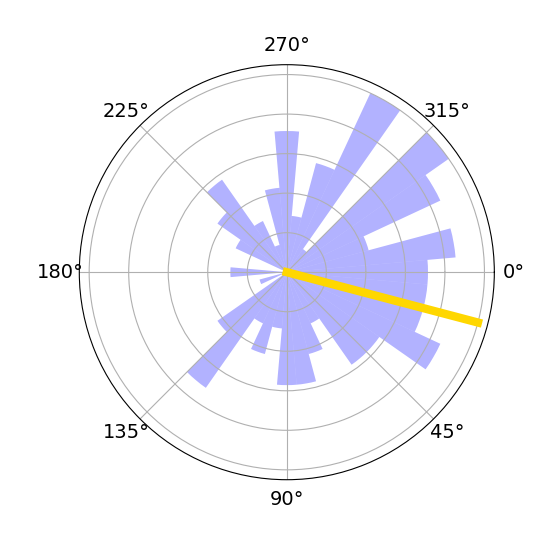}
    \caption{DFN}
    \label{fig:polar_DFN}
  \end{subfigure}
\end{minipage}
\vspace{-10mm} 

\begin{minipage}{0.5\textwidth}
  \centering
  \begin{subfigure}[b]{0.41\textwidth}
    \centering
    \includegraphics[width=\textwidth]{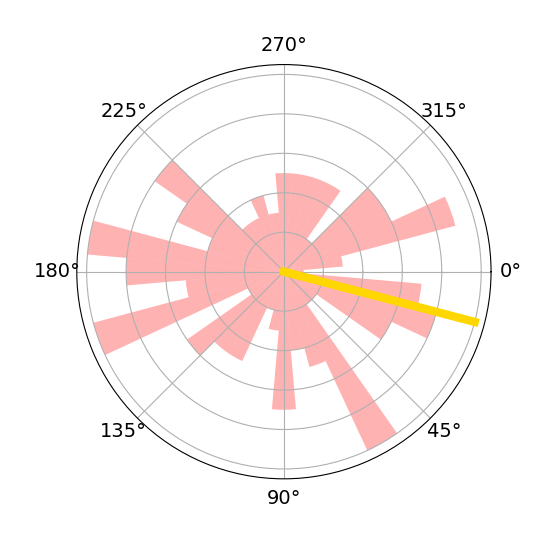}
    \caption{EFN}
    \label{fig:polar_EFN}
  \end{subfigure}
\end{minipage}
\vspace{-10mm} 

\begin{minipage}{0.5\textwidth}
  \centering
  \begin{subfigure}[b]{0.41\textwidth}
    \includegraphics[width=\textwidth]{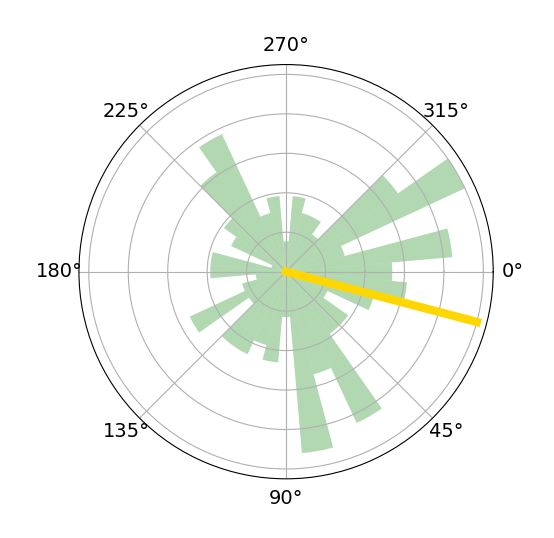}
    \caption{FRIPON}
    \label{fig:polar_FRIPON}
  \end{subfigure}
  \hspace{10mm}
  \begin{subfigure}[b]{0.41\textwidth}
    \includegraphics[width=\textwidth]{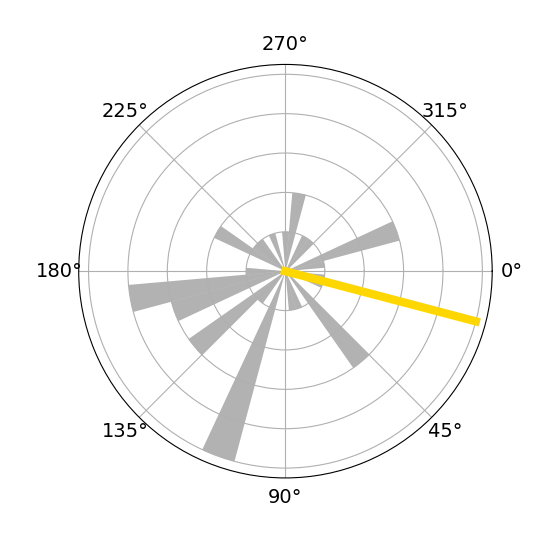}
    \caption{MORP}
    \label{fig:polar_MORP}
  \end{subfigure}
\end{minipage}

\caption{Polar representation of the longitudes of perihelion $\varpi$ for fireballs versus JFCs. The datasets are represented by histograms in unique colors: JFCs (light gray), DFN (blue), EFN (red), FRIPON (green), and MORP (dark gray). A prominent gold line marks Jupiter's longitude of perihelion to demonstrate the concentration of JFC $\varpi$ values towards that of the planet.}
\label{fig:polar_peri}
\end{figure}

There are some inconsistencies in the fireball and JFC populations as we concentrate on the longitude of perihelion. The longitude of perihelion ($\varpi$) is defined as the sum of the argument of perihelion ($\omega$) and the longitude of the ascending node ($\Omega$). For low inclination orbits, it can be thought of as the orientation of the perihelion of an orbit relative to other solar system bodies. The perihelion longitudes of JFCs have long been known to cluster near the $\varpi$ of Jupiter \citep{opik1971comet}, as the encounters are more likely at perihelion and aphelion. In Fig. \ref{fig:polar_peri}, we have plotted a polar histogram of the $\varpi$ values of the sporadic fireballs found within the datasets. The fireball orbits that belonged to a meteor cluster and had a D$_{N}<0.15$ \citep{valsecchi1999meteoroid} were removed from the plot to better compare the general population. A broad concentration of JFCs exists near the value of Jupiter (Fig. \ref{fig:polar_JFCs}). The fireball data, however, mostly shows no correlation except for the DFN, which seems to strongly concentrate towards the planet. Despite this, given the significant variation between the network $\varpi$ distributions, it is difficult to ascertain the significance of this association. 

\begin{figure}[]
    \centering
    \begin{subfigure}[b]{0.45\textwidth}
        \includegraphics[width=\textwidth]{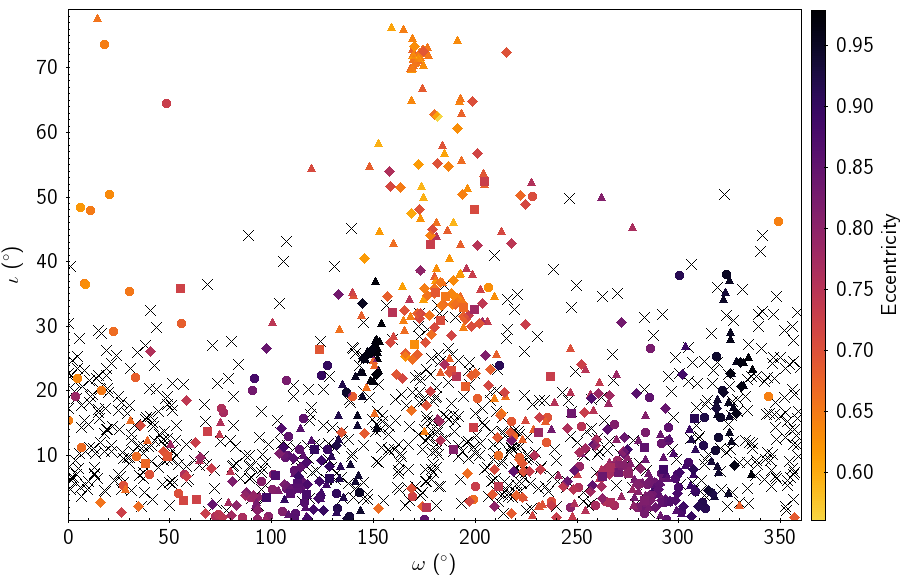}
        \caption{}
        \label{fig:omega_inc_ecc}
    \end{subfigure}
    \vspace{1ex} 
    
    \begin{subfigure}[b]{0.45\textwidth}
        \includegraphics[width=\textwidth]{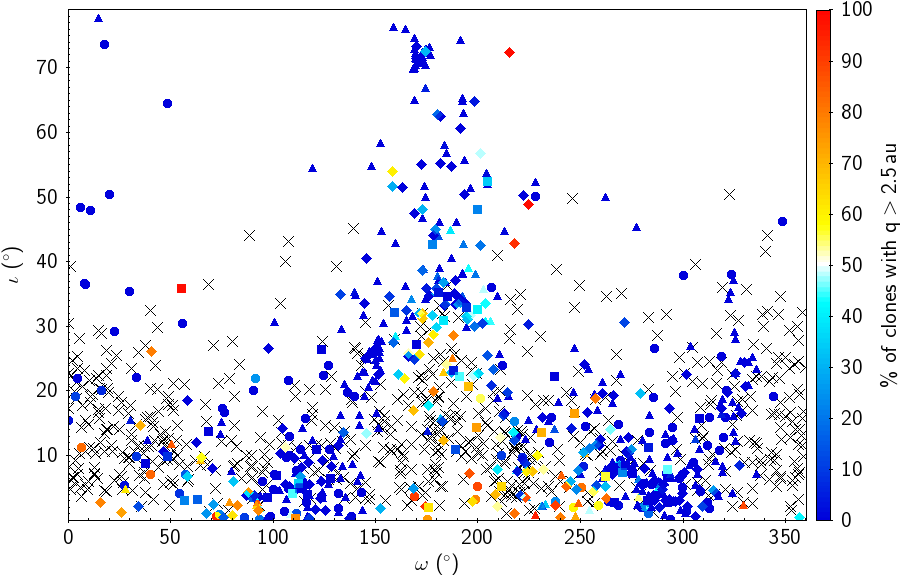}
        \caption{}
        \label{fig:omega_inc_q25}
    \end{subfigure}
    \vspace{1ex} 
    \caption{Argument of perihelion ($\omega$) versus the inclination ($\iota$) of the JFC-like ($2<T_{J}<3$) meteoroid population (cm-m size range) observed by the DFN (circle), EFN (diamond), MORP (square), and FRIPON (triangle). Black crosses denote the 661 JFCs taken from the NASA HORIZONS database. The coloration of subplot (a) is based on the eccentricity of the object, while subplot (b) corresponds to the percentage of particle clones generated within the observational uncertainties of the object that achieved at least a perihelion value of 2.5\,au during a 10\,kyr orbital integration.}
    \label{fig:arg_peri_inc}
\end{figure}

The JFC arguments of perihelion ($\omega$) are also known to concentrate towards 0$^{\circ}$ and 180$^{\circ}$ \citep{dones2015origin} and the inclinations tend to stay $<$30$^{\circ}$ (Fig. \ref{fig:arg_peri_inc}). The 646 fireballs studied here do not show a significant concentration of $\omega$ towards 0$^{\circ}$, and only the northern hemisphere networks show concentrations toward 180$^{\circ}$ (Fig. \ref{fig:arg_peri_inc}). This concentration towards 180$^{\circ}$ is tempting to link to the JFC population distribution, but as can be seen in Fig. \ref{fig:arg_peri_inc}, there is a clear trend toward higher inclinations ($>$30$^{\circ}$) within the fireball dataset when $\omega\sim$\,180$^{\circ}$. Despite the lack of data, there is also a similar trend with only the southern hemisphere fireballs (primarily from the DFN) that trend towards higher inclinations near 0$^{\circ}$. Conversely, the eccentricities tend to decrease for these high-inc portions of the fireball population (as indicated by the color bar in Fig. \ref{fig:arg_peri_inc}). This likely implies that these objects were in a Kozai resonance, and our orbital integrations confirm this. Some showers in the dataset, particularly the high-inc Quandrantid meteor shower (inc$\sim$72$^{\circ}$), support this as other studies have already identified that the meteoroids were moving in a ``Kozai circulation'' \citep{wiegert2005quadrantid}. Otherwise, the high inclinations of these bodies may also be a filtering effect, as the higher inclinations will reduce the amount of time near the equatorial plane and, thus, the chances of a close encounter with Jupiter. This relationship between the probability of an encounter and the inclination is confirmed by our orbital simulations, which (as seen in Fig. \ref{fig:omega_inc_q25}) is higher for the low-inclination meteoroids clustered toward $\omega$ values of 0$^{\circ}$ and 180$^{\circ}$. 

\subsection{Applying the criteria of \citet{tancredi2014criterion}}

For the following analysis, we group all the fireballs detected by the DFN, EFN, FRIPON, and MORP networks. In this section, our focus shifts towards evaluating the efficacy of the simplistic Tisserand's parameter ($T_J$) criterion for JFC identification against the more elaborate classification system proposed by T14. This comparison aims to scrutinize how each criterion fares in discerning cometary from asteroidal origins, especially in light of the orbital integration analyses detailed in subsequent sections.

\begin{figure}[]
    \centering
    \includegraphics[width=0.49\textwidth]{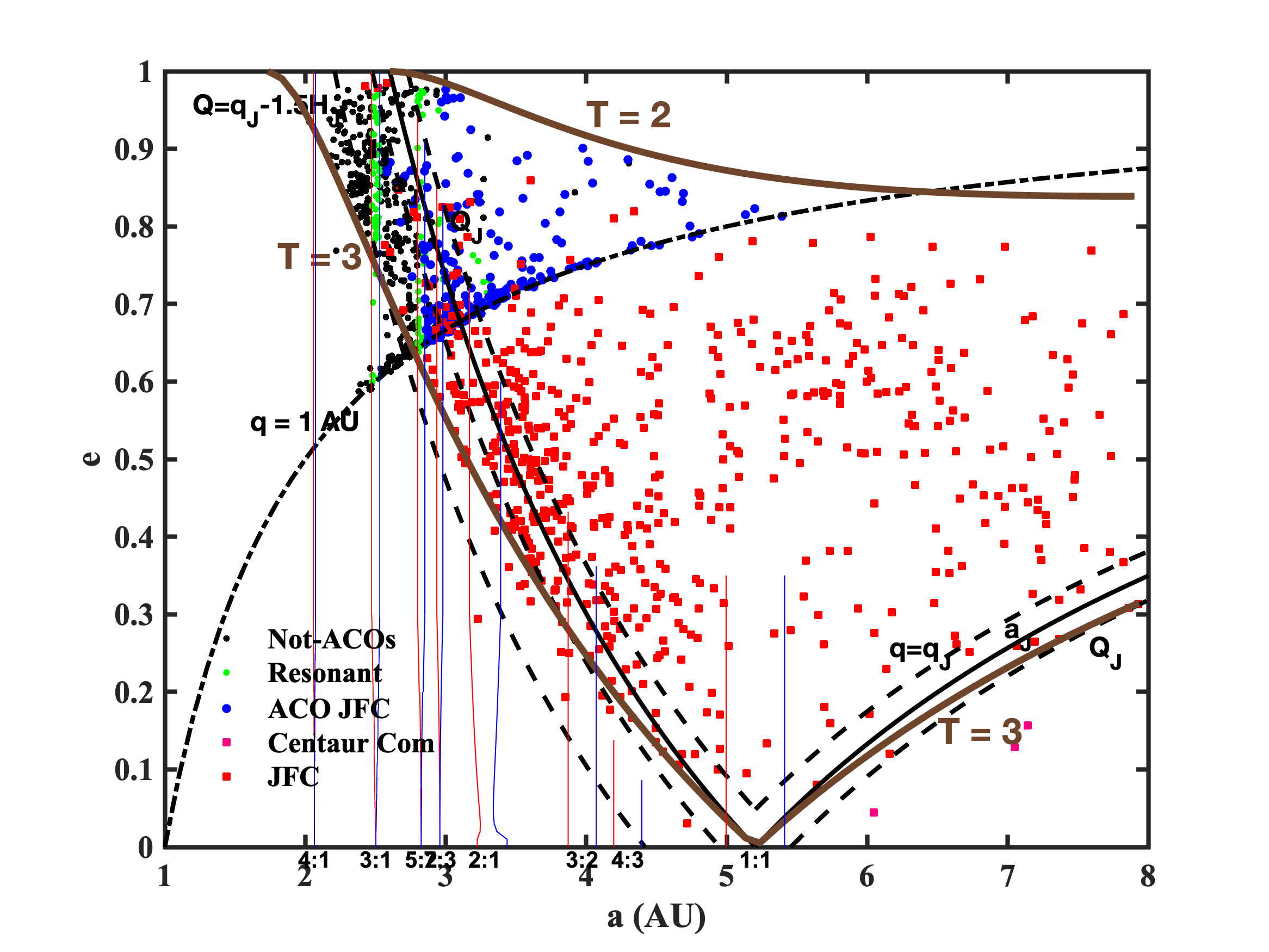}
    \caption{Distribution of fireball and comet orbits in the semi-major axis versus eccentricity space, highlighting the classification according to \citet{tancredi2014criterion} criteria. The data points are: black dots - meteoroids in orbits not similar to cometary orbits, green dots: meteoroids inside resonances, blue dots: meteoroids in Jupiter family orbits, magenta squares: comets in Centaur-type orbits, and red squares: Jupiter family comets. Several relevant lines are drawn.}
    \label{fig:T14_ae_all}
\end{figure}

Our initial dataset comprised 646 fireballs and 661 comets, as defined by having $2 < T_J < 3$. Implementing T14's criteria resulted in the identification of 640 comets within the Jupiter Family and 21 with Centaur-type orbits from the original comet set. Conversely, of the 646 fireballs, 322 were excluded from the ACO classification, and an additional 95 were set aside due to their resonant orbits. Consequently, a substantial 65\% (417 out of 646) of the fireball dataset was deemed non-cometary according to T14's criteria, leaving 230 (35\%) as plausible candidates within the Jupiter family ACO orbits.

Figure~\ref{fig:T14_ae_all} depicts the semi-major axis ($a$) versus eccentricity ($e$) distribution for these groups, accentuated by various markers and lines to delineate the different classifications and relevant dynamical thresholds. The fireballs excluded as cometary candidates (represented by black dots) predominantly occupy regions with $T_J$ values marginally below 3 and feature orbits that do not intersect with Jupiter's path, as indicated by their aphelion distances not surpassing Jupiter's perihelion ($Q \lessapprox q_J$), a delineation refined through the use of minimum orbital intersection distance (MOID) in T14's criteria.

\begin{figure}[]
    \centering
    \includegraphics[width=0.5\textwidth]{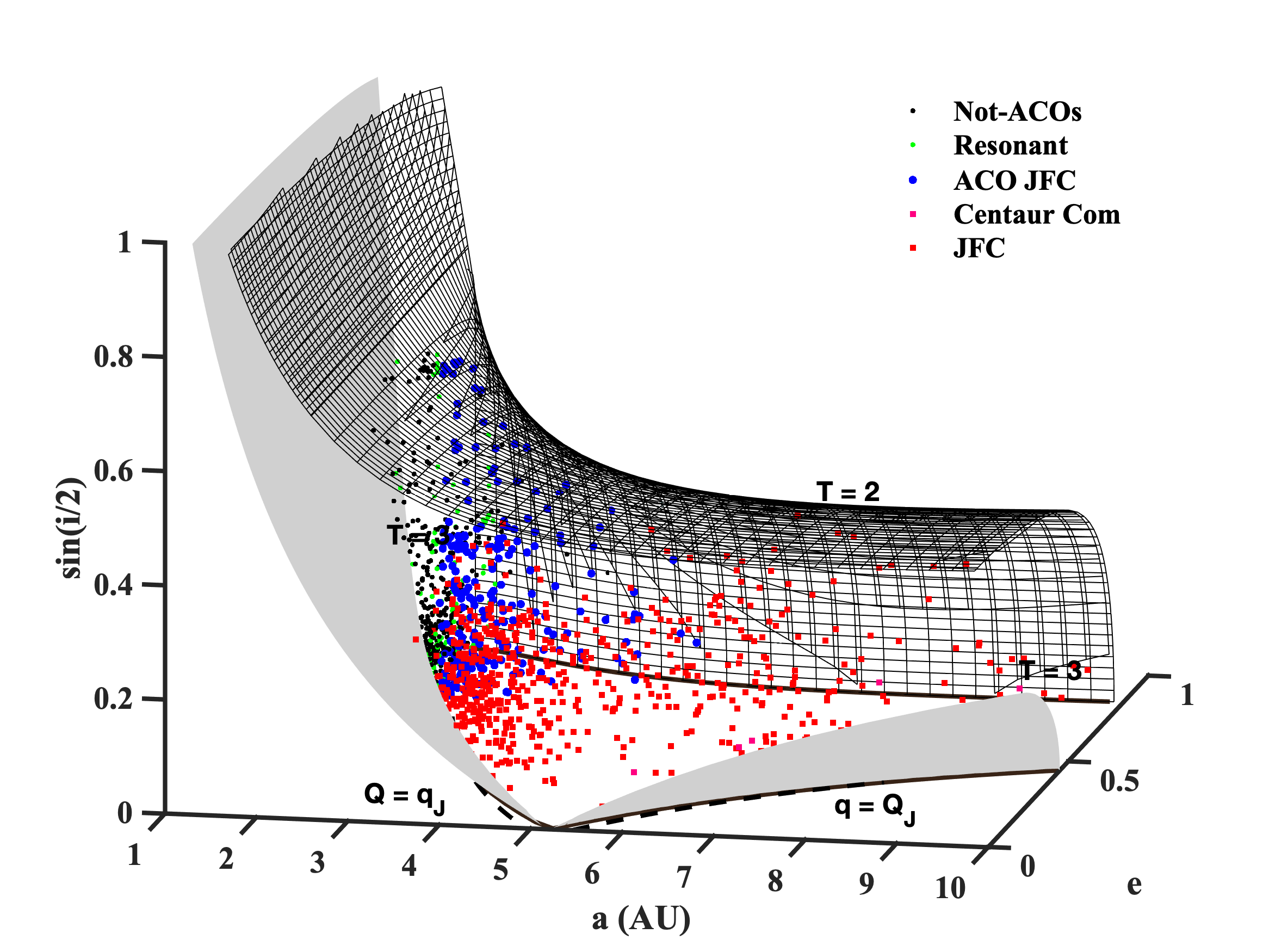}
    \caption{Three-dimensional representation in the $a-e-\sin(i/2)$ space, illustrating the separation between cometary and asteroidal/meteoroidal orbits as per \citet{tancredi2014criterion} criteria. Same symbols as in Fig. \ref{fig:T14_ae_all}.}
    \label{fig:T14_aei_ast_com}
\end{figure}

A clear demarcation between the different object groups emerges within the $a-e-\sin(i/2)$ phase space (Fig.~\ref{fig:T14_aei_ast_com}), showcasing the nuanced separation achieved by T14's classification scheme. Notably, the boundary for $T_J=3$ curves inward for orbits with significant inclinations, excluding them from Jupiter's orbital domain.

The refined analysis provided by T14's criteria markedly narrows down the list of fireball candidates with potential cometary affinities. This subset of fireballs, filtered through the lens of T14's comprehensive classification system, presents a more targeted group for further investigation into their dynamical stability and potential cometary lineage. The ensuing sections will delve into a comparative analysis of these candidates' orbital stability, juxtaposing them with the broader dataset to draw conclusions about their origins and dynamical evolution.

\subsection{Orbital stability}
Simply based on the pre-impact orbital distribution of the DFN, EFN, FRIPON and MORP data, the JFC-like ($2<T_{J}<3$) meteoroids being detected by fireball networks overall do not align well with JFCs. This is further supported by the 10\,kyr orbital integrations of the 661 JFCs and the 646 fireball orbits. 

\subsubsection{Orbital Integrations}
A manual data reduction approach was followed in the study summarized by \citet{shober2021main}. When analyzing the simulations of a handful of objects, the 10\,kyr simulation results were manually examined and the number of JFC-like histories was counted. When we say ``JFC-like histories'', we are describing objects that have several close encounters with Jupiter over a 10,000\,yr timescale, which produce large and unpredictable jumps in the orbital elements (see Fig. 1 from \citet{shober2021main}). This procedure was also undertaken for this study, requiring the analysis and manual interpretation of 646,000\,particle histories. For meteoroids with a semi-major axis still within range of the main belt, the difference between the orbital evolution of an asteroid and a JFC is very clear, and it has been used in several past studies \citep{tancredi2014criterion,fernandez2014assessing,fernandez2015jupiter}. However, as the orbits grow beyond 3\,au, the histories can often begin to defy the simplistic binary classification (``asteroid-like'' vs. ``JFC-like'') due to the gradually increasing chaotic nature of the orbits. The analysis of the dynamics in this study had to evolve from the methodology of \citet{shober2021main} to be more comprehensive, unbiased, and scalable. 

The term "active" is used to describe comets that fall within 2.5\,au of the Sun, as this is approximately when water ice begins to sublimate \citep{levison1997kuiper}. Thus, the rarity of having JFCs $<$2.5\,au is somewhat of a self-fulfilling prophecy because if too many JFCs were below this limit for too long, there would likely not be a steady-state population of active JFCs. After all, the physical lifetimes of these objects are estimated to be only a few thousand years in the inner solar system \citep{fernandez2002there}. Only a very small portion of the JFC population is "active" at any given time, and only $\sim$30\% of the population ever reaches an active phase. Thus, this 2.5\,au criterion was also selected as a metric to compare the fireball population to the JFC population. 

\begin{figure}[]
    \centering
    \includegraphics[width=0.5\textwidth]{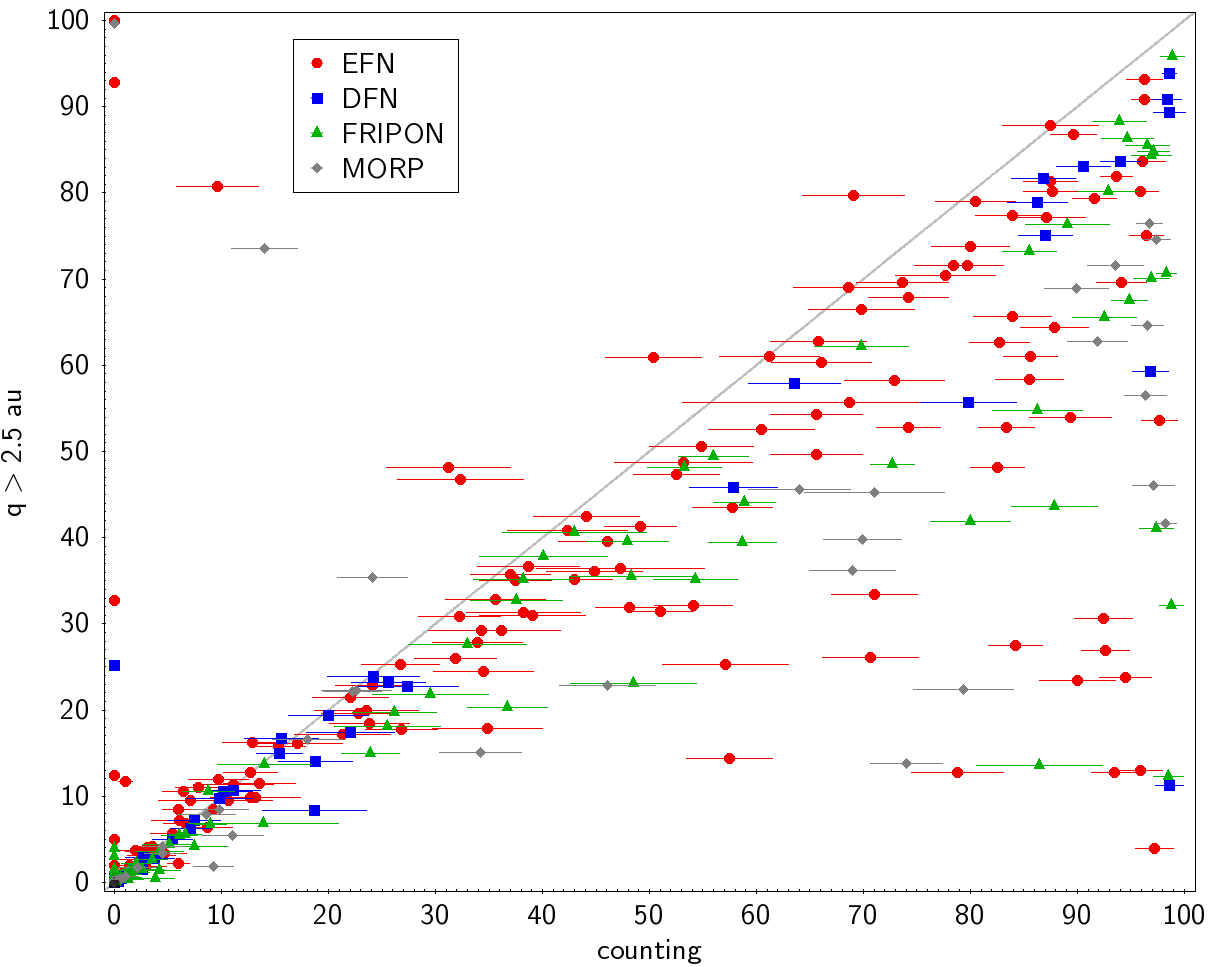}
    \caption{Scatter plot comparing the manual counting method of reviewing integration particle clones histories versus the more simplistic method of identifying the percentage of particle clones that reach a minimum perihelion value of 2.5\,au at least once during the 10\,kyr integrations. The uncertainties of the manual counting method correspond to the standard deviations of the sample means. For each meteoroid, 10 samples of 100 particles were independently counted to gauge the variation in the manual reduction. The line depicts a perfectly 1:1 relationship.}
    \label{fig:count_v_q2_5}
\end{figure}

The manual identification and analysis of the 646,000 particle histories in this study revealed a consistent story with that found by \citet{shober2021main}, with a large majority of the fireballs originating from very stable orbits compared to JFCs. When these manual counts were then compared to the q$>$2.5\,au metric for the same fireball events (Fig. \ref{fig:count_v_q2_5}), the two methods varied very linearly with a Pearson's correlations coefficient of 0.86. This is especially true when the manual counts and the percentage of particle clones that reached q$>$2.5\,au were below 50\%, where the values are nearly identical. Otherwise, for percentages $>$50\%, the manual analysis results seem to be normally over-counting the proportion of "JFC-like" histories. The counting deviates from a 1:1 relationship and only produces a correlation coefficient of 0.31. These over-counts (bottom right quadrant of Fig. \ref{fig:count_v_q2_5}) result from the particles having many distant encounters with Jupiter over the 10\,kyr simulation while never having a close enough encounter to drive them out of the active region. On the other hand, some events have nearly all the particle clones reach 2.5\,au, but were marked as very stable during the manual reduction (top-left quadrant of Fig. \ref{fig:count_v_q2_5}). These, conversely, have very predictable trajectories during the 10\,kyr simulations but can reach higher perihelion orbits through a large q-variation due to a Kozai resonance. 

\begin{figure}[]
    \centering
    \includegraphics[width=0.5\textwidth]{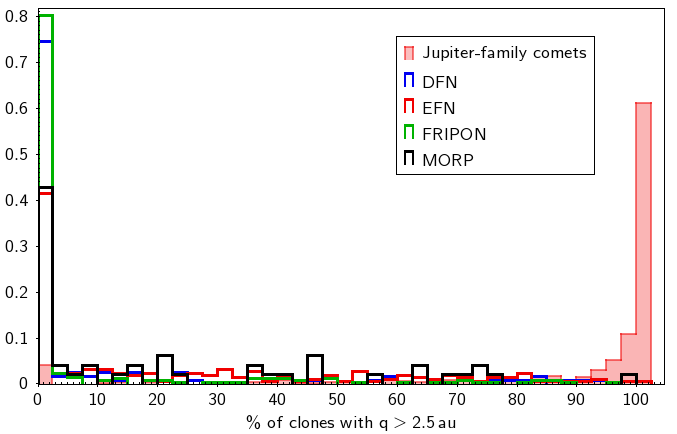}
    \caption{Normalized histogram representing the percentage of particle clones for each object that reach a minimum perihelion value of 2.5\,au at least once during the 10\,kyr integrations for the JFC-like fireballs and the 661 NASA HORIZONS JFCs. The JFCs that have orbits at the observation epoch with q$>$2.5\,au were not included in the histogram.}
    \label{fig:q2_5_hist}
\end{figure}

As seen in Fig. \ref{fig:q2_5_hist}, the q$>$2.5\,au metric is also clearly a very good limit, given that nearly the entire JFC population reaches it on 10\,kyr timescales. Over 90\% of the 661 JFCs included in this study have a $>$50\% chance of reaching q$>$2.5\,au at least once within that time. Given that this minimum perihelion metric is very predicting for JFCs, is much more objective, and prevents the over-counting of some events with only rare events in a Kozai resonance being mischaracterized, this likelihood of q$>$2.5\,au will be used from here forward instead of the manual analysis. This does not change the results in any way, but it does make them more repeatable.  

As seen in Fig. \ref{fig:q2_5_hist}, the fireball distribution drastically differs from the JFC population. The proportion of the JFC-like meteoroids observed by the fireball networks that originate from orbits with $<$50\% chance over reaching q$>$2.5\,au at least once within 10\,kyr is $\sim$90\%. Almost none of the fireballs in JFC-like orbits witnessed by the DFN, EFN, MORP, and FRIPON originate from orbits capable of having close encounters with Jupiter like the JFC population. The N-body modeling confirms the difference between the orbital distributions, demonstrating that these are two distinct populations dynamically. 

However, there are some notable variations and nuances present between the datasets. These differences are likely due to the slightly varied hardware used by each network, resulting in a different limiting magnitude range. These minute differences in the magnitude ranges of the networks here also somewhat modify the size range of objects being detected. Understanding these differences could be the key to better characterizing the physical breakdown of cometary debris and understanding the size-frequency distribution of the results remnants. 

For example, in Fig. \ref{fig:all_networks_orbits_q25} we have plotted the a-e and a-i orbital distribution for all of the fireball networks and modified the coloration to indicate the percentage of particle clones that reach a minimum perihelion value of 2.5\,au. For comparison, the 661 JFCs are also plotted as large crosses. The fireball data technically satisfies the Tisserand's parameter definition of JFC membership ($2<T_{J}<3$; \citealp{carusi1985long}); however, a large proportion of the fireballs lie on orbits hovering just below the perihelion distance of Jupiter. These meteoroids have a dark blue coloration in Fig. \ref{fig:all_networks_orbits_q25}, indicating that they likely have no chance of having a close encounter with Jupiter on 10\,kyr timescales. The further Jupiter-crossing the orbits become, generally the higher percentage of clones have encounters. However, not many fireballs are detected by the FRIPON and DFN networks to originate on orbits larger than 3\,au. On the other hand, the EFN has reported having witnessed many events in comparison coming from $>$\,3\,au (around 30\% more than the DFN or FRIPON), particularly concentrated between 3.2 and 3.5\,au. This abundance of objects originating from beyond 3\,au was claimed by \citet{borovivcka2022_one} to indicate the abundance of cometary material on these orbits; however, no trend in dynamics or longitude of perihelion supports this claim based on our results. Given the lack of close encounters with Jupiter and some semi-major axis values clustering towards 3.2-3.4\,au, we interpret this as likely indicating an origin within the 2:1\,MMR. To verify this, we created a dynamical map where we calculated the Lyapunov exponent of a grid of over 400,000 orbits with $2<T_{J}<3$. The methodology used to calculate the Lyapunov exponents was the same as used for the fireball and JFC data. The orbits were randomly generated within the desired a-e range ([2.9\,au,3\,au], [0.35, 0.99]) and had discrete inclinations of either 0$^{\circ}$, 10$^{\circ}$, 20$^{\circ}$, or 30$^{\circ}$. The results of this chaos mapping can be seen in Fig. \ref{fig:EFN_21_MMR}, where the coloration is indicative of the mean Lyapunov lifetimes and the points are the EFN fireball datapoints. The white regions in Fig. \ref{fig:EFN_21_MMR} contain no points due to not being in range or a lack of convergence. The fireball data points also have the 1-sigma uncertainties plotted, and the pink points indicate that the fireball is within 3-sigma of the 2:1\,MMR. In total, 62 events in the EFN $2<T_{J}<3$ dataset are within 3-sigma of the 2:1\,MMR, which is the maximum limit. This high value, and correspondence with the location of the 2:1\,MMR, is the most likely explanation for the higher number of fireballs with $2<T_{J}<3$ beyond 3\,au. However, exactly why the EFN contains more meteoroids from the 2:1\,MMR is still up for debate. One possible cause of these variations between the fireball networks is the deviation in the fireball absolute magnitude range being observed. 

\afterpage{
\begin{sidewaysfigure*}[p]
    \centering
    \begin{subfigure}[b]{0.24\textwidth}
        \centering
        \includegraphics[width=\textwidth]{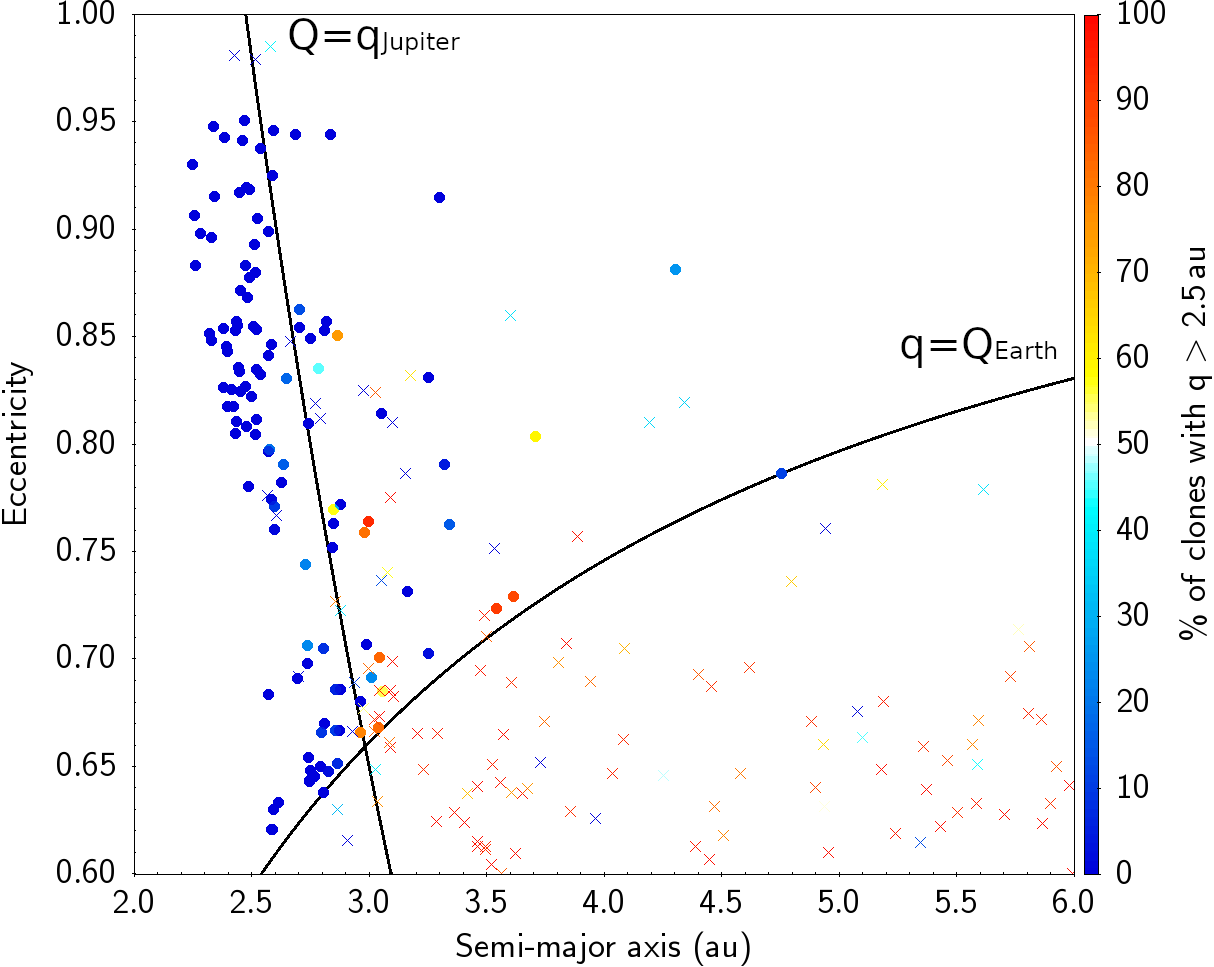}
        \includegraphics[width=\textwidth]{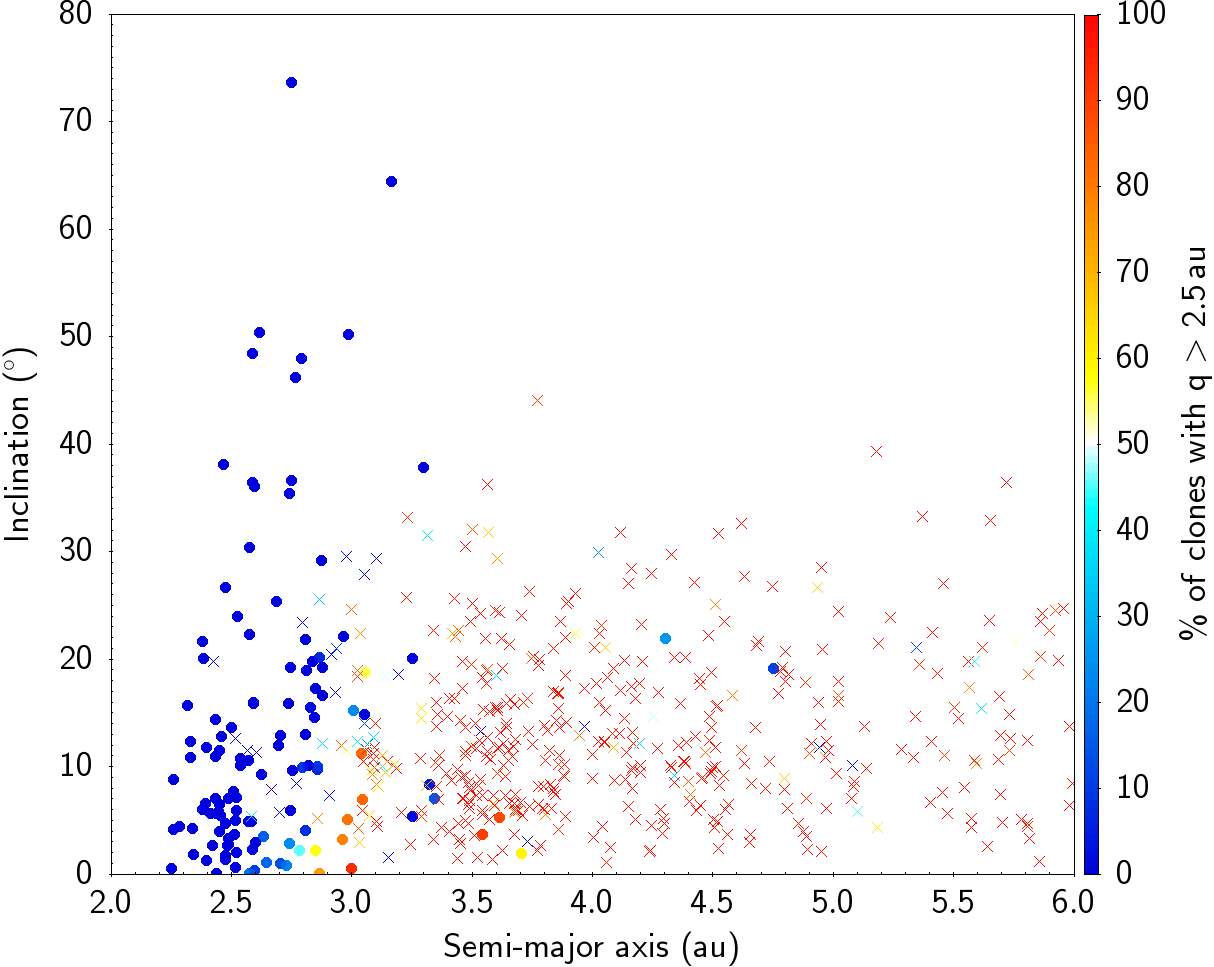}
        \caption{DFN}
        \label{fig:DFN_row}
    \end{subfigure}%
    \begin{subfigure}[b]{0.24\textwidth}
        \centering
        \includegraphics[width=\textwidth]{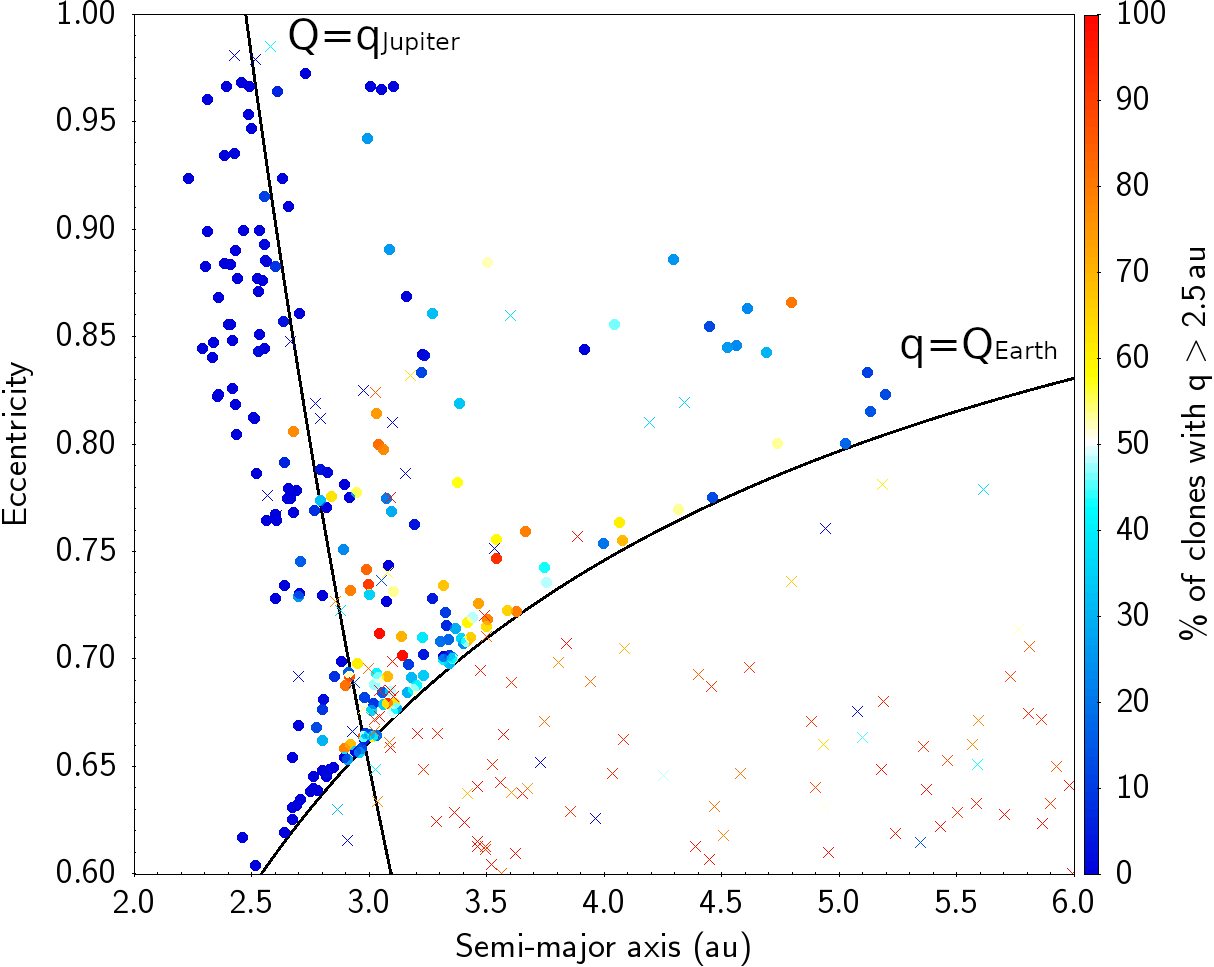}
        \includegraphics[width=\textwidth]{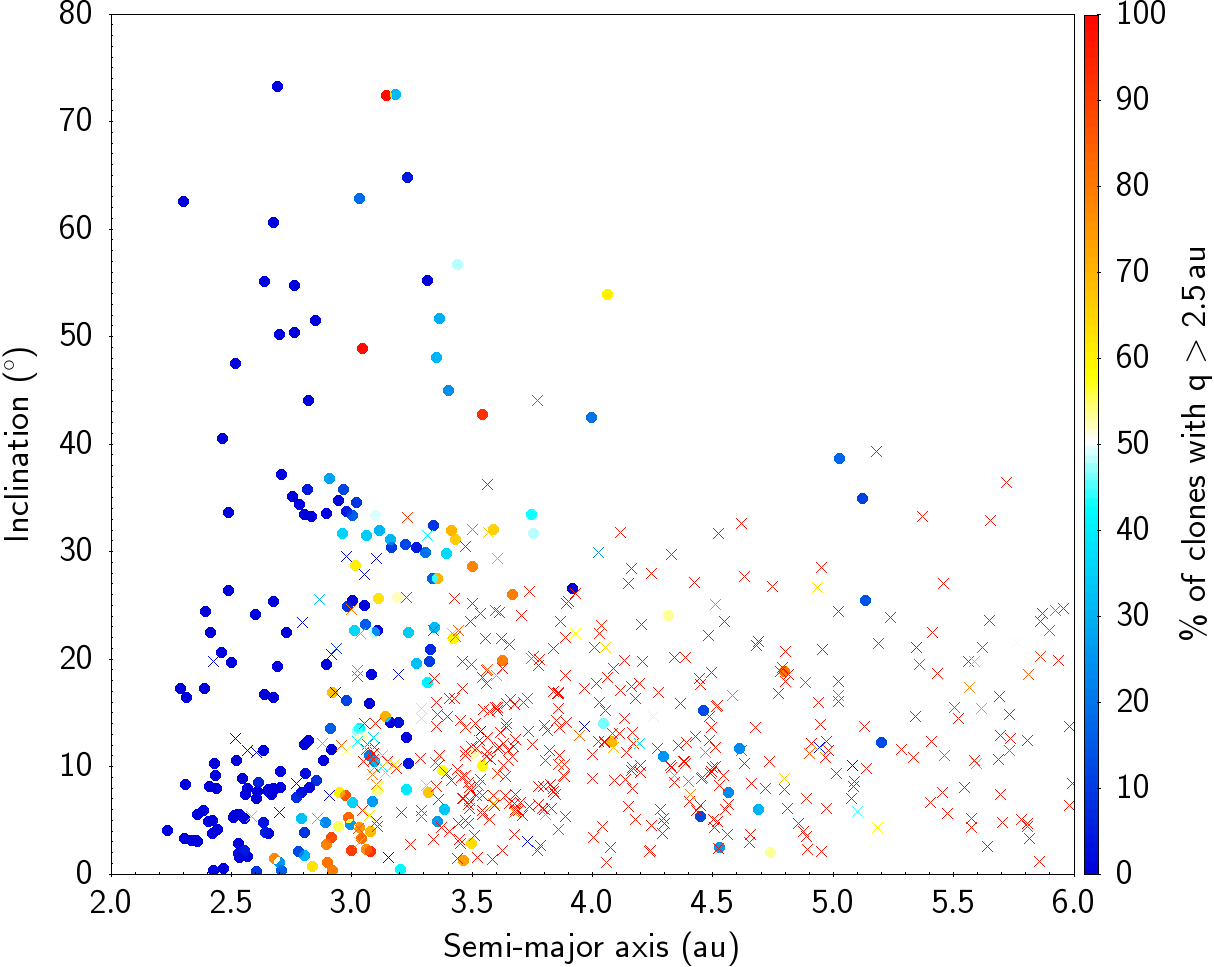}
        \caption{EFN}
        \label{fig:EFN_row}
    \end{subfigure}%
    \begin{subfigure}[b]{0.24\textwidth}
        \centering
        \includegraphics[width=\textwidth]{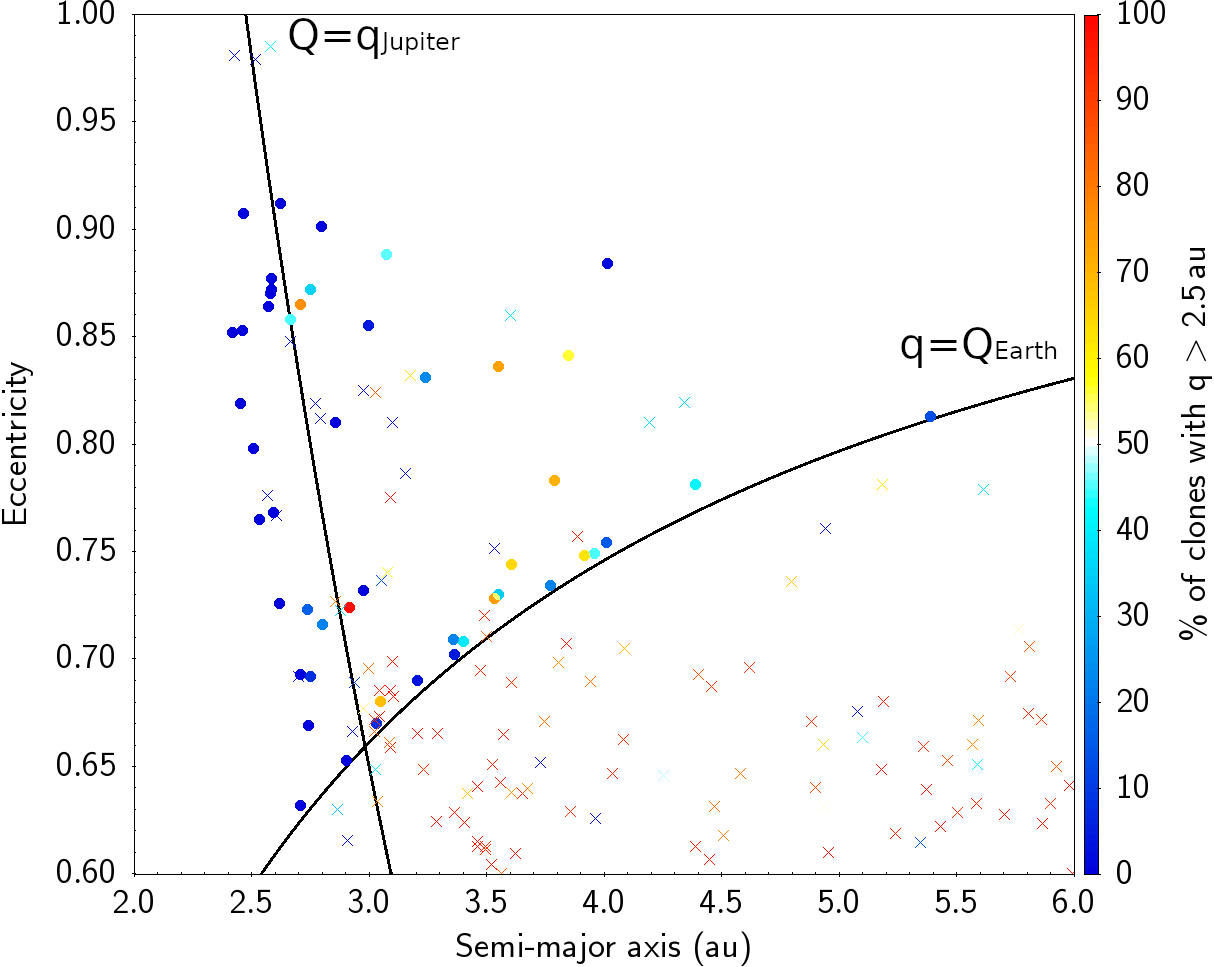}
        \includegraphics[width=\textwidth]{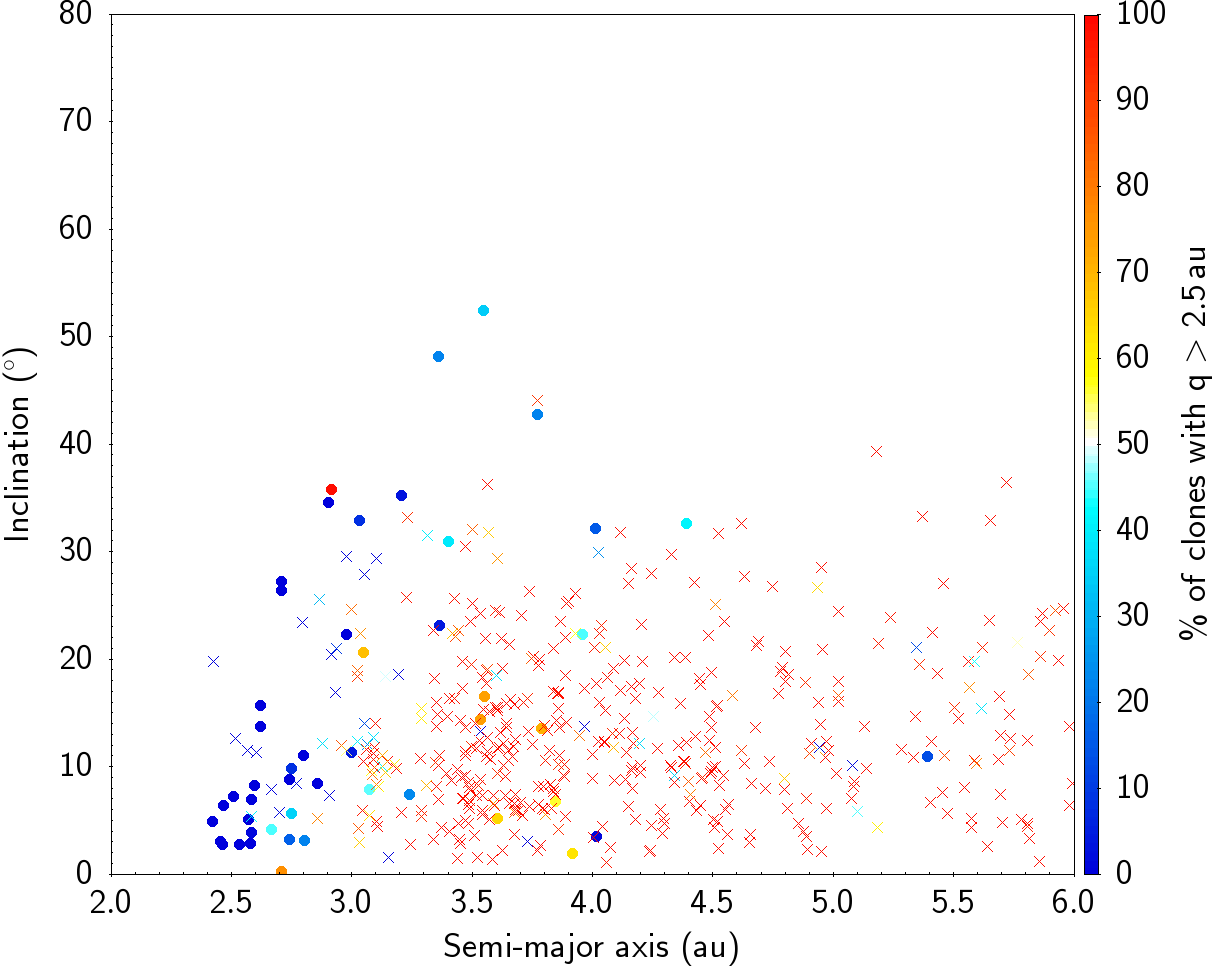}
        \caption{MORP}
        \label{fig:MORP_row}
    \end{subfigure}%
    \begin{subfigure}[b]{0.24\textwidth}
        \centering
        \includegraphics[width=\textwidth]{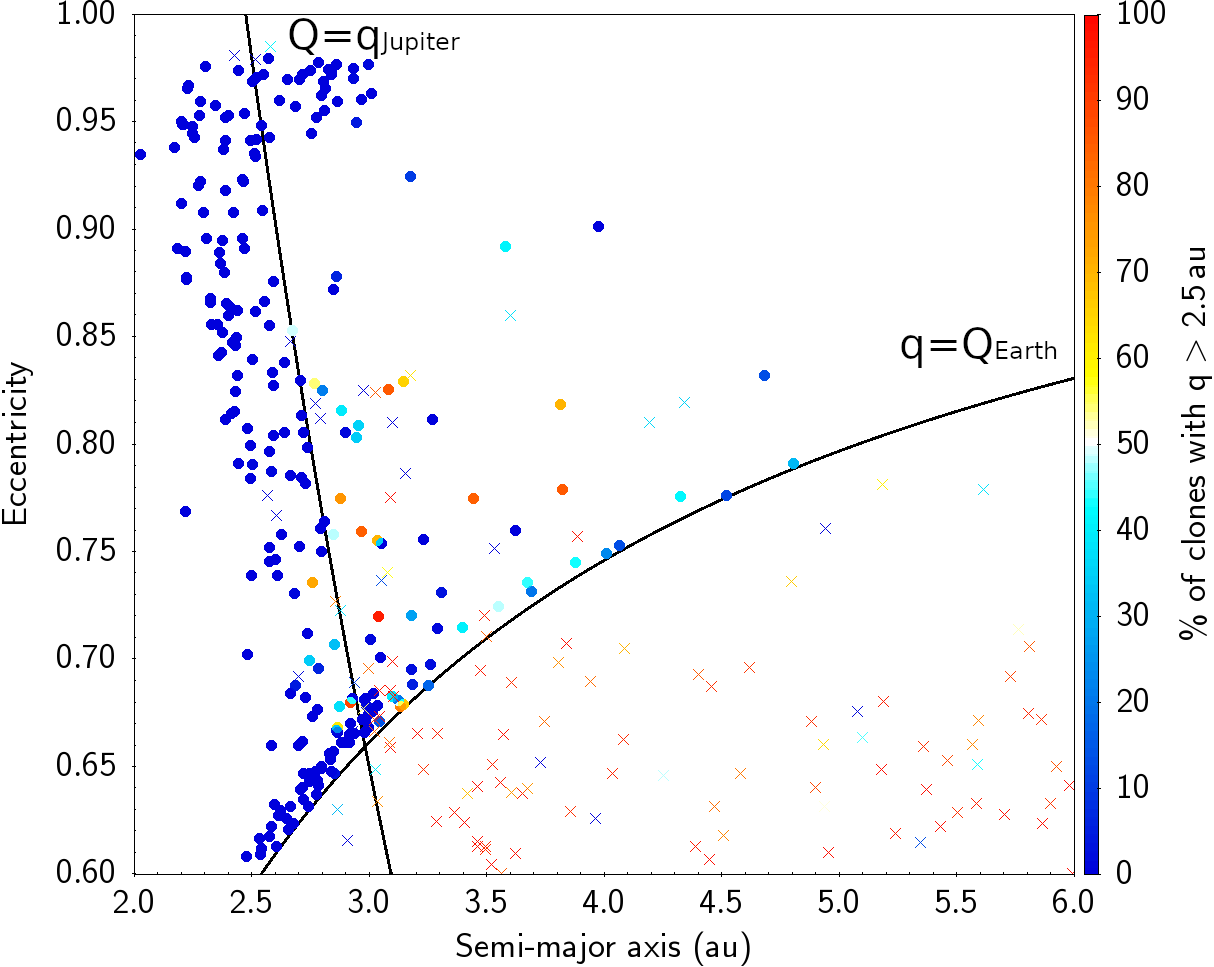}
        \includegraphics[width=\textwidth]{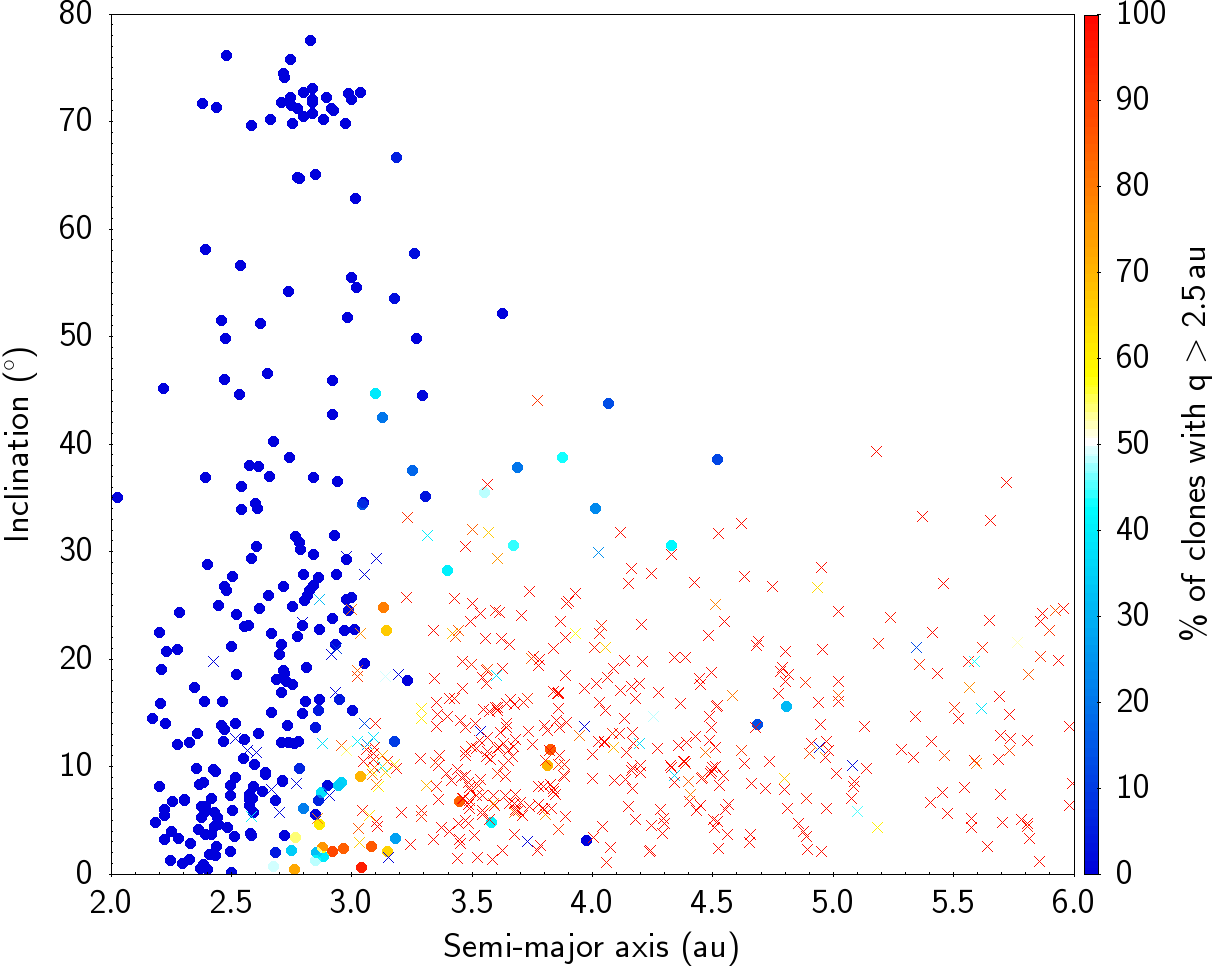}
        \caption{FRIPON}
        \label{fig:FRIPON_row}
    \end{subfigure}

    \caption{Semi-major axis (au) versus eccentricity and inclination ($^{\circ}$) for fireballs observed by each network, along with 661 JFCs from the NASA HORIZONS database. The plot's color scheme represents the outcome of 10 kyr simulations, indicating the percentage of clones per object that underwent significant gravitational interactions with Jupiter, sufficient to increase their perihelion distance to at least 2.5 au. This perihelion distance is critical as it delineates the inner boundary where cometary activity can commence, with water ice starting to sublimate. Objects colored closer to red in this plot suggest a higher dynamical affinity with typical JFC behavior.}
    \label{fig:all_networks_orbits_q25}
\end{sidewaysfigure*}
}

\begin{figure}[]
\centering
\includegraphics[width=0.5\textwidth]{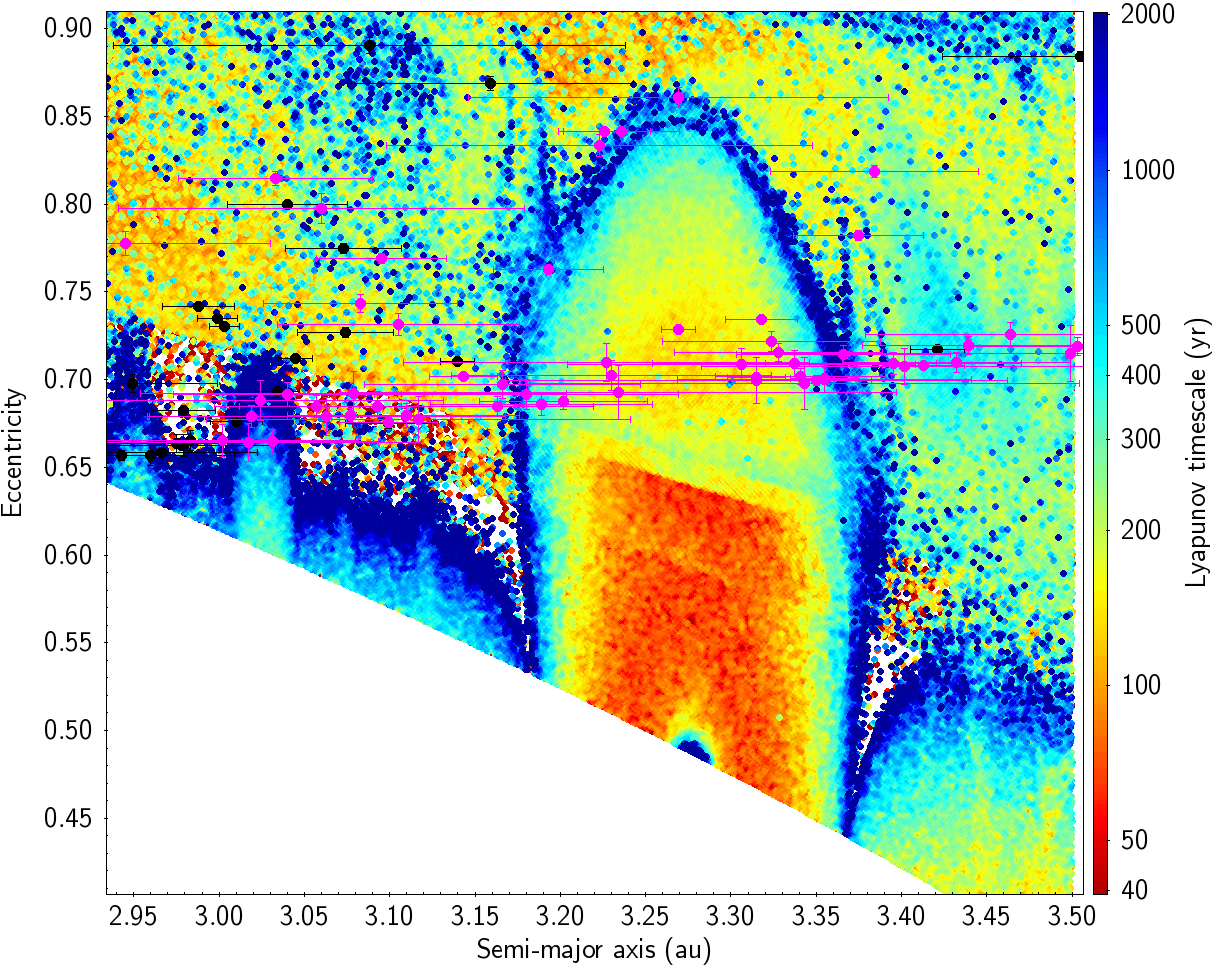} 
\caption{Chaos heat map in a-e space using Lyapunov lifetimes (the inverse of the Lyapunov exponent) calculated from 435,000\, randomized orbits. The coloration indicated the mean Lyapunov lifetime at that point in space, while the individual black points are fireball data points from the EFN network, and the pink points are the fireballs that lie within 3$\sigma$ of the 2:1 MMR. The 2:1\,MMR is clearly visible in the plot centered at around 3.27\,au.}
\label{fig:EFN_21_MMR} 
\end{figure}

\subsubsection{Chaos indicators}

\begin{figure}[]
    \centering
    \begin{subfigure}[b]{0.26\textwidth}
        \includegraphics[width=\textwidth]{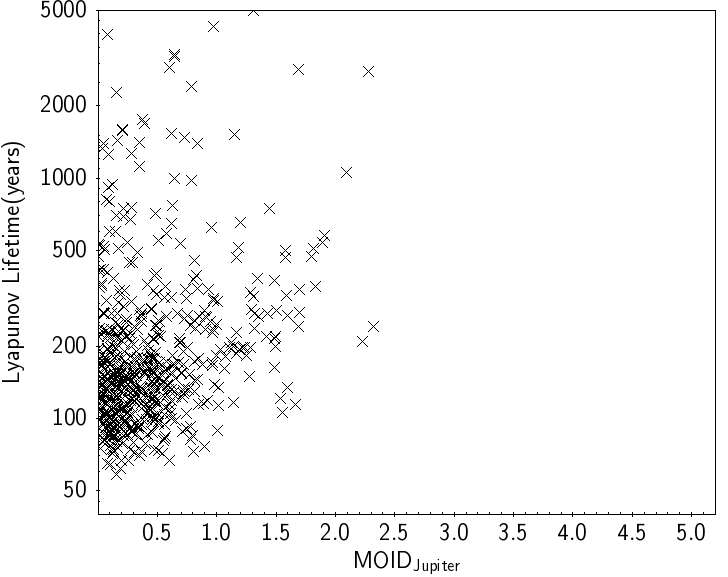}
        \caption{JFC}
        \label{fig:moid_lyapunov_JFCs}
    \end{subfigure}
    
    \begin{subfigure}[b]{0.26\textwidth}
        \includegraphics[width=\textwidth]{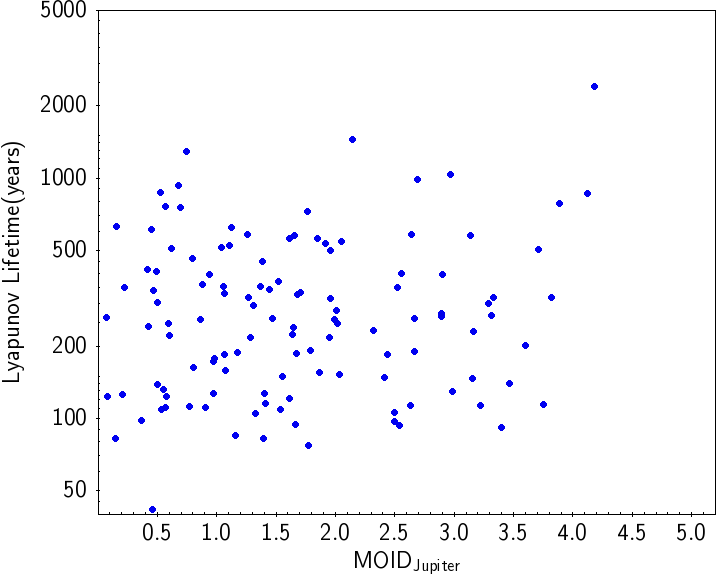}
        \caption{DFN}
        \label{fig:moid_lyapunov_DFN}
    \end{subfigure}

        \begin{subfigure}[b]{0.26\textwidth}
        \includegraphics[width=\textwidth]{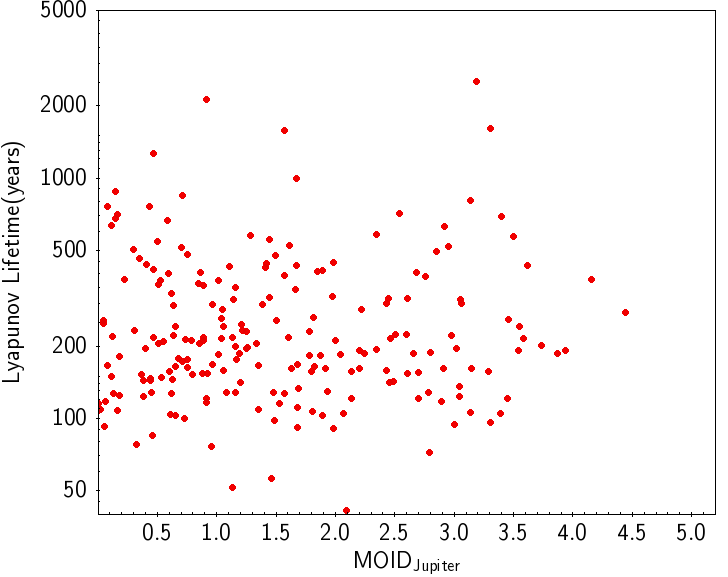}
        \caption{EFN}
        \label{fig:moid_lyapunov_EFN}
    \end{subfigure}
    
    \begin{subfigure}[b]{0.26\textwidth}
        \includegraphics[width=\textwidth]{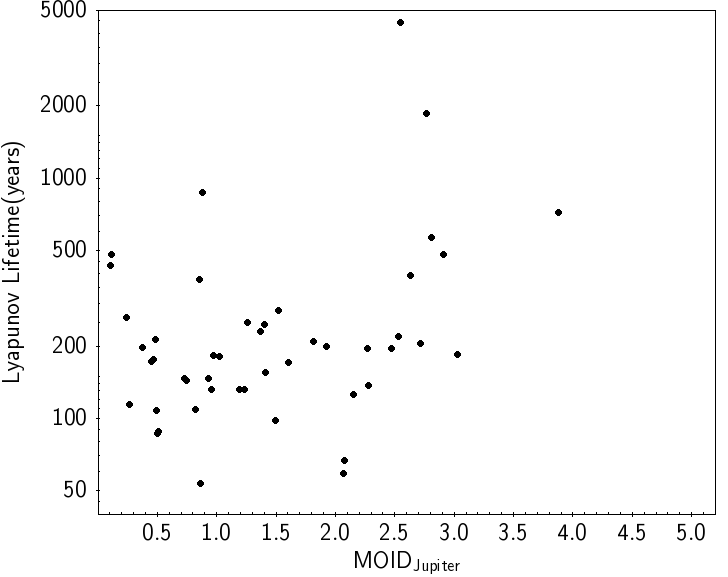}
        \caption{MORP}
        \label{fig:moid_lyapunov_MORP}
    \end{subfigure}

    \begin{subfigure}[b]{0.26\textwidth}
        \includegraphics[width=\textwidth]{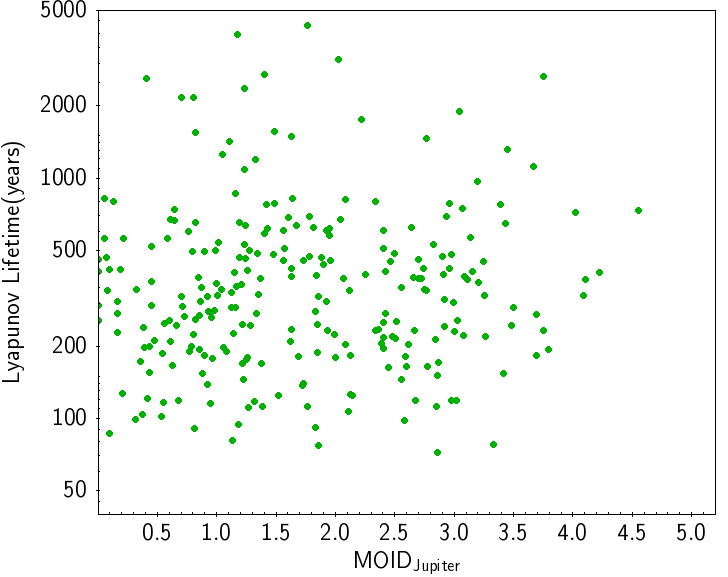}
        \caption{FRIPON}
        \label{fig:moid_lyapunov_FRIPON}
    \end{subfigure}

    \caption{Minimum orbital intersection distances with Jupiter (MOID$_{Jupiter}$; au) for the osculating orbits versus the Lyapunov lifetimes (yrs) for the 661 known JFCs listed in the NASA HORIZONS database compared to the JFC-like fireball datasets of the DFN (blue), EFN (red), MORP (black), and FRIPON (green).}
    \label{fig:moid_lyapunov}
\end{figure}

The use of chaos indicators to characterize the JFC population is not new. The work of \citet{tancredi1995dynamical} and \citet{tancredi1998chaotic} found that the JFC population exhibited a stronger concentration of shorter Lyapunov lifetimes (50-150\,yr), whereas NEOs exhibited a more diffuse distribution. These results showed that the dynamical ``memory'' of JFCs is extremely short due to their frequent close encounters with Jupiter. These encounters produce chaotic scattering and macroscopic changes to the object's orbit.  This pattern found in \citet{tancredi1995dynamical} suggests a common dynamical memory and indicates a connected region of chaos within the JFC population. 

We also calculated the Lyapunov lifetime for the 661 JFCs and compared them to the Lyapunov lifetimes of 646 fireballs from JFC-like ($2<T_{J}<3$) orbits instead. As shown in Fig. \ref{fig:moid_lyapunov}, we have plotted the Lyapunov lifetimes for each fireball network dataset versus the MOID$_{Jupiter}$ for the pre-impact orbit, along with the 661 JFCs. The Lyapunov lifetimes of the JFCs are found to be generally consistent with the results of \citet{tancredi1995dynamical} and \citet{tancredi1998chaotic}. The lifetimes for JFCs concentrate just above 100\,yrs and show minimal spread in values (Fig. \ref{fig:moid_lyapunov_JFCs}). Also, $\sim$90\% of the known JFC population has MOID$_{Jupiter}$ values below 1\,au, consistent with being dynamically linked to Jupiter. Contrarily, the fireball network data show moderately longer lifetimes and larger MOID$_{Jupiter}$ values. A fractional component of the JFC-like fireballs matches the concentration in MOID$_{Jupiter}$ and Lyapunov lifetime, but nowhere to the degree of the JFCs. The fireball data is consistent with the NEO data from \citet{tancredi1998chaotic}, displaying a more diffuse, wider lifetime and MOID distribution. This further supports the theory that the fireball data is mostly indicative of meteoroids originating from the main asteroid belt. 

\subsubsection{Jupiter-family comets}

The 10\,kyr numerical simulations of the 661 known JFCs clearly show most of the JFC population has frequent close encounters with Jupiter, and these encounters tend to keep excursions to the inner solar system to a minimum. This is consistent with the many previous studies of this population, which find that even when JFCs venture onto active (q$<$2.5\,au) orbits, they only do so for $\sim10^{3}$\,yrs \citep{lindgren1992dynamical,fernandez2002there}. This is likely a result of two main coordinating mechanisms: the close encounters with Jupiter which cause chaotic jumps in the orbit, and the physical lifetimes of JFCs in this region which are also thought to be $\sim10^{3}$\,yrs \citep{levison1997kuiper,fernandez2002there}. 

\begin{figure}[]
\centering
\includegraphics[width=0.5\textwidth]{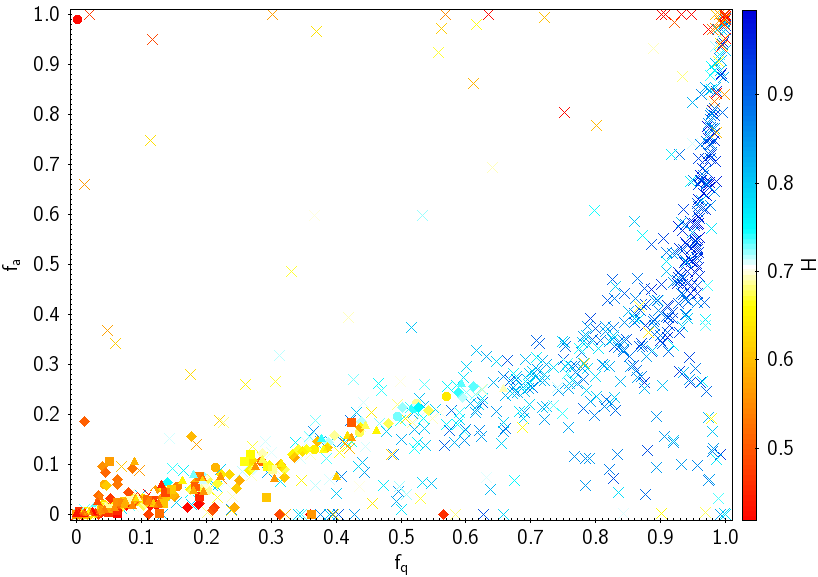} 
\caption{Correlation between the indices $f{q}$ and $f_{a}$ for the 661 JFCs (crosses), and the JFC-like fireball data from the four networks included (DFN, EFN, MORP, FRIPON). The color bar indicates the Kozai energy of JFC or meteoroid, where yellow-red color corresponds to objects that are capable of large cyclic q variation, and blue ones cannot.}
\label{fig:all_fq_fa} 
\end{figure}

Employing the $f_{q}$ - $f_{a}$ indices as defined by \citet{fernandez2014assessing} (refer to Methods), we statistically evaluate the duration for which these bodies typically remain within q$<$2.5\,au and a$>$7.37\,au during the integration period. These indices are crucial for examining this group's overall traits and drawing parallels with the meteoroid population. As depicted in Fig. \ref{fig:all_fq_fa}, JFCs, denoted by crosses, predominantly occupy high-q and high-a orbits for most of the 10\,kyrs. This observation aligns with the findings of \citet{fernandez2014assessing} and \citet{fernandez2015jupiter} on the near-Earth subset of 139 NEAs and 58 JFCs. Our study expands this analysis to include over 600 more JFCs. Also, unlike past studies that concentrated on the near-Earth segment, where a nearly linear correlation was noted, our findings reveal a deviation from linearity for $f_{q}>0.9$, indicative of many JFCs spending most of their existence beyond the terrestrial planets' orbits. This trend is likely rooted in the fact that the majority of these objects often possess a semi-major axis exceeding 7.37\,au and spend minimal time in active orbits \citep{levison1997kuiper,fernandez2002there}. Of the 661 JFCs analyzed, approximately 78\% have $f_{q}$ values over 0.5, suggesting infrequent presence in the active zone ($<$2.5,au). This observation is reinforced by the discovery bias against non-active and high-inclination JFCs \citep{disisto2009population}, hinting at a probable overestimation of their near-Earth residency time. 

Another interesting nuance to this population is that there have been several members of the JFC population previously identified by \citet{fernandez2015jupiter} to have higher than expected $f_{q}$ values given the lack of close encounters witnessed in the N-body simulations conducted. It was found that this q-variation was caused by the coupling of $\omega$ and q, in a Kozai cycle. The Kozai mechanism plays a pivotal role in decreasing the perihelion distance (q) observed in numerical simulations. This dynamical process is understood through the Kozai energy level curves of a celestial body, influenced by its semi-major axis and the parameter $H = \sqrt{1 - e^2} \cos i$, where $H$ is the Kozai energy, $e$ is the eccentricity and $i$ is the inclination. These curves (see Figure 10 in \citealp{fernandez2014assessing}) ensure a long-term orbital evolution that maintains high perihelia for asteroids, both inside and outside the 2:1 MMR with Jupiter, as long as $H \gtrsim 0.85$. Most asteroids meet this condition in the main belt and nearly all JFCs (Fig. \ref{fig:all_fq_fa}). However, a notable shift in these curves occurs when $H$ falls below approximately 0.7, dramatically altering their topology. This change allows for transitions between high and low-q values, paired with oscillations in the semi-major axis and inclination over a few thousand years. The curves are also influenced by whether the asteroid's orbit is resonant or non-resonant, particularly regarding the amplitude and center of libration \citep{gomes2005origin,gallardo2012survey}. The amplitude of these curves is broader for non-resonant bodies, linking large and very low-q values, whereas it is more constrained for resonant bodies with similar $H$ values. In essence, non-resonant bodies experience a stronger Kozai effect. For instance, a resonant asteroid with $H = 0.6$ and $q = 2$\,au shows negligible q variation while in resonance but can drop to 0.7\,au once it leaves the resonance. The Kozai mechanism can also lead some comets and asteroids to sungrazing orbits. Notable examples from our sample include 321P/SOHO and 322P/SOHO, both of which are argued to be asteroids that evolved into sungrazing orbits due to the Kozai-Lidov mechanism. Another notable example not examined in this sample, as the $T_{J}$ is in excess of 3, is asteroid 2003 EH1 one of the proposed parents of the Quadrantid meteor shower. The asteroid has also been shown to be in a strong Kozai resonance, driving the q values to extremely small values in the past \citep{wiegert2005quadrantid,granvik2016super}.

Of the 661 JFCs in this study, $\sim$20\% of them have a $H < 0.7$. Many of these are beyond the orbits of Jupiter; however, 38 are within near-Earth space. Of the JFCs with $H < 0.7$, that are also not beyond Jupiter's orbit ($a < 7.37$\,au), 94\% are within the "active" region and about 50\% are in near-Earth space. Half of the near-Earth JFCs are in a Kozai resonance to some degree. When this is compared to the fireball datasets, this trend becomes overwhelming. Over 90\% of the fireball data have H values less than 0.7, and 100\% are less than 0.75. This indicates that nearly all the meteoroid population on JFC-like ($2<T_{J}<3$) orbits are in some level of Kozai resonance. This is not surprising as the objects that are on Earth-crossing orbits on JFC-like orbits already have high eccentricities (e$>$0.6). This interplay of MMRs and secular oscillations due to the Lidov-Kozai effect have been shown in numerous other studies to be effective together to drive down q-values creating Sun-grazing objects \citep{toliou2023resonant}. 

Similar to \citet{fernandez2015jupiter}, a subset of very "stable" JFCs have also been identified. The particle clones of these JFCs evade the close encounters with Jupiter during the 10\,kyr simulations, resulting in more or less predictable trajectories. The proportion of stable interlopers in the dataset tends to also increase as the initial perihelion decreases. Of the near-Earth JFCs, excluding fragments, (75 in total), 22 ($\sim$30\%) move on extremely stable orbits over the 10\,kyrs timescales according to the histories of 1000 particle clones. These extremely stable JFCs all have $f_{q}$ values less than 0.02 and $f_{a}$ less than 0.005 (Table \ref{tab:22_extremely_stable}), and most of these were also previously identified by \citet{fernandez2015jupiter} for their stability. Of these 22, 16 of the JFCs had zero particles meet either q$>$2.5\,au or the a$>$7.37\,au limits, implying there is a very low probability of close encounters with Jupiter. Moreover, of the 22, 6 are lost comets. Comets 3D/Biela, 5D/Brorsen, and 34D/Gale are thought to have suffered major splitting events and were lost due to their disintegration \citep{fernandez2009s}. Notably, comet 3D/Biela, was one of the first well-observed cometary-splitting events in the observation record, and the debris from this event can still be seen in the Andromedid meteor shower. The other three "defunct" comets are D/1766 G1 (Helfenzrieder), D/1770 L1 (Lexell), and D/1895 Q1 (Swift); these are all only observed for the one perihelion passage, so they were likely lost due to the orbital uncertainty. A breakup of D/1895 Q1 (Swift) was speculated to be the source of the intense meteor storm experienced by Mariner 4 on September 15, 1967; however, the uncertain orbits cast doubt on this explanation \citep{beech1999satellite}. 

Compared with most JFCs, these highly stable objects almost all have small semi-major axis values, with 18 of the 22 within the range of major main-belt resonances (a$<$3.2\,au) and lower Kozai energies. In addition to the three defunct comets with high orbital uncertainties, six others within this extremely stable subset were not previously studied. These include 294P/LINEAR, 321P/SOHO, 322P/SOHO, 384P/Kowalski, 414P/STEREO, and 463P/NEOWISE. Of these newly analyzed comets, only 294P/LINEAR has a semi-major axis beyond 3\,au, however, it is within the range of the 2:1 MMR and was identified as being within the region with the highest concentration of Themis-family asteroids \cite{hsieh2020potential}. The others all have a$<$3\,au, and except for 384P/Kowalski, are well below the Kozai critical energy of 0.7. Comets 321P and 322P are both in extremely low-q orbits, within 0.06\,au of the Sun at closest approach. Such a close encounter with the Sun should produce devastating results for most JFCs; however, these objects are much more likely to be asteroidal in origin.  The orbits of these objects were also found to be consistent with those of small NEAs and their rates of observation correlate with expected rates of small asteroid injection into the near-Sun region \citep{wiegert2020supercatastrophic}. Furthermore, 322P was observed without a coma, bolstering its classification as an asteroid \citep{knight2016comet}. The evidence presented in \citet{wiegert2020supercatastrophic} indicates that objects 321P and 322P, traditionally labeled as comets, might be better classified as asteroids undergoing ``supercatastrophic disruption''. \citet{granvik2016super} found that there exists a paucity of objects expected at extremely low-perihelion values (below around 0.074\,au), due to some form of ``supercatastrophic disruption'' as objects approach the Sun. These objects are also found to oscillate in q-value, going in and out of sungrazing states due to a strong Kozai resonance, with q oscillating between 0.4 and $<$0.1\,au and the inclination correspondingly varying between 3$^{\circ}$ and 45$^{\circ}$ for 321P and between 9$^{\circ}$ and 70$^{\circ}$ for 322P. This is unsurprising as the Kozai energies for 321P and 322P are both below 0.2. The inclinations also reach a maximum as their $\omega$ values start to align with Jupiter (0$^{\circ}$ and 180$^{\circ}$), producing a distribution identical to the fireball distribution seen in Fig. \ref{fig:omega_inc_ecc}, as nearly all the fireballs are in a secular Kozai resonance cycle. This period of high-inclination, low-aphelion, and high-perihelion when the $\omega$ values are at 0$^{\circ}$ and 180$^{\circ}$ decreases the chance of frequent close encounters with Jupiter, as the objects sit significantly above the ecliptic plane and aphelion distance decreases at the encounter point. These meteoroids are crossing the orbit of Jupiter, however, the evolution inside a typical secular Kozai mechanism creates less ideal conditions for a very close encounter due to pushing the orbit out of Jupiter's orbital plane when the orientation is optimal. This ability to protect certain bodies from close encounters with planets was first shown by \citet{kozai1979secular}, where it was found that (1373) Cincinnati avoids close encounters with Jupiter. Meanwhile, (1866) Sisyphus and Midas (1981) avoid close encounters with Mars due to the effect. 

Additionally, there are two non-NEOs with extremely stable orbits as well, meeting the same criteria as the 22 comets in Tab.\ref{tab:22_extremely_stable}. Comets 247P/LINEAR and 421P/McNaught both have moved very predictably over 10\,kyr timescales with 0\% and 1.9\% of the particle clones respectively reaching a q$>$2.5\,au. Both comets are within the active region and have H values between 0.7-0.75, displaying some Kozai cycling. However, both have significantly larger semi-major axis values, 3.96\,au and 5.08\,au, respectively. They are also both stuck in MMRs with Jupiter: 247P/LINEAR is in the 3:2 MMR (a quasi-Hilda orbit) and 421P/McNaught is in 1:1 MMR. Considering these regions of space are dominated by D- and P-type asteroids, making a distinction between the main belt and JFC populations is even more difficult. 

\begin{figure}[]
\centering
\includegraphics[width=0.5\textwidth]{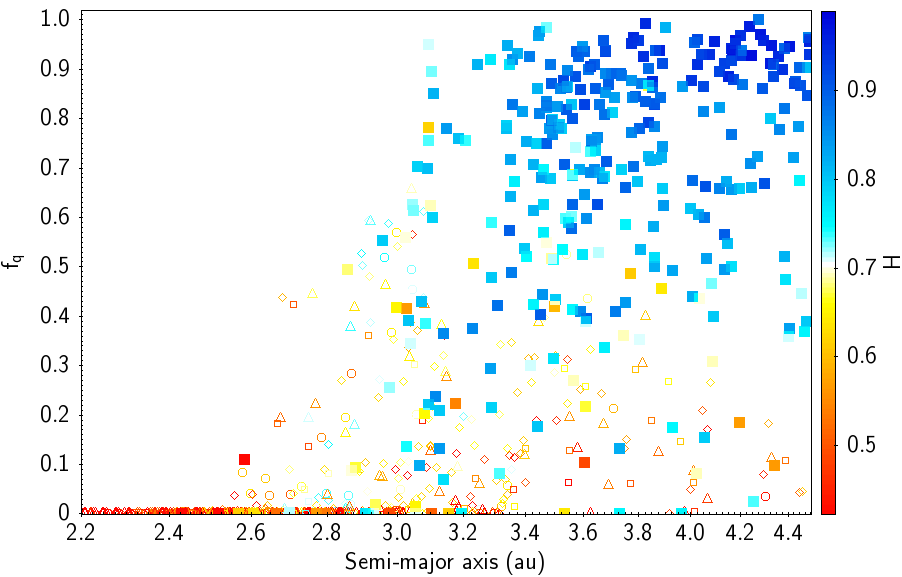} 
\caption{Semi-major axis (au) versus the $f_{q}$ index for 661 JFCs (large square) and 646 JFC-like ($2<T_{J}<3$) fireballs collected by the DFN (hollow circle), EFN (hollow diamond), MORP (hollow square), and FRIPON (hollow triangle) networks. The coloration is indicative of the Kozai energy of the comet or meteoroid orbit at the observed epoch, where 0.7 is the critical limit where Kozai cycling of q - $\omega$ begins \citep{fernandez2014assessing}.}
\label{fig:all_a_vs_H} 
\end{figure}

\begin{table*}[]
    \centering
    \begin{tabular}{|l c c c c c c c|}
    \hline
        Name & a (au) & i ($^{\circ})$ & T$_{J}$ & $MOID_{Jupiter}$ (au) & H & $f_{q}$ & $f_{a}$ \\ \hline 
        3D/Biela & 3.53 & 13.22 & 2.531 & - & 0.642 & 0.0 & 0.0 \\
        5D/Brorsen & 3.1 & 29.38 & 2.467 & - & 0.511 & 0.0 & 0.0 \\
        34D/Gale & 4.94 & 11.73 & 2.291 & - & 0.636 & 0.0 & 0.0 \\
        66P/du Toit & 6.02 & 18.67 & 2.123 & 0.864 & 0.585 & 0.0 & 0.0 \\
        141P/Machholz 2 & 3.05 & 13.98 & 2.709 & 0.56 & 0.656 & 0.015 & 0.004 \\
        162P/Siding Spring & 3.05 & 27.82 & 2.792 & 0.587 & 0.71 & 0.001 & 0.0 \\
        169P/NEAT & 2.61 & 11.3 & 2.888 & 0.978 & 0.629 & 0.0 & 0.0 \\
        182P/LONEOS & 2.93 & 16.91 & 2.846 & 1.007 & 0.713 & 0.0 & 0.0 \\
        189P/NEAT & 2.92 & 20.4 & 2.909 & 0.666 & 0.751 & 0.0 & 0.0 \\
        209P/LINEAR & 2.94 & 20.98 & 2.789 & 0.458 & 0.677 & 0.018 & 0.003 \\
        249P/LINEAR & 2.77 & 8.4 & 2.707 & 0.746 & 0.568 & 0.002 & 0.0 \\
        294P/LINEAR & 3.19 & 18.54 & 2.819 & 0.97 & 0.759 & 0.0 & 0.0 \\
        300P/Catalina & 2.7 & 5.68 & 2.963 & 0.813 & 0.719 & 0.003 & 0.001 \\
        321P/SOHO & 2.43 & 19.74 & 2.396 & 0.678 & 0.184 & 0.0 & 0.0 \\
        322P/SOHO & 2.52 & 12.59 & 2.347 & 1.044 & 0.2 & 0.0 & 0.0 \\
        384P/Kowalski & 2.91 & 7.29 & 2.958 & 0.894 & 0.782 & 0.0 & 0.0 \\
        414P/STEREO & 2.8 & 23.38 & 2.647 & 0.838 & 0.536 & 0.0 & 0.0 \\
        463P/NEOWISE & 2.98 & 29.5 & 2.492 & 0.997 & 0.492 & 0.0 & 0.0 \\
        D/1766 G1 (Helfenzrieder) & 2.66 & 7.87 & 2.705 & - & 0.526 & 0.0 & 0.0 \\
        D/1770 L1 (Lexell) & 3.15 & 1.55 & 2.612 & - & 0.618 & 0.0 & 0.0 \\
        D/1895 Q1 (Swift) & 3.73 & 2.99 & 2.677 & - & 0.757 & 0.0 & 0.0 \\
        P/2003 T12 (SOHO) & 2.57 & 11.48 & 2.894 & 1.019 & 0.618 & 0.0 & 0.0 \\ \hline
    \end{tabular}
    \caption{Near-Earth (q$<$1.3\,au) JFCs that move on extremely stable orbits over $10^{4}$\,yr timescales.}
    \label{tab:22_extremely_stable} 
\end{table*}

\subsubsection{Source region analysis}
Another method to determine source regions of meteoroids from fireball data is the use of debiased near-Earth steady-state models. These models are a mathematical representation of the population of NEOs that accounts for the observational biases inherent in the data collected by various telescopic surveys. Such biases can occur due to factors like the limited sensitivity of telescopes, which can preferentially detect larger or brighter objects, or the positioning and timing of observations, which might miss objects with certain orbits or are visible only at certain times. By accounting for these biases, the model aims to represent the true distribution of NEOs as they exist in space, independent of the limitations of our observations. By comparing the back-tracked orbit calculated from fireball observations with the distribution of orbits in the debiased model, it is possible to identify the most likely source region from which the object originated. This has been used in many studies previously to understand meteorite source regions \citep{bottke2002debiased,granvik2018debiased,2023AJ....166...55N}.

Here, we have used the NEO model described in \citet{2023AJ....166...55N} as another method to determine the source regions of these JFC-like fireballs. However, the problem of size dependence of source regions, which has been identified before, is particularly important to keep in consideration as the model does not consider objects down to the meteoroid cm-m size range. So, when applied to centimeter-meter-sized meteoroid datasets, we had to use the maximum absolute magnitude in the model of H=25.

\begin{figure}[]
\centering
\includegraphics[width=0.5\textwidth]{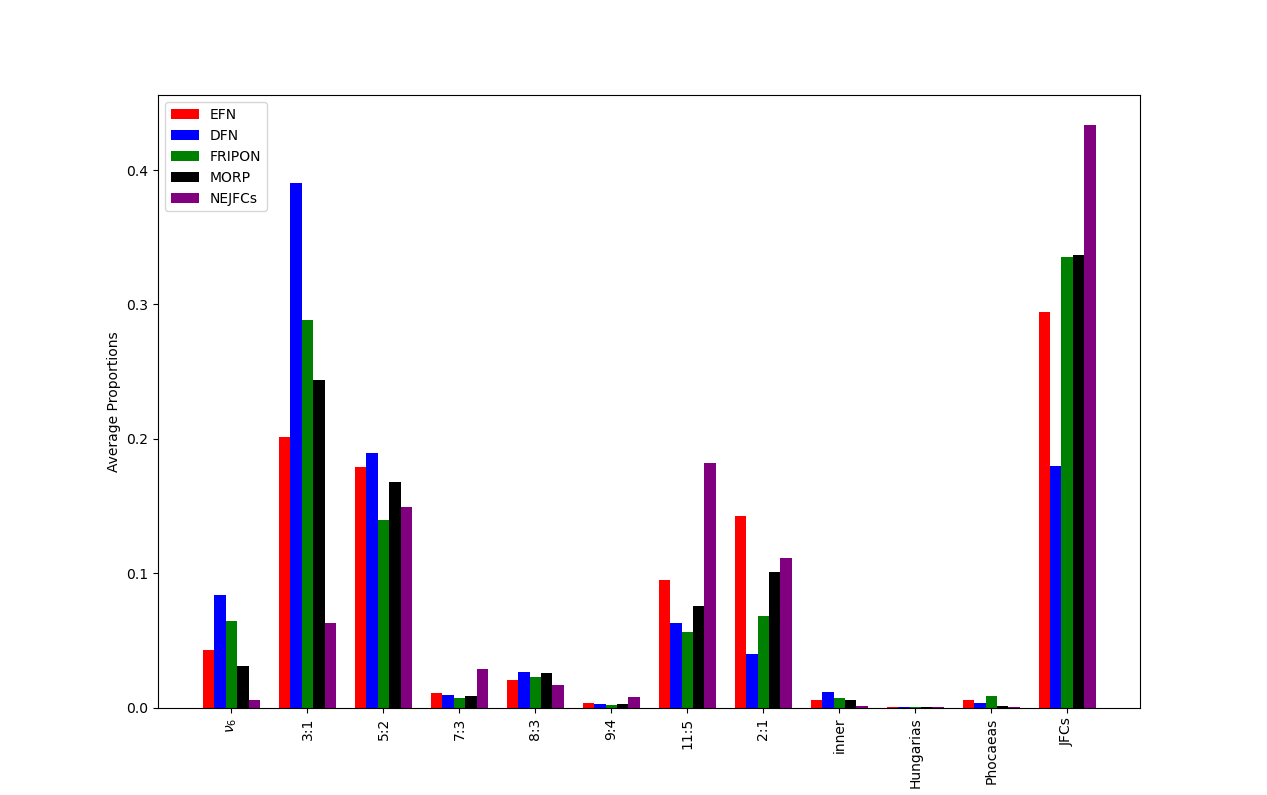} 
\caption{Source region for JFC-like fireballs detected by the DFN, EFN, MORP, and FRIPON determined by the NEOMOD debiased NEO model \citet{2023AJ....166...55N}. The source region probabilities were also determined for the near-Earth portion of the JFC population; the low JFC source region likelihood for these comets is due to the significantly higher number of asteroids from the MB in the same \textit{a-e-i} region.}
\label{fig:source_region_histogram} 
\end{figure}

\begin{sidewaysfigure*}[]
\centering
\begin{subfigure}[b]{0.28\linewidth}
    \includegraphics[width=\linewidth]{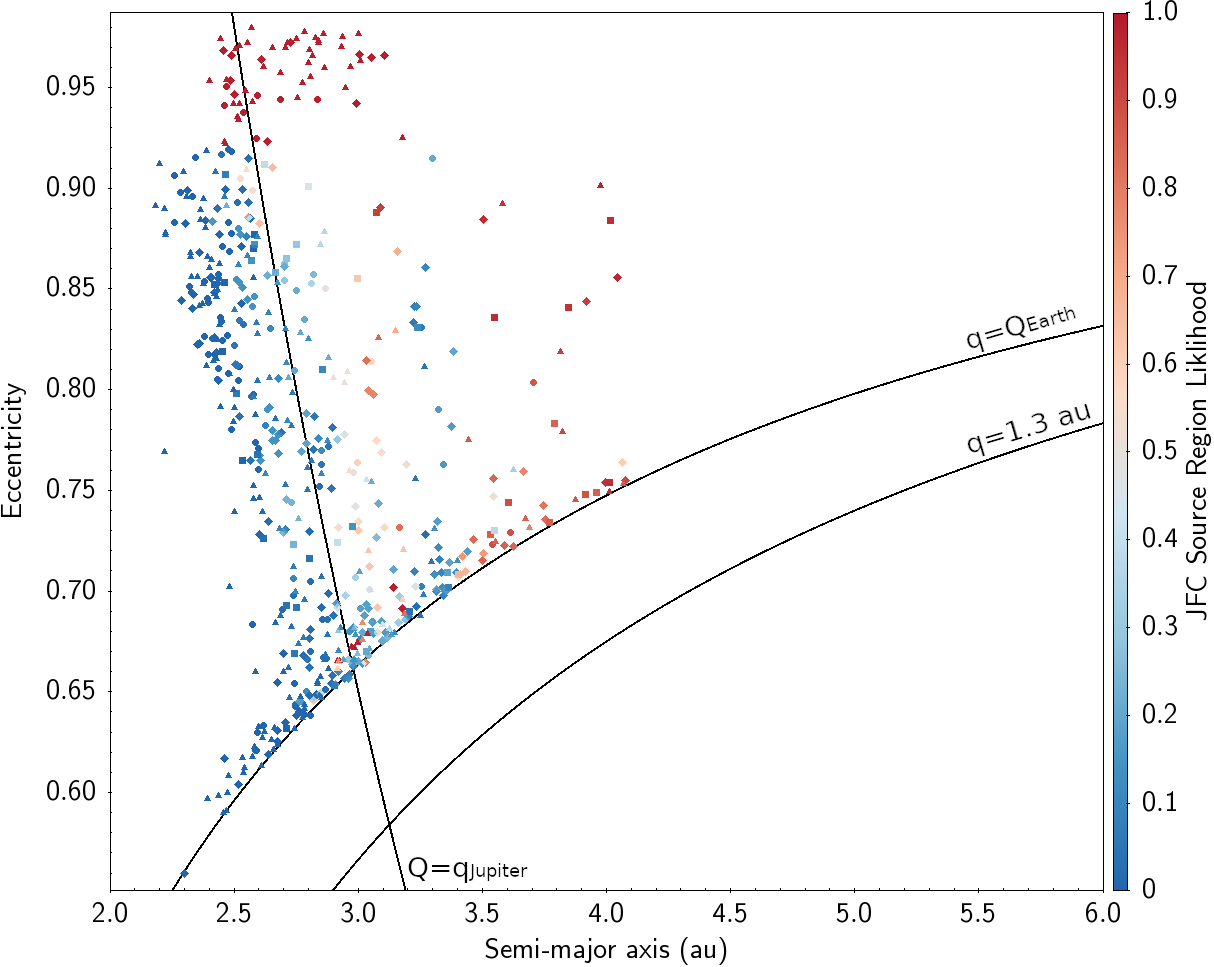}
    \caption{JFC-like Fireballs}
    \includegraphics[width=\linewidth]{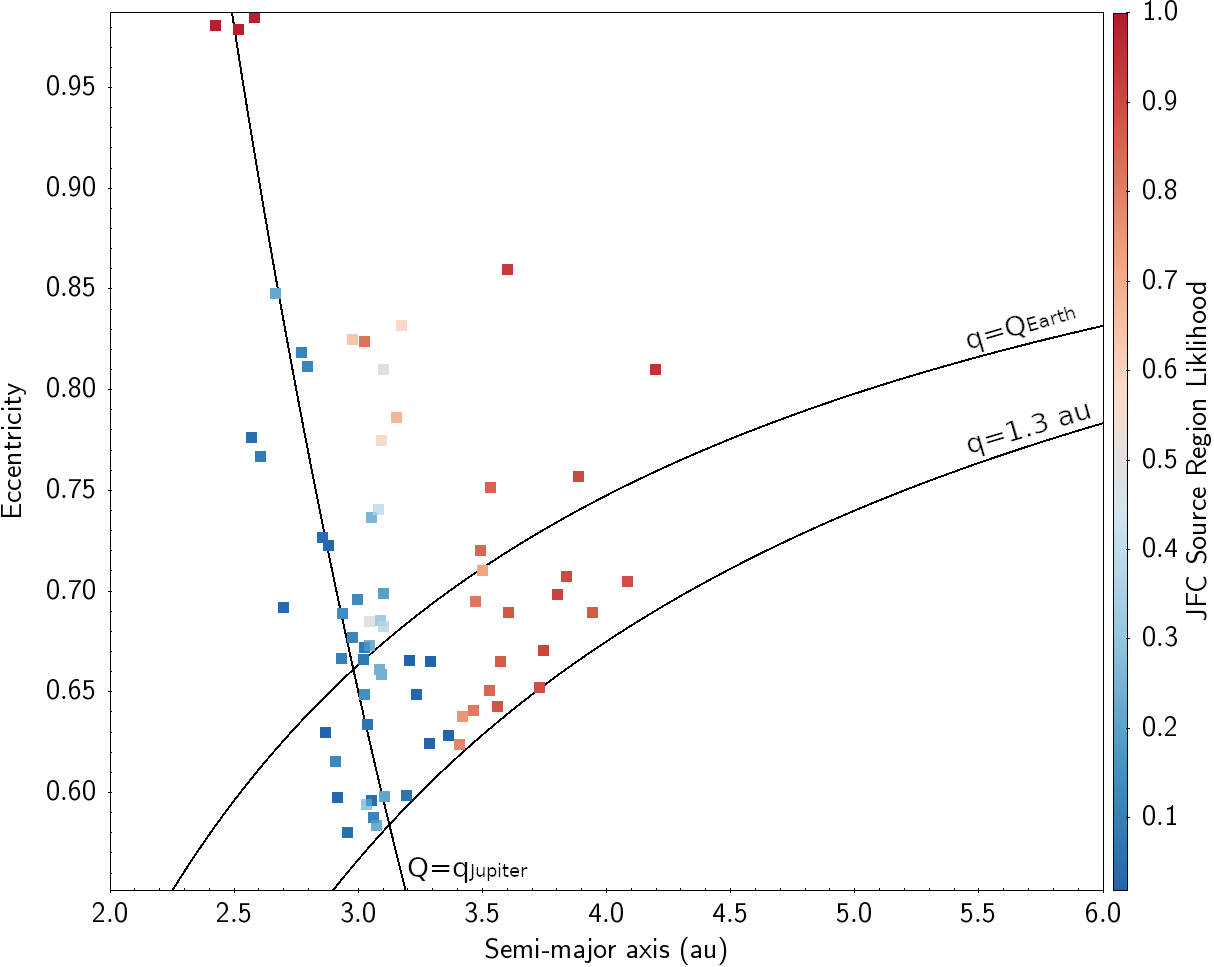}
    \caption{Near-Earth JFCs}
\end{subfigure}
\begin{subfigure}[b]{0.28\linewidth}
    \includegraphics[width=\linewidth]{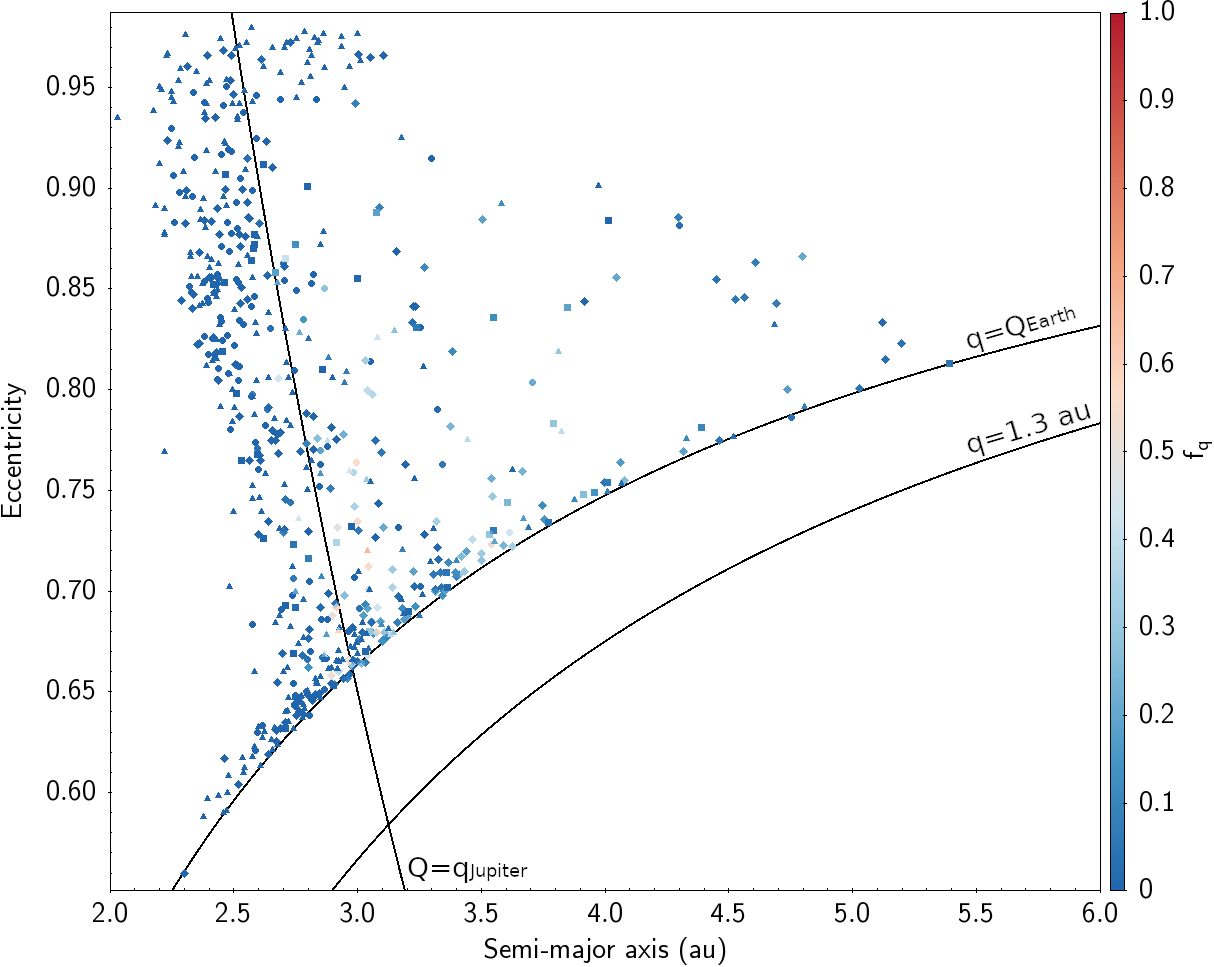}
    \caption{JFC-like Fireballs}
    \includegraphics[width=\linewidth]{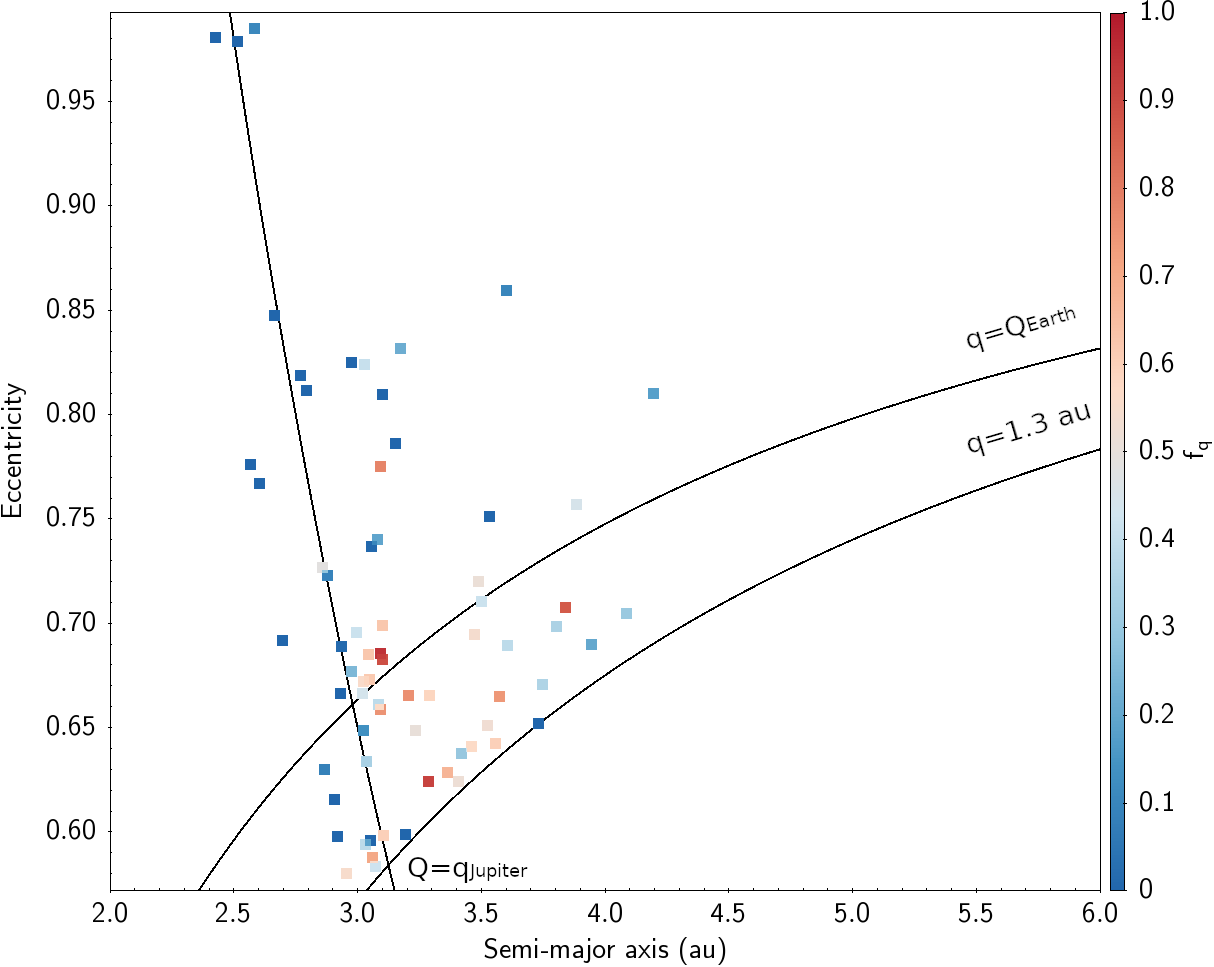}
    \caption{Near-Earth JFCs}
\end{subfigure}
\begin{subfigure}[b]{0.28\linewidth}
    \includegraphics[width=\linewidth]{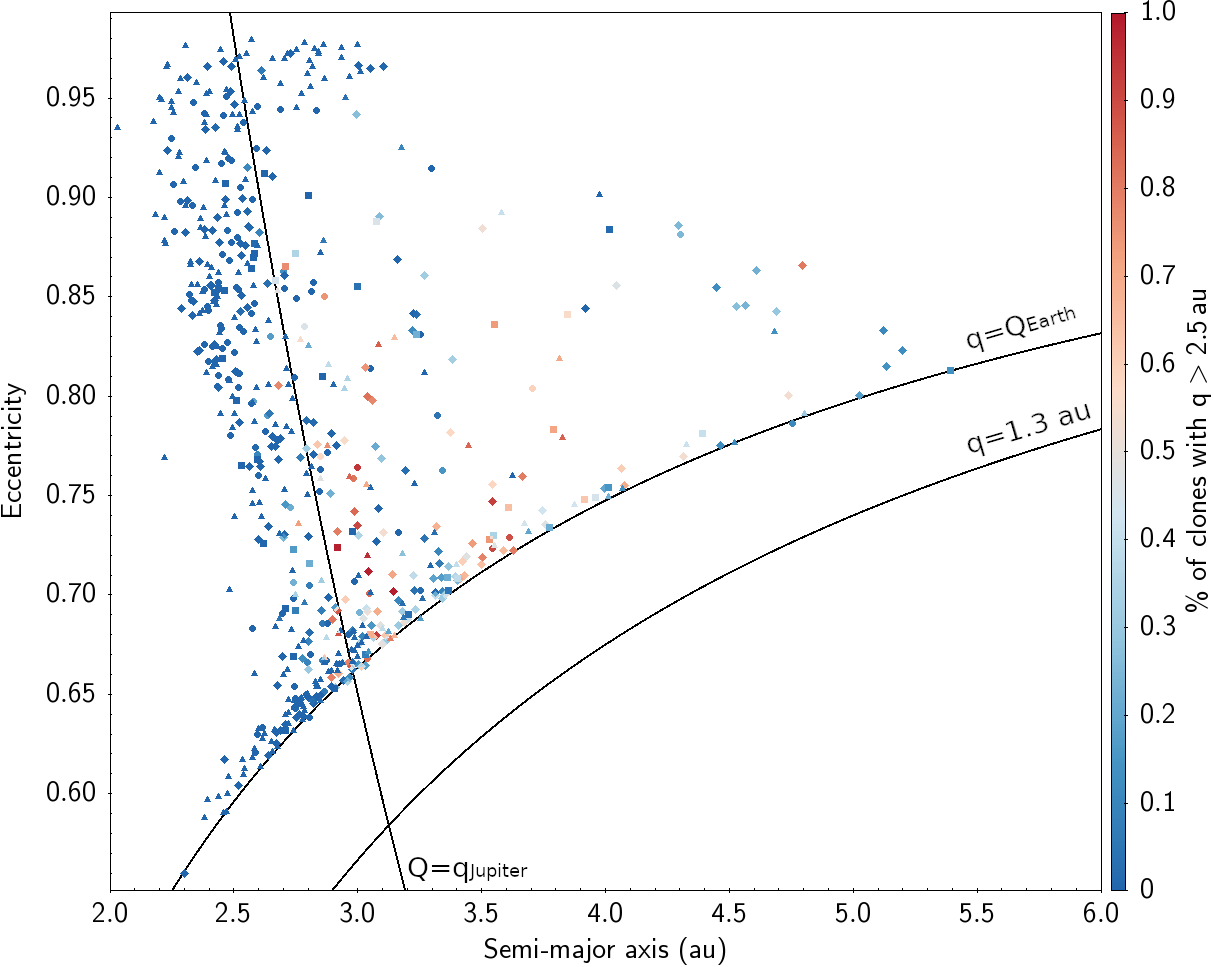}
    \caption{JFC-like Fireballs}
    \includegraphics[width=\linewidth]{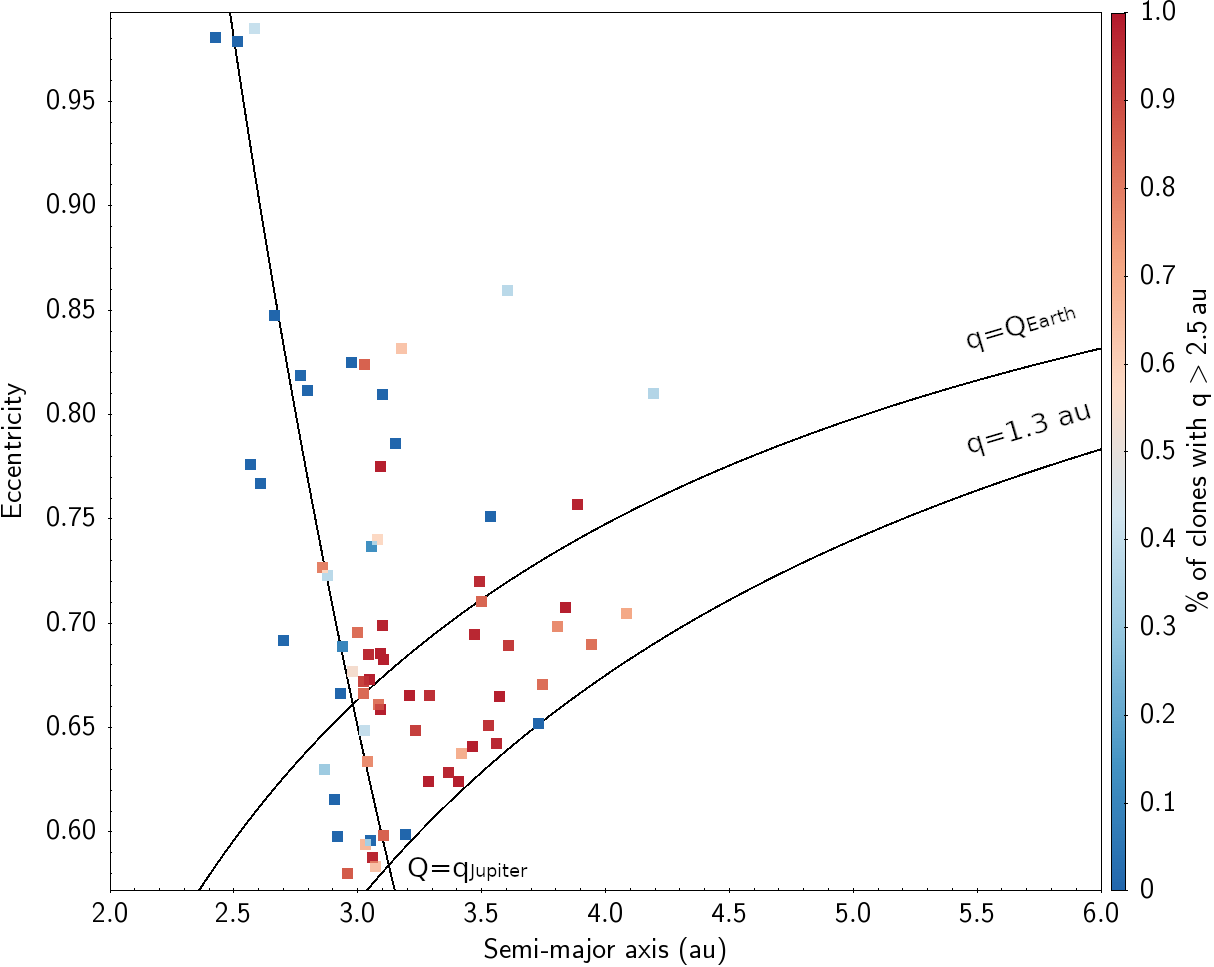}
    \caption{Near-Earth JFCs}
\end{subfigure}
\caption{Comparison of JFC contribution estimated using the NEOMOD model, the $f_{q}$ index, and the percentage of clones that reach q$>$2.5\,au during a 10\,kyr simulation. Subplots (a,c,e) show data for JFC-like ($2<T_{J}<3$) fireball datasets of DFN (circle), EFN (diamond), MORP (square), and FRIPON (triangle). Subplots (b,d,f) show the near-Earth subset of the 661 JFCs used in this study.}
\label{fig:source_region_versus_sim}
\end{sidewaysfigure*}

Based on the NEOMOD model, the JFC population is a more significant source of the JFC-like fireball observations (Fig. \ref{fig:source_region_histogram}) relative to the predictions based on the 10\,kyr dynamics. However, the NEOMOD model still also predicts that a large majority of the fireballs (66-82\%) come from main-belt sources for all the networks. The DFN had the smallest JFC contribution according to the model (18\%), while the MORP and FRIPON networks both had the highest ($\sim$34\%) according to the model. Most surprisingly, it also predicts that only 43\% of the near-Earth JFC population is likely to be sourced from the JFC population, with large proportions coming from the 11:5 (18\%), 5:2 (15\%), and 2:1 (11\%) MMRs. When this is compared to the orbital simulations (Fig. \ref{fig:source_region_versus_sim}), the NEOMOD seems to be overpredicting the contribution of the JFC population to the fireball observations while simultaneously underpredicting for the near-Earth JFC population.

As seen in Fig. \ref{fig:source_region_versus_sim}, NEOMOD predicts very broadly that a large portion of the fireballs beyond 3.5\,au or at higher eccentricities are more likely to come from the JFC source region in the model, a prediction that does not match the stability of these objects during our 10\,kyr simulations. As seen in Fig. \ref{fig:source_region_versus_sim}\,(c,e), all of the high-eccentricity fireballs originate from very stable orbits during the 10\,kyr simulations, which is a strong indicator that these are not directly from the JFC population. The most likely regions to be JFC population in origin tend to concentrate towards the intersection of the line of equal perihelia equal to the Earth's aphelion and the line of equal aphelia corresponding to the perihelion of Jupiter. 

Conversely, the NEOMOD model also does not clearly identify some of the most unstable near-Earth JFCs within our dataset that overlap with the model (69 objects), and underpredicts greatly the instability of the population. However, this is not a mistake of the model. A debiased NEO model works by propagating fictitious particles from several source regions until they leave the simulation, and their period of time in near-Earth space is weighted based on considering the objects discovered by telescopic surveys and their inherent biases. In this way, the model then organizes near-Earth space into bins of \textit{a-e-i}, and by checking the particles within each bin (weighted by observations), the source region likelihoods can be estimated. It is crucially important though to keep in mind that this model considers all objects, not just JFCs. The JFC population is a small minority population within near-Earth space. Thus, even though they are often constrained and defined by having an orbit with a $2<T_{J}<3$, the asteroid population is much more numerous in this region, and thus even a small level of diffusion by outer main belt resonances significantly decreases the overall JFC source region likelihood on these orbits. When only considering the near-Earth JFCs in the Horizons database that are within the bounds of NEOMOD (69 objects), on average, it only predicts a 29-43\% JFC source region contribution (depending on absolute magnitude estimate). To obtain a better estimate, one must include all of the objects, including asteroids, on similar orbits. When the source regions for all 1,860 NEOs on $2<T_{J}<3$ orbits were estimated using the NEOMOD model, we found a much more reasonable estimate of 6-14\%, which corresponds more nicely to our estimate of 8-21\% of the JFC-like fireball population being genetically cometary based on our 10\,kyr simulation results. This demonstrates clearly that although the JFC population is characteristically constrained to these orbits, even a minor diffusion of asteroids from the outer main belt can overwhelm the \textit{a-e-i} statistics. 

\subsection{Meteorite falls}
Another question is whether meteorite-dropping fireballs are contained within the dataset and what kinds of orbits they originate from. The EFN dataset, released by \citet{borovivcka2022_one}, only contains four fireballs from JFC-like orbits ($2<T_{J}<3$) that have a terminal mass estimated to be greater than a gram. Of these four, none originate from beyond 3\,au and $<<$1\% of particles had a close encounter with Jupiter during the 10\,kyr simulations. 

\begin{figure}[]
\centering
\includegraphics[width=0.5\textwidth]{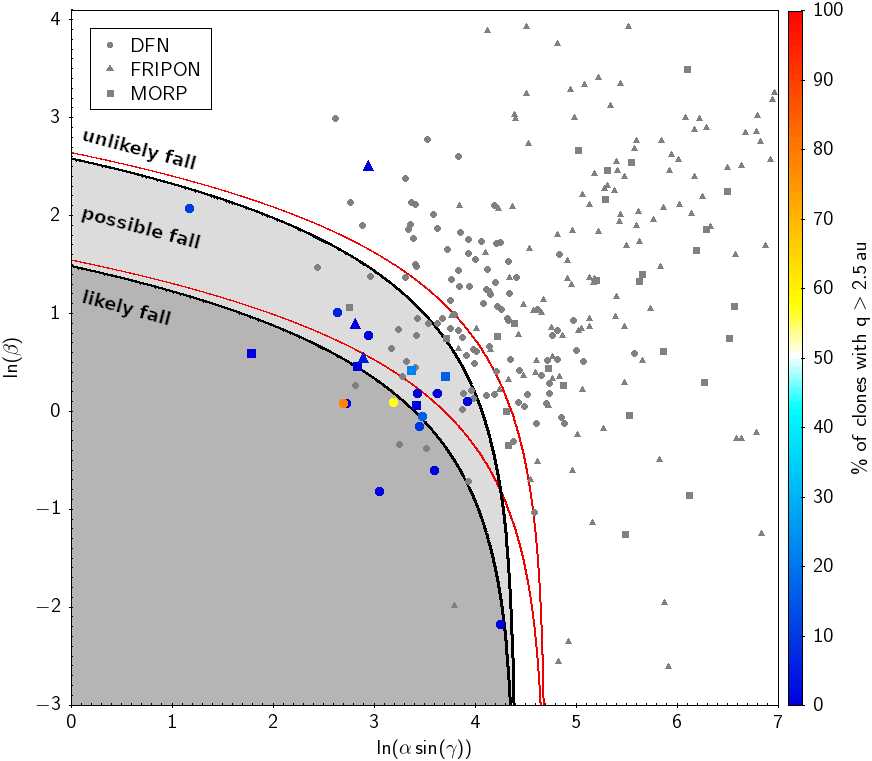} 
\caption{Distribution of $\alpha–\beta$ values for JFC-like fireballs from the DFN, MORP, and FRIPON observation networks. Here $\gamma$ is the trajectory slope relative to the horizontal. If a macroscopic event is considered to have a final mass of $\geq$50 g, assuming $\rho_{m}$ = [2240, 3500] kg m-3 (carbonaceous and ordinary chondrite, respectively) and c$_{dA}$ = 1.5, meteorite-dropping events can easily be identified given the range of possible shape change coefficients ($\mu$). The black lines correspond to the ordinary chondrite density, whereas the red lines denote a carbonaceous meteorite density, described by Eq.\ref{eq:oc_mu0} - \ref{eq:cc_mu23}. The coloration is indicative of the percentage of particle clones within the 10\,kyr simulations that reached perihelion values $>$2.5\,au, and only fireballs that also fulfilled the meteorite-dropping criteria of  V$_{final}$ $<$ 10\,km\,s-1 and H$_{final}$ $<$35\,km are considered likely meteorite droppers (after \citealp{brown2013meteorites}).}
\label{fig:alpha_beta} 
\end{figure}

Otherwise, for the DFN, FRIPON, and MORP networks, we utilized the $\alpha–\beta$ methodology of \citet{sansom2019determining} to identify potential meteorite falls within the dataset. The $\alpha–\beta$ values are easily calculated from the velocity-height data from any fireball and correspond to the ballistic coefficient and mass-loss parameter, respectively. As seen in Fig. \ref{fig:alpha_beta}, a majority of the fireballs are unlikely to drop any meteorite. Assuming an ordinary-chondrite composition, we are expected to have somewhere between 11 and 42 meteorites $>$50\,g from the three fireball networks. Whereas when a carbonaceous chondrite density is used, the estimate grows to between 18 and 62 meteorites $>$50\,g. However, in Fig. \ref{fig:alpha_beta}, only the colored data points that also fall below the limits are considered as the most likely meteorite-dropping events, as they also fulfill the end height/velocity criteria (V$_{final}$ $<$ 10\,km\,s-1 and H$_{final}$ $<$ 35\,km; \citealp{brown2013meteorites}). When combining the end height/velocity criteria with the $\alpha–\beta$ methodology, we find 21 potential meteorite-dropping events. That's 22 potential meteorite-dropping events in total out of the 646 JFC-like fireballs, as only one EFN event is predicted to have a terminal mass $>$50\,g \citep{borovivcka2022_one}.

The coloration in Fig. \ref{fig:alpha_beta} is weighted according to the percentage of particle clones in the 10\,kyr simulations that reach a q$>$2.5\,au, indicating indirectly the probability of a close encounter with Jupiter over the time period. As expected, nearly all of the likely meteorite-dropping events are very stable during the 10\,kyr simulations (dark blue), comprising 20 of the events. However, two DFN events diverge from this trend. DFN events DN210516\_02 (78.9\%) and DN220825\_01 (57.9\%), both have a likelihood of reaching q$>$2.5\,au during the 10\,kyr timescale of over 50\%. However, given that the semi-major axis values of 2.96\,au and 2.85\,au are still within range of the main belt and the longitude of perihelion values of 254.6$^{\circ}$ and 51.0$^{\circ}$ are not near the concentration of JFC values, the meteoroids could have still very easily be transferred from the main-belt to these orbits. Additionally, several studies have demonstrated that debris can migrate out onto these orbits from the main belt \citep{fernandez2015jupiter,shober2020did,shober2020using,hsieh2020potential}. Also, none of the identified likely meteorite-droppers have an association with a meteor shower. 

\subsection{Shower Associations}
As seen in Fig. \ref{fig:shower_Dv}, several meteor showers are present in this dataset. Some of these are edge cases that have drifted over the $T_{J}\sim3$ boundary, while others have likely parent bodies that also inhabit the Jupiter-approaching region. In total, we found nine meteor shower clusters of at least 3 fireballs with a $D_{N}<0.1$, a limit that produces limited false positive rates for the datasets \citep{shober2024generalizable}. There are more possible shower associations beyond just these nine showers, but these nine are the only clear clusters with several fireballs in each. In Fig. \ref{fig:shower_Dv}, we have plotted the minimum $D_{N}$ for each fireball after having compared every fireball to every established shower on the IAU Meteor Data Center's (MDC) database\footnote{\url{https://www.ta3.sk/IAUC22DB/MDC2007/Roje/roje_lista.php?corobic_roje=1&sort_roje=0}}. These showers include: 

\subsubsection{Quadrantids (QUA)}
The Quadrantids meteor shower was only detected by the FRIPON network within this study, with approximately 20\,events matching the shower characteristics. The lack of detections by the other networks could be a result of observational bias against the detection of smaller, fainter meteoroids or due to weather (e.g. \citealp{borovivcka2022_one}) because the shower has a very short peak window (12-14 hours; \citealp{wiegert2005quadrantid}). Despite its short duration, the Quadrantids meteor shower is a strong annual shower. Peaking around early January, it typically displays a zenith hourly rate of approximately 110 to 130 meteors under optimal viewing conditions \citep{abedin2015age}. The parent body of the Quadrantid stream is identified as 2003 EH1, an Amor-type asteroid with a comet-like orbit ($T_{J}=2.06$), indicating its potential as a dormant or extinct comet \citep{jenniskens20042003}. Given the very high inclination of the stream ($\sim72^{\circ}$) and the non-corresponding $\varpi$ of 90-100$^{\circ}$, the proposed parent body (2003 EH1) is currently not in good agreement with the core distribution of JFCs. However, the object exists in a Kozai-resonance, along with other possible related bodies, that could lower the inclination in exchange for eccentricity and render the orbit more resembling the bulk of JFCs. The degree of separation of 2003 EH1 and the Quadrantid stream is most consistent with a quick release of material around 200 years ago, possibly from a fracturing event of a comet nucleus \citep{wiegert2005quadrantid}. 

\subsubsection{Southern $\delta$-Aquariids (SDAs)}
The Southern $\delta$-Aquariids are annually observed from late May to early July, reaching their peak activity when the solar longitude is at $\lambda$ = 126$^{\circ}$ \citep{brown2010meteoroid}. The northern branch, although weaker, displays activity from late July to late August, with its maximum activity occurring at $\lambda$ = 139$^{\circ}$. Previously proposed to be related to the Marsden group comets, the SDA are now generally thought to be related to the 96P/Machholz complex \citep{abedin2018formation}. This complex is also believed to include other major showers and bodies such as asteroid 2003 EH1 and the corresponding QUA shower, a result of a Kozai circulation cycle causing eight intersection points with the Earth's orbits \citep{babadzhanov2008meteor}. The Marsden group of comets was found to be incapable of producing the observed profile features of the SDAs alone \citep{abedin2018formation}. \citet{abedin2018formation} found that the most likely scenario to reproduce the principal activity features of the Quandrantids, Arietids, Southern $\delta$-Aquariids, Northern $\delta$-Aquariids, and several other showers would be the capture of comet 96P/Machholz into a short-period orbit -$\sim$22\,kyr ago. However, they also found that the match to the activity profile could be improved if the comet 96P/Machholz broke up after 100\,CE to form the Marsden group comets, an idea proposed by \citet{sekanina2005origin}. 

For the SDA, there are also 17 fireballs in the dataset with a D$_{N}<0.1$, and they are also almost entirely observed by FRIPON with 16 observations. There are also one associated with the shower observed by the EFN. This abundance of QUA and SDA observations by the FRIPON network demonstrates clearly that these four networks are observing slightly different size ranges. The FRIPON network has the highest limiting magnitude (i.e., it observes the faintest/smallest meteoroids) of the four networks analyzed here. The meteoroids of the QUA and SDA showers are normally only millimeters or less in size, thus limiting the detection rate by the fireball networks. 

\subsubsection{$\alpha$-Capricornids (CAP)}
The EFN, MORP, and FRIPON networks all contribute to the 15 fireballs associated with the $\alpha$-Capricornids within the dataset, with 10, 3, and 2, respectively. The CAP shower is notable for its slow (21-23\,km\,s$^{-1}$) and often bright meteors emanating from the constellation Capricorn. Short-period comet 169P/NEAT (2002 EX12) has been identified as the likely parent body of this shower \citep{jenniskens2010minor}. Comet 169P, as described by \citet{kasuga2010comet}, is a medium-sized (2.3$\pm$0.4\,km) nearly dormant comet. The numerical simulations of \citet{jenniskens20042003} found the object likely had a massive breakup 4000-5000\,yrs ago, losing about half of the comet's original mass. The $\alpha$-Capricornids we observe today are just the outskirts of the resulting dust complex \citep{jenniskens2010minor}. Over the centuries, a fraction of this dust has evolved into longer orbits and is now intersecting Earth's orbit, giving rise to the observed meteor shower. There are also other proposals for associated parents of the $\alpha$-Capricornids. Asteroid 2017 MB1 has been recently proposed as being associated with the stream \citep{ye2018meteor}. Another candidate is P/2003 T12 (SOHO), which, according to \citet{fernandez2015jupiter}, may share a common origin with comet 169P following a fragmentation event around 2900 years ago. However, \citet{fernandez2015jupiter} also argued this common origin may have originated in the main asteroid belt based on the stability of the orbits, a result supported by the 10\,kyr simulation results of this study (Fig. \ref{fig:shower_q25}). 

\subsubsection{Taurid Complex - Northern Taurids (NTAs) and Southern Taurids (STAs)}
The Taurid meteor shower, encompassing the Northern and Southern branches, is a complicated, puzzling shower of long duration and wide breadth. Many bodies have been speculated to be associated with the complex based on their similar orbital history and proximity \citep{egal2021dynamical}. However, Comet 2P/Encke is widely recognized as the primary parent of the shower. The Taurid complex is characterized by its low inclination and perihelia between 0.2 and 0.5\,au, influenced by planetary perturbations, resulting in its very diffuse structure \citep{matlovivc2017spectra}. The entire mass of the Taurid stream, estimated to be around $10^{14}$\,kg, is thought to have been released over the last 6 to 30\,kyrs, indicating a mass loss rate higher than that of typical Jupiter family comets \citep{jenniskens2006meteor}. Also, this large stream of bodies contains a diversity of size ranges, even observed to contain meter-scale debris \citep{brown2013meteorites,spurny2017discovery,devillepoix2021taurid}. However, based on the spectra and fragmentation characteristics, typical break up around 0.02–0.10 MPa, the meteoroids seem to be quite weak and possibly carbonaceous-like \citep{matlovivc2017spectra}. Most of the bodies are detached from Jupiter ($T_{J}>3$), but some can be observed on $T_{J}<3$ orbits, as seen by 21 fireballs in the study. These fireballs were observed by MORP, EFN, and FRIPON, consisting of 15 NTA fireballs (5 MORP, 6 EFN, 4 FRIPON) and 6 STA fireballs (4 EFN, 2 MORP). 

\subsubsection{$\eta$-Virginids (EVI)}
The $\eta$-Virginids is a minor shower that occurs annually during February and March. It is known to have stronger meteoroids that penetrate further into the atmosphere, similar to the Geminids \citep{jenniskens2016established,borovivcka2022_two}. This study has 12 associated with the shower, 4 DFN, 3 EFN, and 4 FRIPON events. This is the only shower distinctly present within the DFN JFC-like ($T_{J}>3$) fireball database. 

\subsubsection{August Draconids (AUD) and $\kappa$-Cygnids (KCG)}
A few minor showers are often overshadowed by the Perseids, which also occur in August. Two of these are found within the fireball datasets: the August Draconids and the $\kappa$-Cygnids. First reported in 1877, KCG is known for its activity between August 3 and 31, peaking around August 18 \citep{jenniskens2006meteor}. The asteroid 2008 ED69 is proposed to be the parent body of the stream, originating during a breakup event 2–3 nutation cycles ago (3600–6000 yrs ago; \citealp{jenniskens2008minor}). Additionally, \citet{jones2006kappa} found that asteroids 2001 MG1 and 2004 LA12 have similar orbits to the KCG. Within the JFC-like databases examined here, we also find a diffuse cluster of 12 fireballs that may be associated with the August Draconids or $\kappa$-Cygnids, with 6 for each respectively. Most of AUD fireballs, with eight in total, were observed by the EFN with the rest observed by FRIPON. The KCG fireballs were observed by the EFN (3), FRIPON (2), and MORP (1). 

\subsubsection{October Draconids (DRA)}
Lastly, we have three October Draconid fireballs observed by the EFN. Considering the low $D_{N}$ values of $\sim$0.01, and the similar dynamics of the meteoroid orbit, these are very likely genuine despite the low numbers. The Draconid meteor shower, associated with the JFC 21P/Giacobini-Zinner, is known for its occurrence in early October and is notable for the low velocity of its meteors at around 20 km/s. The Draconids are distinctive due to their fragile composition and low density \citep{borovivcka2007atmospheric}. The Draconid meteoroids show varying grain compositions; some are coarse-grained while others are fine-grained. This variance in grain size impacts the meteoroid's interaction with the atmosphere, influencing the light curves and deceleration patterns observed. Elements such as Na, Mg, and Fe are present in nearly chondritic proportions, but the differential ablation causes a preferential loss of sodium at the beginning of the trajectory. These findings align with the understanding of cometary dust as a complex mixture of materials with diverse physical and chemical properties \citet{borovivcka2007atmospheric}.

\subsubsection{Shower dynamics}
\begin{figure}[]
    \centering
    \begin{subfigure}[b]{0.5\textwidth}
        \includegraphics[width=\textwidth]{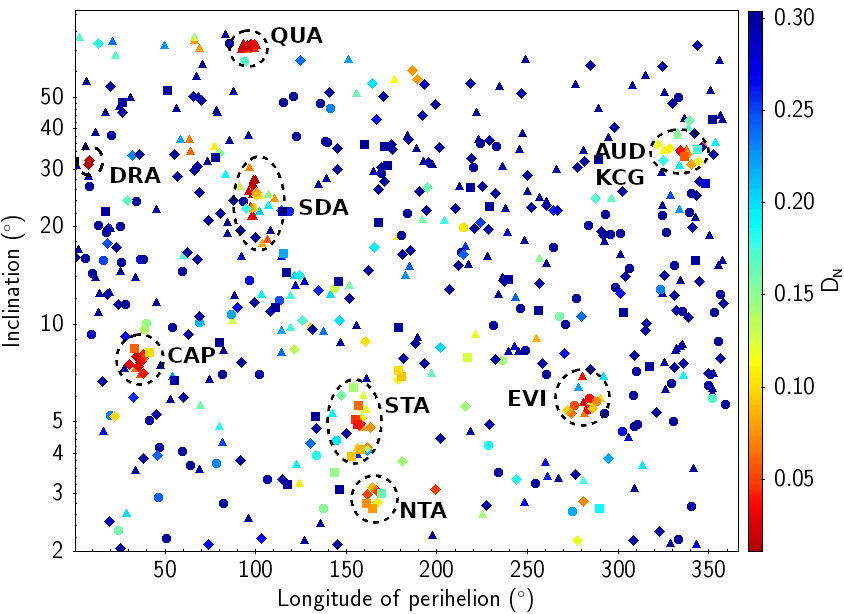}
        \caption{}
        \label{fig:shower_Dv}
    \end{subfigure}
    \begin{subfigure}[b]{0.5\textwidth}
        \includegraphics[width=\textwidth]{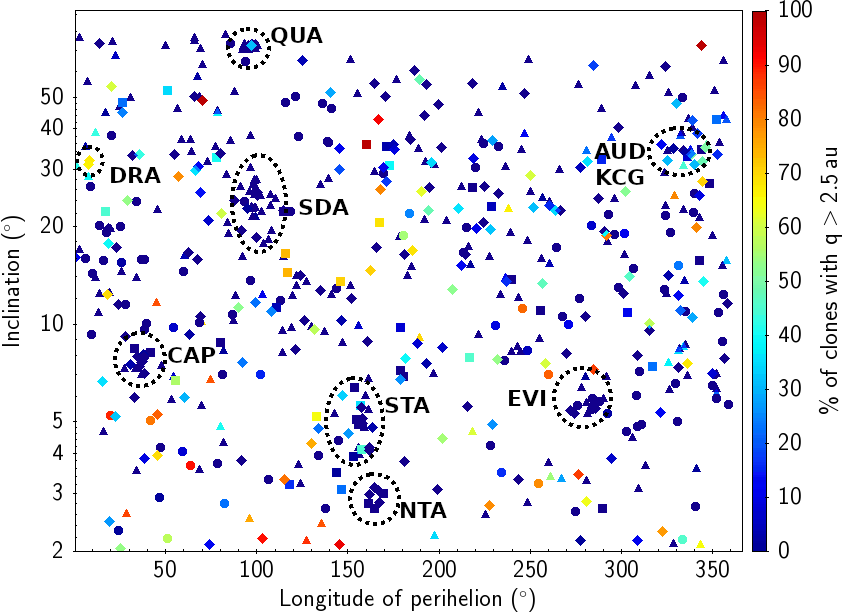}
        \caption{}
        \label{fig:shower_q25}
    \end{subfigure}
    \caption{Meteor showers shown in longitude of perihelion ($^{\circ}$) versus inclination ($^{\circ}$) distributions for JFC-like fireballs detected by DFN (circle), EFN (diamond), MORP (square), and FRIPON (triangle). Meteor shower clusters of 3+ fireballs are denoted with their official abbreviations. The coloration of subplot (a) corresponds to the minimum $D_{N}$ calculated relative to all of members of IAU MDC's established shower list. The coloration of sublot (b) corresponds to the number of particle clones generated within the observational uncertainties of each fireball that reached a perihelion of 2.5\,au or greater during a 10\,kyr N-body simulation.}
    \label{fig:showers}
\end{figure}

Surprisingly, despite the evident presence of showers in Fig. \ref{fig:shower_Dv} in the dataset, it is clear that, in fact, nearly all of the showers elude close encounters with Jupiter that send them on higher perihelion orbits. When compared to the sporadic background, in Fig. \ref{fig:shower_q25}, all the showers are dark blue in color, signifying there was a low percentage of particle clones that reached q$>$2.5\,au during the 10\,kyr simulations. However, there is one shower that stands out as being very consistent with JFC dynamics: the October Draconids. Although there are only three EFN Draconid events, all three have a 65-70\% probability of reaching a q$>$2.5\,au within a 10\,kyr period. There are multiple other showers mention here also thought to have a cometary parent body; however, these parents tend not to be explicitly classified as JFCs. Furthermore, often, there seem to be several associated objects on nearby orbits or those with similar dynamics, and these bodies are asteroids, comets, or something in between. For example, comet 2P/Encke, which is widely accepted as the primary parent of the Taurid Complex, is detached from Jupiter dynamically and displays a mixture of characteristics of asteroids and comets. To be clear, all of these objects exhibit chaos at some level and timescale. Nevertheless, the chaotic behavior of JFCs is primarily dominated by very close encounters with Jupiter which cause macroscopic changes in the orbits on millennium timescales, and only the three Draconids here are consistent with that diagnostic feature of the population. 

\subsection{Case studies}
Over the previous decades, there have been a handful of large JFC-associated fireball case studies reported by several networks across the globe. These individual fireball observations often claim to have witnessed large, strong material on a JFC-like orbit, indicating the presence of durable material and the possibility of JFC meteorites. Here, we have done the same 10\,kyrs numerical simulations based on the reported orbits of three of such JFC-like fireballs (Table~\ref{tab:case_studies}), generating 1000 particle clones within the orbital uncertainties provided. 

\begin{table*}
    \centering
    \begin{tabular}{|c|c|c|c|c|c|c|c|}
        \hline 
        Network & a (au) & e & $\iota$ ($^{\circ}$) & \% of clones q$>$2.5 & $f_{q}$ & $f_{a}$ & Ref. \\
        \hline
        \hline
        SPMN & 4.3$\pm$0.7 & 0.79$\pm$0.04 & 7.1$\pm$0.1 & 17.3 & 0.07 & 0.08 & \citet{madiedo2014trajectory} \\
        SPMN & 2.89$\pm$0.27 & 0.737$\pm$0.024 & 29.9$\pm$0.6 & 18.9 & 0.05 & 0.03 & \citet{trigo2009observations} \\
        SACN & 3.58$\pm$0.02 & 0.725$\pm$0.002 & 9.06$\pm$0.02 & 39.7 & 0.15 & 0.08 & \citet{hughes2022analysis} \\
        \hline 
    \end{tabular}
    \caption{Orbital elements (a-e-i) and 10\,kyr simulation results for three very bright fireballs reported to be originating from the JFC population observed by the Spanish Meteor Netork (SPMN) and the Spalding Allsky Camera Network (SACN).}
    \label{tab:case_studies}
\end{table*}

As seen in Table \ref{tab:case_studies}, all three of these events have low probabilities of their q values reaching 2.5\,au on 10\,kyr timescales. This is not surprising for the two SPMN fireballs, as both have extremely high orbital uncertainties and can easily be argued as main belt in origin. The published semi-major axis values for the fireball reported by \citet{madiedo2014trajectory} is extraordinarily high, however, the uncertainties are so high that it is within 3\,$\sigma$ of the entire main belt. Meanwhile, the very bright (19-magnitude) fireball reported by \cite{hughes2022analysis} has the highest likelihood of being a genuine JFC, but it still had $<$50\% of the particle clones in the simulation reach a perihelion above 2.5\,au, which is not statistically consistent with JFCs. Also, while the orbit reported by \citet{hughes2022analysis} is likely very precise as a multi-parameter fit was used \citep{Gural_2012M&PS}, the uncertainties are likely vastly underestimated. The uncertainties would be expected to be larger, given that only two observations were used for the triangulation \citep{shober2023comparison}. Also, both observations were over both 200\,km away with a convergence angle of less than 5$^{\circ}$. The initial velocity uncertainty expected for such a low convergence angle should be $>$200\,\si{\meter\per\second} according to the work of \citet{jansen2020dynamic} (Figure. 2) -- four times greater than the $\pm$50\,\si{\meter\per\second} uncertainty quoted in \citet{hughes2022analysis}.


\section{Discussion}

\subsection{Orbital stability and dynamics}
The analysis of the orbital trajectories of 661 JFCs retrieved from the NASA HORIZONS database\footnote{\url{https://ssd.jpl.nasa.gov/horizons/}} shows a consistent trend with previous studies, with a majority of the population characterized by their frequent close encounters with Jupiter over 10,000 years. These close encounters rapidly change the orbits, particularly the perihelion distance on hundred to thousand-year timescales. In addition, we also find that the JFC population tends to concentrate toward low Lyapunov lifetimes (50-150\,yrs). This result is also consistent with previous studies of the chaos present in this population, which tends to reduce the dynamical memory of these objects \citep{tancredi1995dynamical,tancredi1998chaotic}. However, as we start to examine JFCs in active orbits (q$<$2.5\,au) or even in near-Earth space, this dynamical fingerprint begins to get blurry. The study by \citet{fernandez2015jupiter} found many stable JFCs within near-Earth orbits, and this result is replicated here where we find 24 extremely stable JFCs on "active" orbits (22 of which are NEOs). These objects could be asteroids that have diffused out from the main asteroid belt or possibly comets that are in an anomalously longer period of orbital stability. Contamination from the main belt would not be completely unexpected, as JFCs seldom spend time in the inner solar system, and those periods are often only a few thousand years \citep{fernandez2002there}. Furthermore, there is only expected to be 0.1-1\% of objects transferred from the JFC region to detached orbits \citep{hsieh2016potential}, a value 4-40 times lower than the rate described here if all the bodies in Table~\ref{tab:22_extremely_stable} are detached comets. This low percentage rate of detached orbits could still be interesting as the dynamical lifetimes are much longer on these orbits that no longer encounter Jupiter regularly. However, we would also need to have a highly physically evolved object with an insulating layer present in order to survive the longer timescales in the inner solar system, where JFCs are believed to have physical lifetimes of a few thousand years \citep{levison1997kuiper,disisto2009population}. This is believed to possibly be the case for comet 2P/Encke, which is the parent body of the very active and large Taurid meteor shower complex. The study by \citet{levison2006origin} found that comet 2P/Encke's detachment from Jupiter could have resulted from gravitational interactions with JFCs and terrestrial planets alone, however, these transitions are significantly longer than the typical physical lifetimes of comets ($\sim 10^{3}$\,yrs). The study suggests that 2P/Encke likely became dormant after being influenced by Jupiter and reactivated when its perihelion distance was reduced by the $\nu_{6}$ secular resonance. The research also highlights the need for considering non-gravitational forces in explaining 2P/Encke's orbit, while warning against using over-simplified constant non-gravitational forces to achieve shorter timescales.  

However, given the considerable number of JFCs in near-Earth orbits found to move on stable trajectories by \citet{fernandez2015jupiter} and in this study, the contamination of dark, hydrated outer main belt asteroids in JFC-like orbits is also quite considerable. It has since been demonstrated that asteroids from the outer main belt are capable of diffusing out due to eccentricity excitation by the 2:1 MMR and subsequent terrestrial planets encounters \citep{hsieh2020potential}. The objects from the outer main belt that are primarily dominated by D- and P-type asteroids, could diffuse onto comet-like orbits and be very difficult to discern from pristine comets from the trans-Neptunian regions. 

In the 661 JFCs studied here, we found a significant number of objects with stable trajectories on $10^{4}$\,yr timescales. Of the JFCs in the NASA HORIZONS database on near-Earth orbits, we find that 22 of the 75 (approximately 30\%) move on extremely stable orbits, with nearly all of the 1000 particle clones per object avoiding close encounters with Jupiter. Six of these were designated as lost comets. Comets 3D/Biela, 5D/Brorsen, and 34D/Gale suffered major splitting events and were lost due to their physical breakdown \cite{fernandez2009s}. The other three are D/1766 G1 (Helfenzrieder), D/1770 L1 (Lexell), and D/1895 Q1 (Swift), which were all observed for only one perihelion passage and were lost due to their orbital uncertainty. The three comets lost to disintegration and splitting events might indicate that the longer stability of their orbits in near-Earth space might be a contributing factor. Their arrival into a more stable orbit, spending more time in the `active' region compared to most JFCs, could explain their physical demise. Thus, ignoring the three comets lost to high uncertainties and considering the comets lost to splitting events as genuine JFCs this gives us a final estimate of approximately 22\% (16 out of 72 objects) as potentially being asteroidal interlopers. This is consistent with the findings of \citet{fernandez2015jupiter}, which found that between 14-33\% of near-Earth JFCs have asteroid-like trajectories on 10\,kyr timescales. In addition, 15 of the 16 are found to have a$<$3.2\,au, which strengthens the argument for an origin in the main belt. The only very stable body with a considerable semi-major axis is 66P/du Toit (a$\sim$6.02). Comet 66P/du Toit was similarly identified by \citet{fernandez2015jupiter} for being highly stable, and they found the object to be trapped in a 4:5 MMR with Jupiter with critical angle librating around 180$^{\circ}$. Our estimate could be slightly higher as these are just the objects on extremely stable orbits, and there were other objects in our analysis that had lower-than-average likelihoods of close encounters with Jupiter. 

The study by \citet{ye2016dormant} found by analyzing 13,567,542 meteor orbits recorded by Canadian Meteor Orbit Radar (CMOR), which captures the ionized trails of mostly submillimeter-sized objects hitting the atmosphere, comets likely disrupt more often than they become dormant. Their study used these meteor observations to identify dormant comets within the near-Earth object (NEO) population, and they identified five statistically significant meteor showers, suggesting a dormant comet fraction of at least 2.0 $\pm$ 1.7\% in the NEO population. The findings indicate a low dormancy rate of approximately $10^{-5} yr^{-1}$ per comet and support the hypothesis that disruption and dynamical removal, rather than dormancy, are the prevalent end states for near-Earth JFCs. These studies have estimated that there could be as many as 20-30 dormant comets in near-Earth space \citep{mommert2015exploreneos,ye2016dormant}. Other similar estimates have been made for the proportion of dormant comets in the NEO population, assuming the number on $T_{J}>3$ orbits is negligible, and they range between 2-14\% \citep{bottke2002debiased,fernandez2005albedos,mommert2015exploreneos,granvik2018debiased}. There are only 75 near-Earth JFCs, of which a quarter have very asteroid-like dynamics. At the same time, there are currently 1,875 NEAs with JFC-like orbits, according to the NASA HORIZONS database. If this is the case, it is not surprising to find an abundance of centimeter-to-meter debris with asteroid-like dynamics on $2<T_{J}<3$ orbits, as there are two orders of magnitude more objects at 100\,m to km scale from the main belt compared to JFCs. 

Fireball networks also often record events originating from these JFC-like orbits, and these have been used to better understand the JFC population. However, based on the dynamics results of \citet{shober2021main}, only 3 of the 50 sporadic JFC-like fireballs detected by the DFN displayed comet-like frequent encounters with Jupiter. \citet{shober2021main} interpreted this finding as indicating that almost none of the sporadic JFC-like meteoroids impacting the Earth were originating from JFCs. More recently, \citet{borovivcka2022_one} examined the 824 fireballs detected by the EFN between 2017-2018 and disputed these claims. They argued that the use of dynamic masses (i.e., mass estimates based on deceleration profiles) in the methodology of \citet{shober2021main} prohibited the use of low-deceleration fireballs and thus severely biased the data. This use of dynamic masses is certainly a caveat, and the analysis of \citet{borovivcka2022_two} correctly describes the inability to use them for the PE criterion -- a criterion used to describe the strengths of meteoroids based on a fireball's ablation characteristics. However, \citet{borovivcka2022_two} assumed that low-deceleration events were removed, and this is untrue, as can be seen in the data shared by the study where 29 of the 50 fireballs (58\%) have less than 20\% deceleration. In contrast, the EFN JFC-like dataset contains 66\% events with low deceleration. When a chi-squared test is applied to see whether the EFN has statistically significantly more low-deceleration events as claimed by \citet{borovivcka2022_two}, a P-value of 0.373 is found, i.e., the datasets do not differ significantly as claimed. 

Still, there are significant differences between the orbital datasets of the EFN and DFN due to the hardware of the networks themselves. As explained in \citet{howie2017submillisecond}, the absolute timing of the DFN is recorded very differently compared to the EFN, as the timing is encoded into the image itself in a de-Bruijn sequence using a liquid-crystal shutter. By doing this, the DFN can achieve submillisecond precision on timing. However, the fireball events need to be at least one second long to have enough points in the de Bruijn sequence to get a unique solution. This one-second minimum for the DFN was viewed as acceptable as the primary objective of the network is to recover meteorites; an objective achieved as the team has now used observations in the recovery of 15 meteorites around the world\footnote{\url{https://dfn.gfo.rocks/meteorites.html}}. 

\begin{figure}[]
\centering
\includegraphics[width=0.5\textwidth]{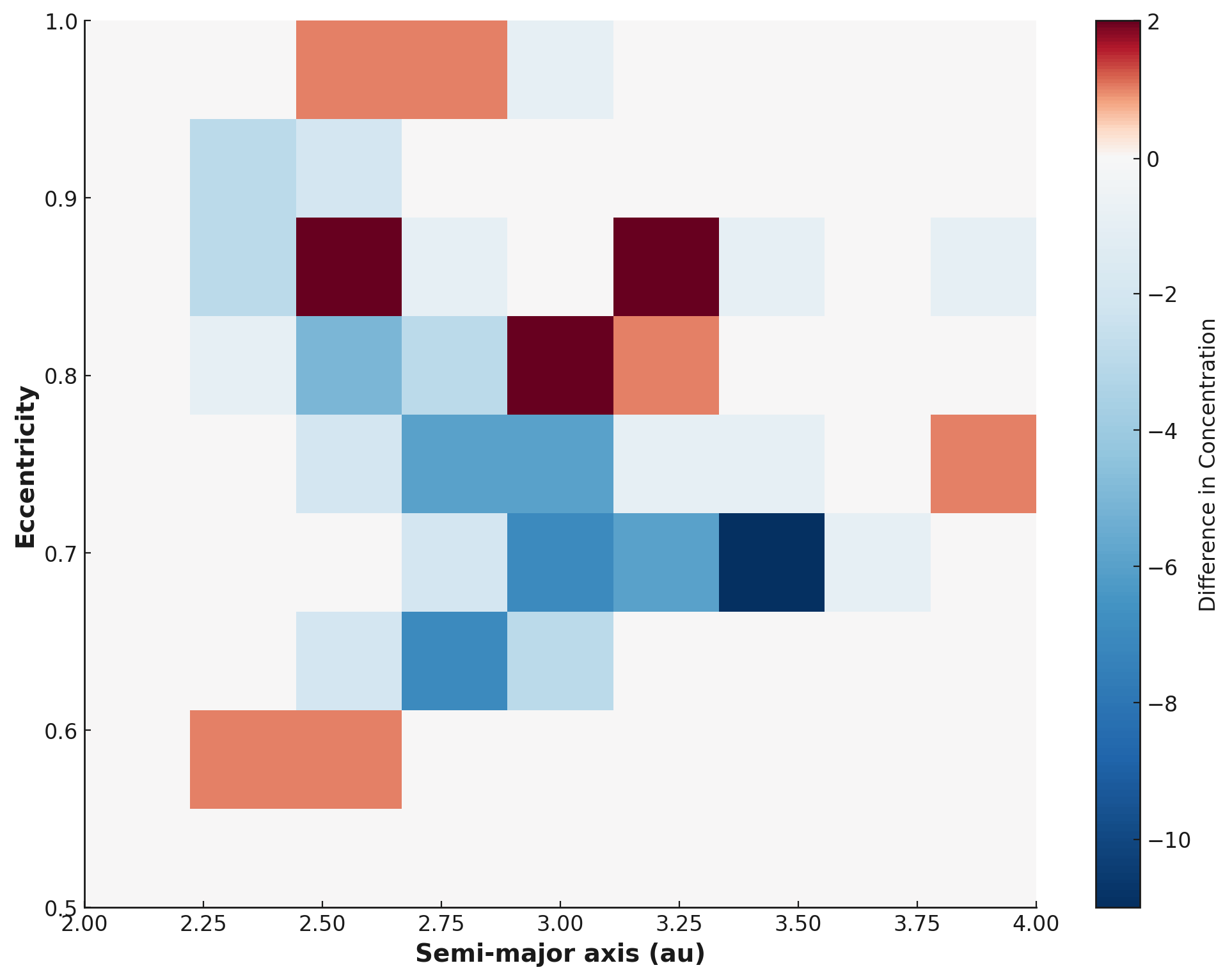} 
\caption{Semi-major axis (au) versus eccentricity heat map, marking the differences in concentration of fireballs detected by the EFN on JFC-like orbits \citep{borovivcka2022_one} for fireballs of less than one second and greater than one second. Blue areas indicate regions where $<$1\,second fireballs have a higher concentration of data points compared to the number of fireballs $>$1\,second, and the red is the inverse}
\label{fig:1sec_diff_EFN} 
\end{figure}

Thus, a bias is present in the DFN data against short-duration fireballs. To see how this bias would affect the data, assuming the EFN dataset is comparable in every other way, we compared the orbital distribution of fireballs $<$1\,sec (29\%) and $>$1\,sec (71\%) found within the EFN dataset \citep{borovivcka2022_one}. As seen in Fig. \ref{fig:1sec_diff_EFN}, this bias within the EFN dataset seems to have the greatest effect on reducing the number of fireballs originating from between 3.0-3.5\,au. This coincides nicely with the range described as being the ``comet domain'' in \citet{borovivcka2022_two}, seeming to support their claim that the results of \citet{shober2021main} are missing the cometary component (however, for a different reason than they claimed). This breaks down however, as a vast majority of JFC-like EFN fireballs from \citet{borovivcka2022_one} and 66\% of those beyond 3\,au have relatively low probabilities ($<$50\%) of close encounters with Jupiter on 10\,kyrs timescales (Fig. \ref{fig:all_fq_fa}) and have more dispersed and larger Lyapunov lifetimes compared to the JFC population (Fig. \ref{fig:moid_lyapunov}). Additionally, this ``comet domain'' indicated by \citet{borovivcka2022_two} to mostly contain weak cometary material with a subset of strong asteroidal objects, does not account for the 2:1 MMR which can produce asteroids on comet resembling orbits \citep{hsieh2020potential}. As seen in Fig. \ref{fig:EFN_21_MMR}, 62 of the 213 JFC-like EFN events are within 3$\sigma$ uncertainty of the resonance, likely indicating it is a source of these weaker meteoroids on stable orbits. 

Based on our 10\,kyr numerical simulations, the DFN, EFN, FRIPON, and MORP have 9\%, 21\%, 8\%, and 18\% respectively of their JFC-like fireball population likely ($>$50\%) to reach q$>$2.5\,au over the integration time. This indicates that only a minor fraction of the JFC-like fireballs are pristine cometary material delivered from the trans-Neptunian region. The variations between the datasets are likely due to a mixture of hardware and data processing choices which influence the composition of the datasets. While we are still uncertain of all the exact causes of the differences between the four datasets, based on the dynamics and fireball observations, the EFN network's abundance of shower and weaker meteoroids likely indicates that the EFN JFC-like subset contains some of the smallest, faintest objects. The FRIPON network contains much more meteors within the dataset, but given the importance of orbital uncertainties in analyzing the dynamics of these fireballs, the FRIPON observations were limited to fireballs observed by four or more cameras. This likely introduces another bias into the dataset; however, based on preliminary analysis of FRIPON event with only two or three observations \citep{shober2023comparison}, this choice is a reasonable one. Understanding these differences and accurately characterizing the biases of the datasets is critically important to better understand the meteoroid population as a whole. As equipment to conduct fireball and meteor observations continues to become more affordable, we need to start sharing our data reduction methods and hardware specifications more openly to compare and ensure the data being collected by the community is translatable \citep{shober2023comparison}.

\begin{figure}[]
\centering
\includegraphics[width=0.5\textwidth]{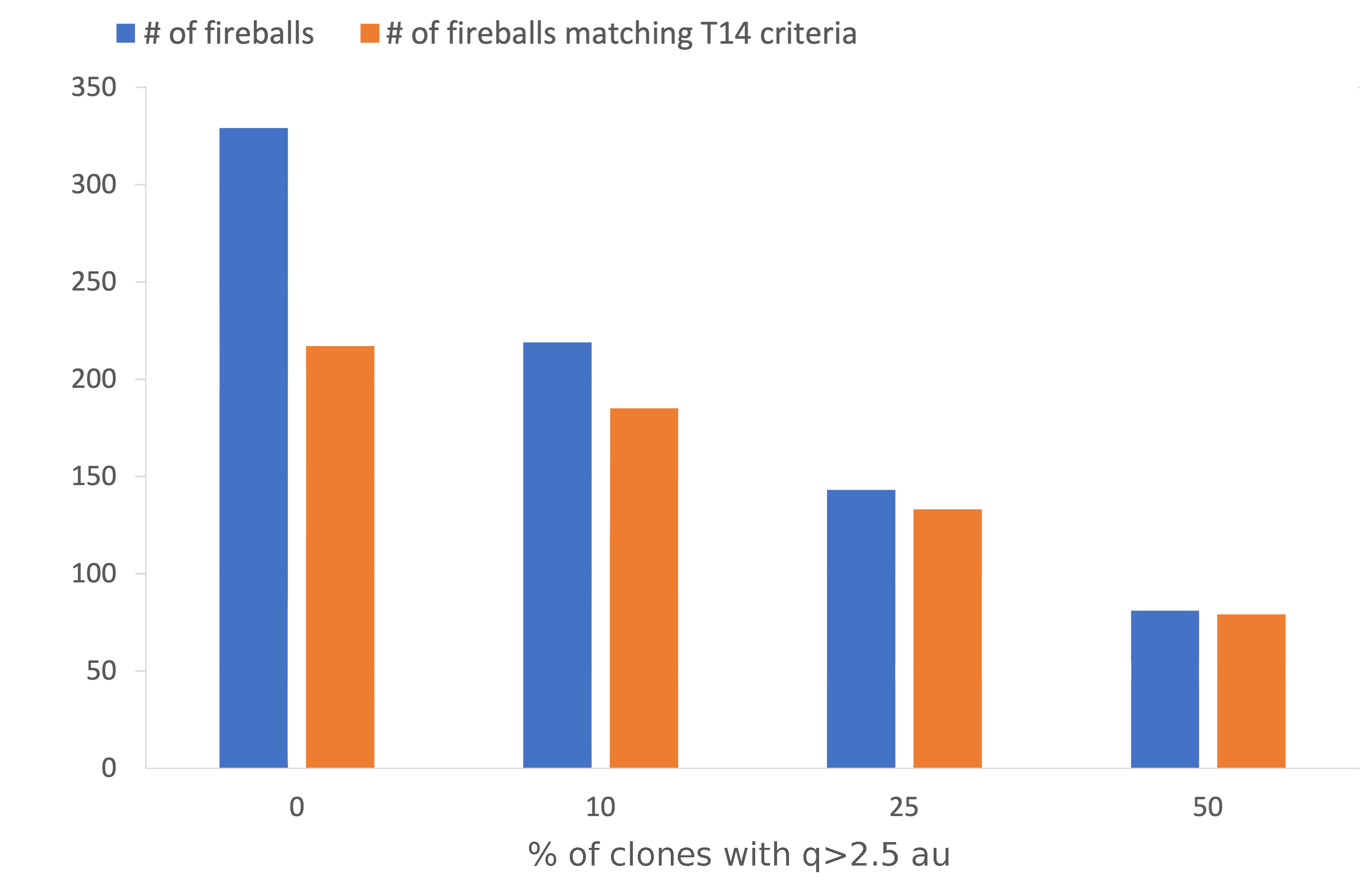} 
\caption{Comparison of the proportion of the 646 ``JFC-like'' ($2 < T_J < 3$) fireball dataset that also fulfills the requirements of the stricter criteria defined by \citet{tancredi2014criterion} for cometary orbit membership as a function of the percentage of particle clones that reach q$>$2.5\,au orbits during our 10\,kyr integrations. The blue bar is the total number of meteoroids where the fraction of clones reaching q$>$2.5\,au is above a certain value (0\%, 10\%, 25\% \& 50\%), whereas the orange bar is the subset of the meteoroids that also fulfill T14's cometary orbit classification. As the likelihood of having a significant encounter(s) with Jupiter to drive the perihelion values up increases, the correspondence between the T14 criterion and the dynamics is more evident.}
\label{fig:T14_comparison} 
\end{figure}

When we compare the various criteria for classifying fireball orbits as potential JFC-like candidates, we notice that while 646 orbits match the simple criterion ($2 < T_J < 3$), there are only 237 fulfilling the more stringent T14's criteria, and 329 where at least one clone q$>$2.5\,au over the integration time. A detailed comparison between the T14's orbital criteria and the \citet{fernandez2015jupiter}'s dynamical criteria can be done by comparing the T14's classification of the fireballs as a function of the fraction of clones reaching q$>$2.5\,au over the integration time. A large fraction of clones correspond to unstable orbits, typically of JFCs. A comparison of these criteria is shown in Fig.~\ref{fig:T14_comparison}: the blue bars are the number of meteoroids with an increasing fraction of clones reaching q$>$2.5\,au (e.g. there are 219 objects with a fraction of clones larger than 10\%); the orange bars are among the corresponding set, the number of meteoroids that fulfill the JFC criteria of T14 (in the previous example there are 185). The numbers are converging for a higher fraction of clones; for a fraction $>$50\%, the numbers are almost identical (81 and 79). We conclude that the sets selected by the orbital and dynamical criteria converge for the events with high probabilities of reaching q$>$2.5\,au orbits during our integrations. 

\begin{figure}[]
\centering
\includegraphics[width=0.5\textwidth]{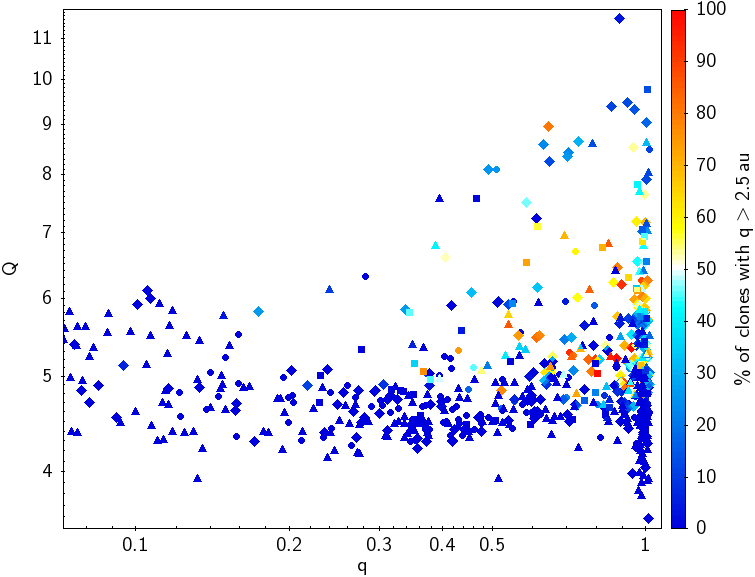} 
\caption{Perihelion distance (q; au) versus aphelion distance (Q; au) for 646 JFC-like fireballs observed by DFN (circle), EFN (diamond), MORP (square), and FRIPON (triangle) data points. The coloration is derived from 10\,kyr simulations of every object and represents the percentage of particle clones within the simulations that have experienced significant gravitational interactions with Jupiter, raising their perihelion to at least 2.5\,au. This threshold is significant as it marks roughly the distance beyond which water ice can begin sublimation, a process that renders JFCs `active'.}
\label{fig:Q_q_plot} 
\end{figure}

\begin{figure}[]
    \centering
    \begin{subfigure}[b]{0.5\textwidth}
        \includegraphics[width=\textwidth]{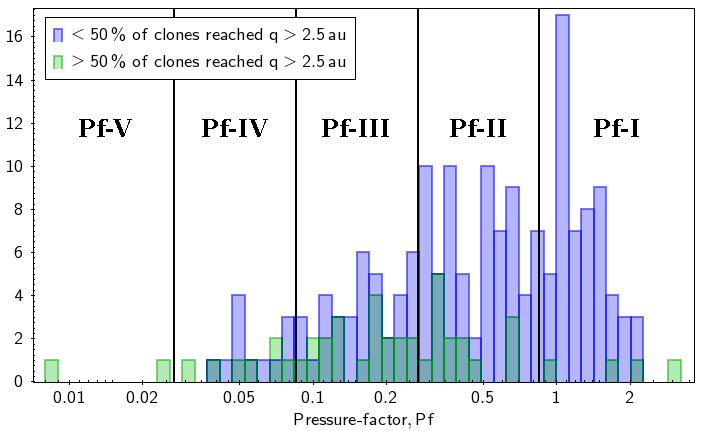}
        \caption{}
        \label{fig:pf_q25}
    \end{subfigure}
    \vspace{1ex} 
    
    \begin{subfigure}[b]{0.5\textwidth}
        \includegraphics[width=\textwidth]{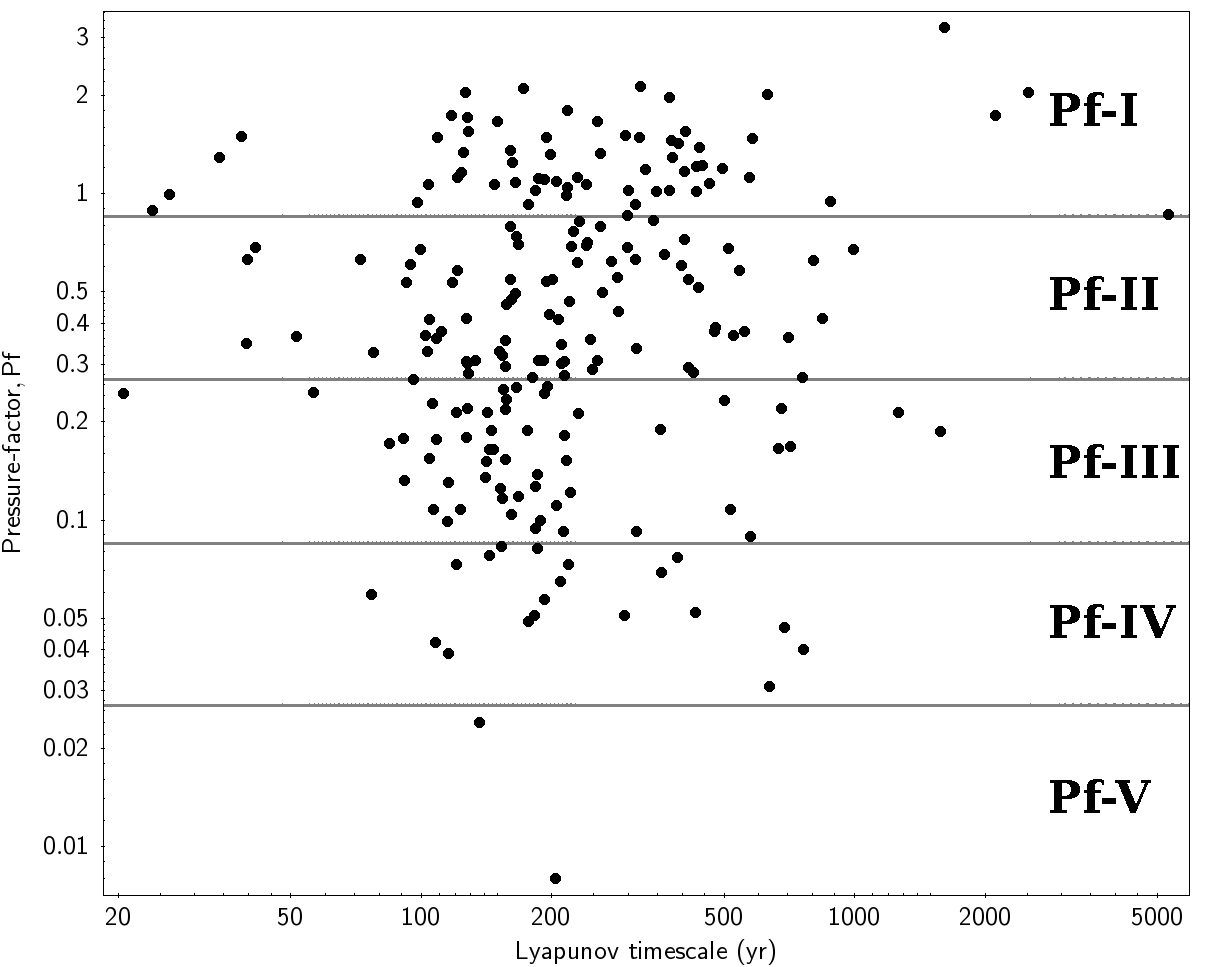}
        \caption{}
        \label{fig:pf_lyapunov}
    \end{subfigure}
    \vspace{1ex} 

    \caption{Comparison of the pressure-factor (Pf), introduced by \citet{borovivcka2022_two}, and orbital dynamics of the 213 JFC-like fireballs detected by EFN. Subplot a is a histogram plot comparing the Pf values of the fireballs that had greater or less than 50\% chance of the particle clones generated within the observational uncertainties of the object achieved at least a perihelion value of 2.5\,au during a 10\,kyr orbital integration. Subplot b is a scatter plot showing the Lyapunov lifetime of the fireball orbit based on 20\,kyr N-body simulations versus the percentage of particle clones that reached q=2.5\,au at least once.}
    \label{fig:EFN_pf_v_stability}
\end{figure}

When we compare the meteoroid strengths of the EFN dataset, using the `pressure-factor' (Pf; \citealp{borovivcka2022_two}), we see a small trend associated with the dynamic stability of the JFC orbits. As seen in Fig. \ref{fig:EFN_pf_v_stability}, the Pf factor tends to be lower for most of the unstable meteoroids within the EFN dataset (21\% of the dataset), concentrating towards values of Pf-II or Pf-III. Whereas the more `stable' subset contains nearly all of the strongest Pf-I material. Conversely, there is no trend at all when the Pf factor is compared to the Lyapunov lifetimes of the EFN fireballs (Fig.~\ref{fig:pf_lyapunov}), however, this is not as conclusive as the Lyapunov lifetimes can vary significantly for individual objects, making the comparison to Pf difficult. 

\begin{figure}[]
    \centering
    \includegraphics[width=0.5\textwidth]{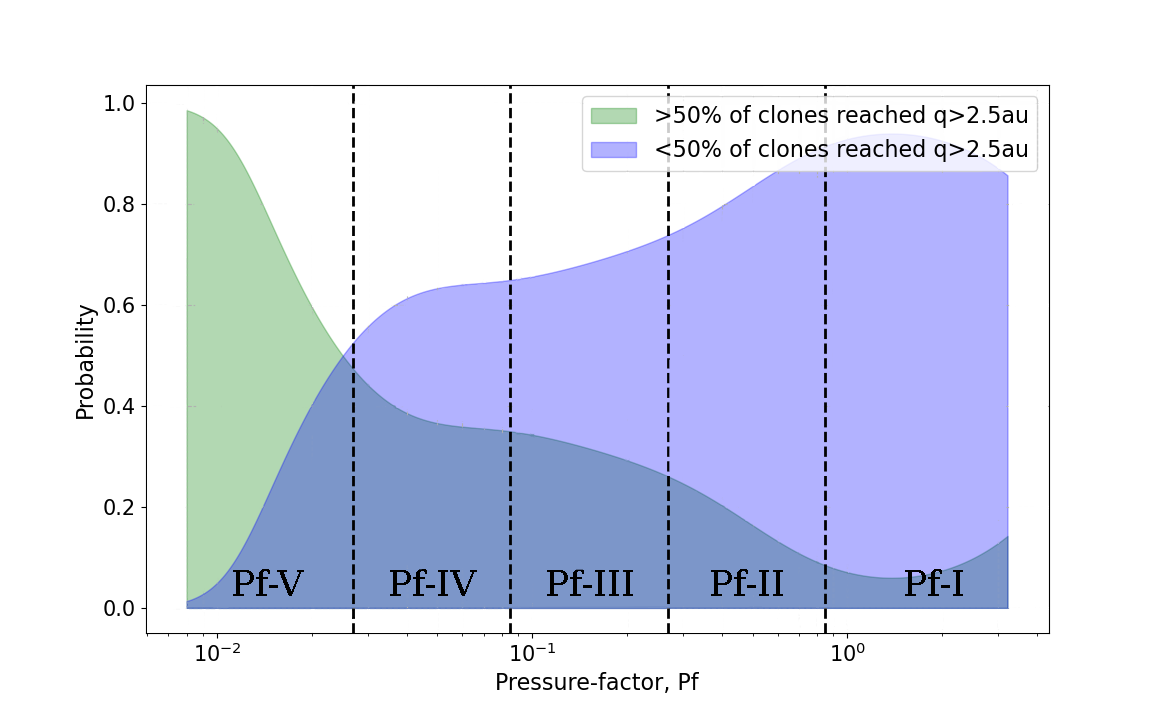}
    \caption{Normalized probability density functions of Pf values for EFN meteoroids on JFC-like orbits, segmented into those with more than 50\% and less than 50\% of clones reaching q$>$2.5\,au. The bold vertical lines at Pf values of 0.85, 0.27, 0.085, and 0.027 are the limits for the Pf classes. The probabilities sum to 1.0 for each Pf value, emphasizing the comparative likelihood of belonging to either distribution (seen in Fig.~\ref{fig:pf_q25}) and providing a probabilistic framework for understanding meteoroid origins in the context of strengths. While the distributions are significantly different, their overlap does not make Pf values a good discriminating mechanism between the stable and unstable comet-like meteoroids in the dataset.}
    \label{fig:Pf_source_probs}
\end{figure}

When performing orbital integrations of fireball trajectories, we find the least stable meteoroids tend to be weaker on average, and the strongest meteoroids are no longer good candidates as cometary debris. Despite this, there is still a significant Pf overlap of the `stable' and `unstable' groups. This does not signal that the asteroidal and cometary populations supplying the JFC-like ($2<T_{J}<3$) meteoroid component do not have different physical strengths, but these strengths alone cannot be used to identify them. As seen in Fig.~\ref{fig:Pf_source_probs}, we have plotted the probability of originating from an orbit with a greater or less than 50\% chance of having significant encounter(s) of Jupiter over 10\,kyr. While many of the strongest meteoroids (Pf-I) in the EFN dataset are on considerable stabler orbits, most of the stable meteoroids ($>$65\%) are Pf-II or weaker. The more unstable meteoroids on "JFC-like" orbits do have significantly weaker Pf-values within the EFN dataset, however, the larger contribution of stable meteoroids on similar orbits is overwhelming. Only in the Pf-V class, does the likelihood of originating from a more chaotic orbit become dominant, albeit given a low statistic of just two  fireballs. 

Weaker objects can also be expected for outer main belt asteroids. \citet{hsieh2020potential} showed that asteroids from the Themis group in and around the 2:1 MMR could enter JFC-like orbits. This family is predominately composed of B-type and C-type asteroids given their spectral characteristics \citep{jiang2021thermophysical}, consistent with hydrated carbonaceous material. This is also true for the outer main asteroid belt in general, as many low-albedo-hydrated asteroids dominate the landscape \citep{takir2012outer}. These low-albedo hydrated asteroids, with composition resembling CI/CM chondrite meteorites, is one likely explanation for the abundance of meteoroids on stable orbits beyond 3\,au and concentrating in or near the 2:1 MMR (Fig. \ref{fig:ae_all}). Another reason for this apparent large overlap in strengths of these two dynamically separated groups of meteoroids could also indicate that even a portion of meteoroids from the more chaotic, unstable trajectories could be sourced from the main belt as well. A diffusion of large-Q main belt objects onto chaotic orbits has been theorized and observed previously \citep{fernandez2002there,shober2020did,shober2020using}.

On the other hand, there is a clear trend in Pf factor as the q-value decreases, with only the strongest meteoroids reaching low-q values. This trend as q values decrease ties into the overall $\omega$ cycling distribution (Fig. \ref{fig:omega_inc_ecc}), indicating that weaker meteoroids that enter lower-q values near the Earth are in a secular Kozai cycle of $\omega$ which periodically lowers their q even further, removing the weakest meteoroids from the population.  This would not be the first time strengths have been associated with q-value. In \citet{toliou2021minimum} a probabilistic analysis of the minimum perihelion distance (q) reached by NEAs and meteorite falls reveals that carbonaceous chondrites typically exhibit short dwell times at small q, while ordinary chondrites have dwell times ranging from 10,000 to 500,000 years, supporting the notion of `supercatastrophic' disruption of asteroids at small perihelion distances. This disintegration of weaker material at lower q values likely explains the correlation between q and Pf within the EFN dataset. 

\subsection{Source region analysis}

In addition to the 10\,kyr orbital simulations and the 20\,kyr simulations to calculate the Lyapunov lifetimes for the JFC-like fireballs, we also used the debiased NEO model described in \citet{2023AJ....166...55N} to estimate the source regions of these meteoroids. As seen in Fig. \ref{fig:source_region_histogram}, the majority of meteoroids are still predicted to originate from main-belt sources (66-82\%), although the degree of which is slightly lower for all the fireball networks compared to the dynamics of the clones in the 10\,kyr simulations. This, however (as seen in Fig. \ref{fig:source_region_versus_sim}) is due to the model predicting almost all of the objects with orbits larger than 3.5\,au or particularly with high eccentricities have a high probability of originating from the JFC population. Many of these objects are very stable on 10\,kyr timescales, thus their inclusion significantly erroneously increases the JFC contribution to the fireball population. For example, the JFC contribution for FRIPON based on NEOMOD including these high-eccentricity meteoroids is about 34\%. This estimate is reduced to 8\% if the high-eccentricity FRIPON fireballs are discarded. This seems to indicate a strong size-dependent bias influencing the source region prediction for the fireball observations. For example, \citet{granvik2018identification}, surprisingly found that two H5 chondrites with measured orbits, Ejby and Ko\v{s}ice, showed comparatively strong probabilities of being from JFCs according to the NEO model described in \citet{granvik2018debiased}. The NEO model used within this study described by \citet{2023AJ....166...55N}, NEOMOD, seems to have improved in that there are no longer any meteorites with orbits with JFC population likelihoods greater than 4\%. This improved source region prediction may be attributed to their use of the dynamical model of JFCs of \citet{nesvorny2017origin}, which took into account galactic tides, passing stars, different fading laws, and the early movement of Neptune's orbit. 

Another key consideration when using the NEOMOD model on certain groups of objects is to ensure that all the objects within the same \textit{a-e-i} space are included in order to get accurate source region estimates. As the JFC contribution seems to be underestimated by the model when compared to 10\,kyr simulations, but when all 1,860 NEOs (asteroids and comets) on $2<T_{J}<3$ orbits are considered, the JFC contribution is only estimated to be 6-14\%. This is in line with our results from our 10\,kyr simulations of JFC-like fireballs, which found that only 8-21\% of the meteoroids have a $>$50\% likelihood of reaching q$>$2.5\,au. If this estimate of NEOMOD is correct, there should be $\sim$120-260 JFCs in near-Earth space -- in range of the 151 JFCs (with cometary fragments included) discovered and the 20-30 dormant comets estimated to exist based on meteor radar observations \citep{mommert2015exploreneos,ye2016dormant}. Supporting these observations, the work of \citet{martino_s_2019} identified a significant number of asteroids on cometary orbits, likely representing dormant or quasi-dormant comets, with only a minimal fraction exhibiting activity. These observations suggest that transitioning from an active to an inert state is an infrequent occurrence, further affirming the notion that comet dormancy is a less common outcome than disintegration.

This estimate from the NEOMOD model underscores the primary reason fireball networks do not observe many fireballs from the JFC-region. Despite JFCs being characterized by having $2<T_{J}<3$ orbits, in near-Earth space, JFCs are a small minority in the region as highlighted by several models \citep{levison1994long,fernandez2002there}. So, even if the diffusion of material from the main belt is less efficient, it is a much larger and closer reservoir to near-Earth space. Thus, we see that the likely sources for this debris observed on JFC-like orbits by DFN, EFN, MORP, and FRIPON are the outer main-belt resonances, in particular the 11:5, 2:1, 5:2 MMRs for material beyond 3\,au. 

\subsection{Meteor showers}

Unlike the study of \citet{shober2021main}, which only focused on the sporadic JFC-like fireballs detected by the DFN, here we have identified and analyzed several meteor showers present in the JFC-like subset of the DFN, EFN, MORP, and FRIPON datasets. In total, we  identified nine established showers with at least a few fireballs represented within the four datasets. These showers were identified using the $D_{N}$ criterion \citep{valsecchi1999meteoroid}, a geocentric-based criterion, and can be seen in Fig. \ref{fig:shower_Dv}. The nine showers identified include: Quadrantids (QUA), Southern-$\delta$ Aquariids (SDA), $\alpha$-Capricornids (CAP), Northern Taurids (NTA), Southern Taurids (STA), $\eta$-Virginids (EVI), August Draconids (AUD), $\kappa$-Cygnids (KCG), and October Draconids (DRA). If considering a limiting $D_{N}$ value of 0.1, on average the networks contain about 17\% shower-associated fireballs. However, this is simplistic as the number of shower associations varies significantly. Despite all four networks being ``fireball'' networks, the size range detected by each of them varies enough to significantly change the composition of the datasets. Out of the four networks, the DFN has the lowest number of shower-associated fireball events, with only about 4\% of the JFC-like dataset with a $D_{N}<0.1$, which are mostly $\eta$-Virginids. Whereas the other datasets each contain relatively similar levels of shower-associated data, with 16\% of EFN, 20\% of MORP, and 22\% of FRIPON JFC-like fireballs. The seemingly drastic paucity of showers in the DFN database is likely due to the 1\,second cutoff present within the dataset. However, it could also be a side-effect of being the only southern hemisphere network in the study. The showers that are present in the other datasets are more likely to be observed in the northern hemisphere. Either way, there also seems to be fewer shower associated fireballs in general for fireballs originating from JFC-like orbits, which is curious. 

The significant shower contribution difference for JFC-like fireballs in the DFN versus the other three networks, however, seems to have resulted in very little difference in the dynamic analysis results. Despite having several times the number of shower-associated fireballs, the other networks (EFN, MORP, FRIPON) seem to have similarly low levels of dynamically unstable meteoroids. In \citet{shober2021main}, the shower component was removed because the assumption was made that fireballs associated with JFC-like showers should be cometary and should be mostly very unstable. To the contrary, we find here the opposite to be true. As clearly seen in Fig. \ref{fig:shower_q25}, the shower component of the dataset is much more stable, even more than the background. This is because almost all of these showers originate from objects that are not classified as traditional JFCs. 

The Quadrantids, for example, are linked to asteroid 2003 EH1, an Amor-type asteroid with a comet-like orbit, suggesting it could be a dormant or extinct comet. This parent body is currently not in alignment with the core distribution of JFCs with a very high inclination (70.8$^{\circ}$) and low $T_{J}$ value ($\sim$2.06). The body is also influenced by Kozai-resonance, causing an exchange between the eccentricity and inclination but more importantly, placing it with a $\omega$-circulation. This circulation of $\omega$ values has linked asteroid 2003 EH1 to several other meteor showers and parent bodies, which are also in this circulation. Additionally, the research indicates that the Quadrantid complex might be the result of a series of cometary disintegrations starting more than 5000 years ago, leading to the formation of various large and small bodies \citep{wiegert2005quadrantid}. These bodies are believed to circulate in the Kozai resonance, contributing to the complexity of the Quadrantid stream. However, the authors caution that while 2003 EH1's orbit closely resembles that of the core Quadrantid stream, it may not be the direct parent of the entire stream \citep{wiegert2005quadrantid}. 

The southern $\delta$-Aquariids, observed from late May to early July, are now widely believed to be linked to the 96P/Machholz complex, as delineated by \citet{abedin2018formation}. This complex, including the Quadrantids (QUA) and asteroid 2003 EH1, is influenced by a Kozai circulation cycle, resulting in eight intersection points with Earth's orbit \citep{babadzhanov2008meteor}. \citet{abedin2018formation} suggested that capturing comet 96P/Machholz into a short-period orbit around 22,000 years ago could account for the primary activity patterns of various showers, including the Quadrantids, Arieds, both Northern and Southern $\delta$-Aquariids, and others.

Other showers detected, including the $\alpha$-Capricornids (CAP), Taurid Complex (NTA and STA), $\eta$-Virginids (EVI), August Draconids (AUD), $\kappa$-Cygnids (KCG), are all also very stable and avoid close encounters with Jupiter during our 10\,kyr N-body simulations. The $\alpha$-Capricornids are associated with short-period comet 169P/NEAT, with evidence suggesting a massive breakup event 4000-5000 years ago. The comet 169P/NEAT was identified as one of the extremely stable (Table\ref{tab:22_extremely_stable}) in this study as well as in the previous analysis of \citet{fernandez2015jupiter}. Also, notably, the Taurid Complex is primarily linked to Comet 2P/Encke, which is completely detached from Jupiter and the dynamics have been extensively studied. 

Interestingly, of all the shower components detected, only one shower displays orbital characteristics in line with those of the majority of JFCs: the October Draconids (DRA). The October Draconids stand out for being consistent with JFC dynamics, having a high probability of reaching higher perihelion distances within 10,000 years. The parent body of the Draconids, comet 21P/Giacobini-Zinner, has also been shown to be consistent with JFC dynamics within our simulations, with 85\% of the clones reaching q$>$2.5\,au within the 10\,kyr simulation. This finding contrasts with other showers, whose parent bodies are not explicitly classified as JFCs and often include associated objects that are either asteroids, comets, or a mixture of both. The October Draconids meteor shower has demonstrated variable activity levels, with reports of both weak and intense outbursts. While some instances may be characterized as anomalously weak, other observations have indicated unexpected intense activity, highlighting the dynamic nature of this meteor shower. Moreover, in \citet{borovivcka2007atmospheric}, the Draconids were identified to be distinctive due to their fragile composition and low density. The dynamical unpredictability and physical weakness, along with a parent body that also experiences close encounters with Jupiter on short time scales, robustly demonstrates the type of shower expected from a pristine object of the JFC population. 

\subsection{What is happening on JFC-like orbits on the km-scale versus the cm-m scale}
The research presented in this paper reveals significant insights into the nature of objects on JFC-like orbits at different scales. By examining the orbital distribution, stability, and dynamics of these objects, we gain an understanding of their origins and the processes affecting them.

\subsubsection{Kilometer scale: Realm of the JFCs and asteroid contaminants}

At the kilometer scale, represented by the 661 JFCs from the NASA HORIZONS database, we observe a dynamic population predominantly influenced by close encounters with Jupiter on short thousand-year timescales. Most of these JFCs exhibit frequent close encounters with Jupiter, leading to rapid and significant changes in their orbits, particularly the perihelion distances. These encounters typically limit their excursions to the inner solar system, confining them to short, transient active phases. However, interestingly, a subset of JFCs on near-Earth orbits shows remarkably stable trajectories over 10,000 years, suggesting a potential asteroidal contamination. This contamination theory is supported by previous studies of the population \citep{fernandez2015jupiter,hsieh2020potential}. These objects, comprising about $\sim$30\% of near-Earth JFCs, could be dark, hydrated asteroids from the outer main belt that have migrated onto comet-like orbits. However, questions remain such as: could a hydrated asteroid produce enough activity to be compatible with observations? Furthermore, the intricate relationship between chaotic and partially stable regions in the orbital elements phase space warrants deeper investigation to fully understand the nature and origins of these bodies. 

\subsubsection{Centimeter-to-meter scale: Meteoroids on stable orbits}

A different picture emerges at the centimeter-to-meter scale, represented by the meteoroids detected by the DFN, EFN, MORP, and FRIPON networks. The vast majority of these meteoroids, although on JFC-like orbits, do not align dynamically with the typical JFC population. Instead, they often originate from stable orbits, with only between 8-21\% likely to have experienced close encounters with Jupiter in our 10\,kyr simulations. These findings disagree with the notion that the cm-m scale meteoroids on JFC-like orbits predominantly originate from JFCs. This is likely due to the fact that the primary reasoning for the cometary classification was based on the strength of the meteoroids \citep{borovivcka2022_two}.

\citet{borovivcka2022_two} also discussed the surprising prevalence of resistant material on excited short-period orbits, defined by aphelia greater than 4.9\,au, semi-major axes less than 5\,au, and either high eccentricities (e\,$>$\,0.9) or high inclinations (i\,$>$40$^{\circ}$). This resistance to disintegration is particularly notable among meteoroids in orbits with semi-major axes between 2.6 and 3.8\,au, suggesting a zone rich in dynamically excited yet physically robust material. The robust nature of these materials, as highlighted in the flat distribution of their pressure-factor (Pf) values, predominantly falls within Pf-I and Pf-II classes, underscoring their unexpected resilience \citep{borovivcka2022_two}. The sodium-deficient spectra of specific meteoroids, particularly those akin to EN310718\_213217 and EN111118\_221402, with high inclinations and semi-major axes greater than 3\,au, signal a potential alteration or loss of volatiles through close solar encounters or prolonged solar exposure. However, our analysis of the orbital stability and close encounter frequency with Jupiter shows that the contribution of asteroidal sources is much more significant than the strengths alone imply.

As shown in Fig. \ref{fig:EFN_pf_v_stability}, the Pf values are lower (i.e., weaker meteoroids) for the more unstable meteoroids in the EFN dataset. Thus, the more unstable `comet-like' trajectories also tend to be statistically weaker. However, over 65\% of the meteoroids on more stable trajectories that do not encounter Jupiter also have `weaker' characteristics (Fig.~\ref{fig:Pf_source_probs}). As seen in Fig.~\ref{fig:Pf_source_probs}, the strength alone cannot be used to identify source regions very well, even for Pf-II, III, and IV meteoroids there is only a 10-50\% chance of coming from an unstable Jupiter-encountering orbit based on the strength alone. This is consistent with previous studies that found the strengths can be greatly influenced by macro-scale features in the meteoroid such as cracks or porosity \citep{popova2011very}, and there are several meteorite falls with Pf-III classification \citep{borovivcka2022_two}. This is not to say there are no variations in strengths due to dynamics, but the strengths alone do not predict the source regions well. Even within the region denoted the `comet domain' in \citet{borovivcka2022_two} (a$<$5\,au; Q$\geq$4.9\,au; $\iota\leq$40$^{\circ}$), only 34\% of the EFN meteoroids are likely to leave the `visible' (q$>$2.5\,au) region within 10\,kyr. While the cometary meteoroids concentrate the most in this region, the non-JFC contamination is revealed to be significantly more only after the dynamics are considered. The weaker meteoroids beyond 3\,au are more likely to be dominated by weaker carbonaceous debris from the outer solar system resonances (particularly the 2:1 and 11:5 MMRs), which could also reproduce these features, including the dynamic stability of the meteoroid population. 

In addition to the mostly stable dynamics of the JFC-like fireball dataset, we have also found that nearly all of the fireballs are in some form of Kozai-induced $\omega$-circulation. We deduce that this circulation is the likely cause of the stability of the fireballs, despite their JFC-like Tisserand parameters. The circulation of $\omega$ values, aligns the periods of high-inclination, low-aphelion, and high-perihelion when the $\omega$ values are at 0$^{\circ}$ and 180$^{\circ}$, i.e. when the orbits are in the best orientation to have an encounter with Jupiter. This effect decreases the chance of frequent close encounters with Jupiter. This protective feature of the secular Kozai resonance cycle from close encounters with certain planets was first shown by \citet{kozai1979secular}. Many of the showers found in the JFC-like datasets of the DFN, EFN, MORP, and FRIPON are within some $\omega$-circulation, protecting them from these close encounters Jupiter. Thus, many of the more `stable' meteoroids in the fireball dataset could also be sourced from 96P/Machholz complex bodies or Marsden group comets, which display similar dynamics. While these bodies show significant differences from the JFC population and their provenance is still ill-determined, they could provide a cometary source for a significant portion of meteoroids. Regardless, based on the dynamics; the October Draconids, in fact, may be the only "pristine" JFC material in our sample based on the extreme fragility of the meteoroids and their chaotic orbits that frequently intersect Jupiter. 

\subsubsection{Comparing the scales: Implications and observations}

It is crucial to note that in our analysis, we only examined the dynamics of 661 km-scale bodies that meet this $T_{J}$ criterion and have cometary designations by IAU. However, over 27,000 objects in the NASA HORIZONS database are falling within this $T_{J}$ range. Of these, we found only 2,745 are classified as ACOs using the criteria of \citet{tancredi2014criterion}. This discrepancy underscores that many objects classified based on $T_{J}$ alone may not exhibit traditional cometary behavior. Thus, it is unsurprising that many objects at the cm-m scale have significantly different dynamics than JFCs. Jupiter's gravity heavily influences kilometer-scale JFCs, leading to a uniquely chaotic dynamic signature, short active phases, and short physical lifetimes in near-Earth space. Meanwhile, the smaller meteoroids exhibit an overwhelming abundance of stable and Kozai-resonant bodies. This stark dynamical distinction between size scales suggests varied evolutionary processes for solar system debris, further complicated by observational biases between telescopically observed and Earth-impacting populations.

There are likely two solutions to the results of our analysis. Firstly, the stability of the meteoroid population is a result of the transfer of small bodies from the outer main asteroid belt onto JFC-like orbits. This transfer of outer main-belt debris would likely be physically weaker due to the dominance of carbonaceous C- D and P-type asteroids in the outer main belt. Additionally, as shown in \citet{hsieh2020potential}, some bodies could even become Jupiter-crossing and be almost impossible to distinguish from genuine JFCs. The Kozai resonance could serve as the protection mechanism for these objects, and their longer dynamic lifetimes would cause them to dominate the fireball population. The reduction in q would make them Earth-crossing but also more stable and longer-lasting in near-Earth space.

Moreover, from a purely population-size standpoint, the JFCs are the minority on Earth-crossing JFC-like orbits, with 1,875 NEAs on JFC-like orbits and only 75 JFCs, according to NASA HORIZONS. Even if there are an estimated 20-30 dormant comets in near-Earth space \citep{mommert2015exploreneos,ye2016dormant}, the number of asteroids diffused out from the main belt is overwhelming. The expected extremely short physical lifetimes of cometary debris should likely also play a significant role. The continuous breakdown and disintegration of cometary meteoroids and boulders are necessary as the mass influx from normal JFC activity is several times lower than necessary to maintain the ZC \citep{nesvorny2010cometary,nesvorny2011dynamical,rigley2022comet}. Thus, in addition to the reduced contribution expected in the JFC-like fireball population from JFCs due to purely orbital instability in near-Earth space, the bodies that do evolve onto near-Earth orbits for significantly extended periods likely also break down physically into smaller-sized meteoroids before ever having the opportunity to impact the Earth. This explanation aligns with the reasoning laid out in \citet{shober2021main}. 

However, based on the analysis of \citet{genge2020micrometeorites}, most micrometeorites larger than 50\,$\mu m$ are believed to originate from carbonaceous asteroids, with a significant portion of smaller micrometeorites likely coming from comets. This would entail that the physical breakdown of centimeter to meter-sized JFC debris is not the source of the larger micrometeorite population. The stability of the debris we see from fireball network JFC-like orbits may then indicate that there is not much JFC material within this size range on Earth-crossing orbits in general. However, studying micrometeorites is challenging; changes in mineralogy and composition during atmospheric entry, along with biases in collection methods, weathering, contamination, and erosion, complicate the accurate assessment of their sources. 

Our extended analysis here, which has built on the work of \citet{fernandez2015jupiter} and \citet{shober2021main}, has also found that almost all of the meteor showers in the JFC-like dataset and their corresponding parent bodies are in very stable orbits on $10^{4}$\,yr timescales; leading to a possible second source of this material. Many of these objects are speculated to be from the JFC population; others are presumably asteroids, but all are related through their $\omega$-circulation induced by the Lidov-Kozai effect. The fireballs detected in this study are almost entirely in some level of a secular Kozai circulation, protecting the bodies from close encounters like the suspected parent bodies such as 2003 EH1, and other objects in the 96P/Machholz Complex \citep{abedin2018formation}. Thus, the provenance of the JFC-like fireballs analyzed here could be dependent on a better understanding of these transitional asteroid and comet bodies that may be an evolved population derived from JFC sources or the main belt. For instance, the association of meteoroids with the Machholz interplanetary complex, which includes objects like 96P/Machholz, SOHO comets, and asteroid 2003 EH1, signifies a common origin potentially rooted in cometary fragmentation and evolutionary transitions to asteroid-like bodies \citep{sekanina2005origin,borovivcka2022_two}. These bodies exhibit a range of dynamical stability yet share a connection to a lineage that has experienced significant physical processing and dynamical migration. Objects contained in the main belt, which display cometary features or activity and may originate from the outer solar system, could also be connected to the fireballs observed here \citep{hsieh2006population}.  

In either scenario, the JFC-like fireballs detected by fireball networks are not consistent with pristine JFC debris, except for the October Draconids. The objects we observe with fireball networks are either asteroidal contamination enhanced by survival bias or meteoroids related to the 96P/Machholz Complex or Marsden group comets.  



\section{Summary}
The orbital evolution of centimeter- and meter-scale JFC-like ($2<T_{J}<3$) meteoroids should be exactly the same as the 100\,m or kilometer-sized JFCs, as non-gravitational forces are not strong enough to decouple the meteoroids from having close encounters with Jupiter. However, we find that these populations contain significant variations, indicating a larger degree of contamination within the meteoroid population compared to the km-scale population. Only a small fraction of the JFC population is on Earth-crossing orbits, so inherently, tthe fireball population we are considering here (if indeed of JFC origin) would be a small biased subset. Our major results can be summarized as follows: 

\begin{itemize}
    \item The 10\,kyr orbital integrations of 661 JFCs indicate that the trajectories are heavily influenced by close encounters with Jupiter. These encounters lead to rapid and significant changes in their orbits, particularly affecting the perihelion distances on hundred to thousand-year timescales.

    \item The Lyapunov lifetimes of the 661 JFCs tend to concentrate towards lower values ($\sim$120\,yrs), consistent with the results of \citet{tancredi1995dynamical} and \citet{tancredi1998chaotic}. However, the JFC-like fireball population has a much more diffuse Lyapunov lifetime distribution, consistent with near-Earth asteroids. 

    \item When excluding JFCs lost after splitting events (3D/Biela, 5D/Brorsen, and 34D/Gale) or those lost to higher orbital uncertainties (D/1766 G1 (Helfenzrieder), D/1770 L1 (Lexell), and D/1895 Q1 (Swift)), we found that 22\% (16 out of 72) JFCs in near-Earth orbits move on extremely stable orbits. 

    \item The NEOMOD model, described in \citet{2023AJ....166...55N}, finds the most likely source regions for JFC-like debris are the outer main belt resonances. The 3:1, 5:2, 2:1, and 11:5  modes all likely contributed significantly to the meteoroids detected by DFN, EFN, MORP, and FRIPON. 
    
    \item The vast majority of centimeter- to meter-scale meteoroids detected by DFN, EFN, MORP, and FRIPON fireball networks are on orbits with $2<T_{J}<3$, however, they are found to originate from very stable orbits. Using the more restrictive criteria for considering a JFC-like orbits presented by \cite{tancredi2014criterion}, only 35\% of the studied meteoroids are in this category. Additionally, based on our 10\,kyr simulations, only 8-21\% are likely to have experienced close encounters with Jupiter enough to raise their perihelia above 2.5\,au within 10\,kyr, indicating a different source or evolution compared to kilometer-scale JFCs. Considering the size of each dataset, this implies that only 1\%, 5\%, 4\%, and 1\% of the DFN, EFN, MORP, and FRIPON fireball datasets are dynamically consistent with JFC dynamics. These estimates are consistent or slightly lower than the 1.7-10\% estimate for the JFC contribution to the population of small NEOs \citep{bottke2002debiased,granvik2018debiased,2023AJ....166...55N}.

    
    \item An analysis of meteor showers within the JFC-like fireball datasets reveals that most showers and their associated parent bodies are in stable orbits, often protected from close encounters with Jupiter due to Kozai-resonance induced circulation of the argument of perihelion ($\omega$), aligning periods of low-aphelia and high-inclination when the argument of perihelia is at 0$^{\circ}$ or 180$^{\circ}$. The prevalence of Kozai-circulation for all the meteoroids, regardless of fireball network or shower association, may indicate that objects like 96P/Machholz, SOHO comets, and asteroid 2003 EH1 may be the key to understanding the source of these meteoroids.
    
    \item The physical strengths of the EFN JFC-like meteoroids, measured by their pressure factor (Pf), are slightly weaker for the unstable comet-like meteoroids in the study. However, due to the significant overlap of Pf ranges for stable and unstable meteoroids, we find that strength alone is not a reliable indicator of discriminating between asteroidal and cometary source regions.  Additionally, there is a slight concentration of dynamically stable, weaker meteoroids being observed in orbits 3.0-3.4\,au; this is currently best explained by the influx of weaker carbonaceous debris from the outer solar system resonances, particularly the 2:1 and 11:5 MMRs with Jupiter.
    
    \item Observational biases, including the detection limits and targeted surveys, are likely influencing our understanding of the JFC-like population, particularly in the context of size-dependent dynamical evolution and physical processes.
    
    \item The short physical lifetimes of cometary debris, coupled with rapid physical evolution, might result in a lack of smaller cometary debris surviving long enough to be observed as fireballs, further complicating the interpretation of the observed fireball population.
    
    \item The October Draconids were identified as potentially the only pristine JFC material in the observed fireball sample, with their parent body, comet 21P/Giacobini-Zinner, exhibiting dynamics consistent with typical JFCs. Thus indicating that the "anomalously" weak meteoroids may be more consistent with what we might expect from debris from JFCs.  

    \item Individual fireball case studies of "cometary" fireballs do not provide an accurate view of the population. All of the single-event ``JFC fireball'' cases reviewed here either had extremely high orbital uncertainties or highly underestimated orbital uncertainties. Even with detailed observations of a strong object originating from an orbit with $T_{J}<3$, the chaotic nature of these orbits renders a single data point less meaningful. 
    
\end{itemize}

\section{Acknowledgments}
This project has received funding from the European Union’s Horizon 2020 research and innovation programme under the Marie Skłodowska-Curie grant agreement No945298 ParisRegionFP. This work was also funded by the Australian Research Council as part of the Australian Discovery Project scheme (DP170102529). The authors also acknowledge financial support from CSIC-Udelar and Agencia Nacional de Investigación e Innovación ANII (Uruguay). SSTC authors acknowledge institutional support from Curtin University.  

FRIPON was initiated by funding from ANR (grant N.13-BS05-0009-03), carried by the Paris Observatory, Muséum National d’Histoire Naturelle, Paris-Saclay University and Institut Pythéas (LAM-CEREGE). VigieCiel was part of the 65 Millions d’Observateurs project, carried by the Muséum National d’Histoire Naturelle and funded by the French Investissements d’Avenir program. FRIPON data are hosted and processed at Institut Pythéas SIP (Service Informatique Pythéas), and a mirror is hosted at IMCCE (Institut de MécaniqueCéleste et de Calcul des Éphémérides / Paris Observatory)

This research made use of TOPCAT for visualization and figures \citep{taylor2005topcat}. This research also made use of Astropy, a community-developed core Python package for Astronomy \citep{robitaille2013astropy}. Simulations in this paper used the REBOUND software package\footnote{\url{http://github.com/hannorein/REBOUND}} \citep{rebound2012A&A...537A.128R}. The authors declare no competing interests. All data needed to evaluate the conclusions in the paper are present in the paper and/or at \url{https://doi.org/10.5281/zenodo.4710556}.

The authors would also like to thank David Nesvorn\'y, for his insightful comments through email correspondence and help to interpret the NEOMOD model introduced in \citet{2023AJ....166...55N}.






\bibliographystyle{aa}
\bibliography{main} 

\begin{thebibliography}{156}
\expandafter\ifx\csname natexlab\endcsname\relax\def\natexlab#1{#1}\fi

\bibitem[{Abedin {et~al.}(2015)Abedin, Spurn{\`y}, Wiegert, Pokorn{\`y},
  Borovi{\v{c}}ka, \& Brown}]{abedin2015age}
Abedin, A., Spurn{\`y}, P., Wiegert, P., {et~al.} 2015, Icarus, 261, 100

\bibitem[{Abedin {et~al.}(2018)Abedin, Wiegert, Janches, Pokorn{\`y}, Brown, \&
  Hormaechea}]{abedin2018formation}
Abedin, A., Wiegert, P., Janches, D., {et~al.} 2018, Icarus, 300, 360

\bibitem[{Agarwal {et~al.}(2016)Agarwal, A'Hearn, Vincent, G{\"u}ttler,
  H{\"o}fner, Sierks, Tubiana, Barbieri, Lamy, Rodrigo,
  {et~al.}}]{agarwal2016acceleration}
Agarwal, J., A'Hearn, M.~F., Vincent, J.-B., {et~al.} 2016, Monthly Notices of
  the Royal Astronomical Society, 462, S78

\bibitem[{A'Hearn {et~al.}(1995)A'Hearn, Millis, Schleicher, Osip, \&
  Birch}]{ahearn1995ensemble}
A'Hearn, M.~F., Millis, R.~C., Schleicher, D.~G., Osip, D.~J., \& Birch, P.~V.
  1995, Icarus, 118, 223

\bibitem[{{Anderson} {et~al.}(2022){Anderson}, {Towner}, {Fairweather},
  {Bland}, {Devillepoix}, {Sansom}, {Cup{\'a}k}, {Shober}, \&
  {Benedix}}]{Anderson2022ApJ}
{Anderson}, S.~L., {Towner}, M.~C., {Fairweather}, J., {et~al.} 2022, The
  Astrophysical Journal Letters, 930, L25

\bibitem[{Anghel {et~al.}(2019)Anghel, Birlan, Nedelcu, \&
  Boaca}]{anghel2019photometric}
Anghel, S., Birlan, M., Nedelcu, D.-A., \& Boaca, I. 2019, RoAJ, 29, 191

\bibitem[{Babadzhanov {et~al.}(2008)Babadzhanov, Williams, \&
  Kokhirova}]{babadzhanov2008meteor}
Babadzhanov, P., Williams, I., \& Kokhirova, G. 2008, Monthly Notices of the
  Royal Astronomical Society, 386, 2271

\bibitem[{Baluev \& Mikryukov(2019)}]{baluev2019fast}
Baluev, R.~V. \& Mikryukov, D.~V. 2019, Astronomy and Computing, 27, 11

\bibitem[{Beech {et~al.}(1999)Beech, Jehn, Brown, \&
  Jones}]{beech1999satellite}
Beech, M., Jehn, R., Brown, P., \& Jones, J. 1999, Acta astronautica, 44, 281

\bibitem[{Beech \& Nikolova(2001)}]{beech2001endurance}
Beech, M. \& Nikolova, S. 2001, Planetary and Space Science, 49, 23

\bibitem[{Bland {et~al.}(2012)Bland, Spurn{\'y}, Bevan, Howard, Towner,
  Benedix, Greenwood, Shrben{\'y}, Franchi, Deacon,
  {et~al.}}]{bland2012australian}
Bland, P., Spurn{\'y}, P., Bevan, A., {et~al.} 2012, Australian Journal of
  Earth Sciences, 59, 177

\bibitem[{Boehnhardt(2004)}]{boehnhardt2004split}
Boehnhardt, H. 2004, Comets II, 745, 301

\bibitem[{{Borovicka}(1990)}]{Borovicka_1990BAICz}
{Borovicka}, J. 1990, Bulletin of the Astronomical Institutes of
  Czechoslovakia, 41, 391

\bibitem[{Borovi{\v{c}}ka {et~al.}(2007)Borovi{\v{c}}ka, Spurn{\'y}, \&
  Koten}]{borovivcka2007atmospheric}
Borovi{\v{c}}ka, J., Spurn{\'y}, P., \& Koten, P. 2007, Astronomy \&
  Astrophysics, 473, 661

\bibitem[{Borovi{\v{c}}ka {et~al.}(2022{\natexlab{a}})Borovi{\v{c}}ka,
  Spurn{\`y}, \& Shrben{\`y}}]{borovivcka2022_two}
Borovi{\v{c}}ka, J., Spurn{\`y}, P., \& Shrben{\`y}, L. 2022{\natexlab{a}},
  Astronomy \& Astrophysics, 667, A158

\bibitem[{Borovi{\v{c}}ka {et~al.}(2022{\natexlab{b}})Borovi{\v{c}}ka,
  Spurn{\`y}, Shrben{\`y}, {\v{S}}tork, Kotkov{\'a}, Fuchs, Kecl{\'\i}kov{\'a},
  Zichov{\'a}, M{\'a}nek, V{\'a}chov{\'a}, {et~al.}}]{borovivcka2022_one}
Borovi{\v{c}}ka, J., Spurn{\`y}, P., Shrben{\`y}, L., {et~al.}
  2022{\natexlab{b}}, Astronomy \& Astrophysics, 667, A157

\bibitem[{Borovi{\v{c}}ka {et~al.}(2013)Borovi{\v{c}}ka, T{\'o}th, Igaz,
  Spurn{\'y}, Kalenda, Haloda, Svore{\v{n}}, Korno{\v{s}}, Silber, Brown,
  {et~al.}}]{borovivcka2013kovsice}
Borovi{\v{c}}ka, J., T{\'o}th, J., Igaz, A., {et~al.} 2013, Meteoritics \&
  Planetary Science, 48, 1757

\bibitem[{Bottke~Jr {et~al.}(2002)Bottke~Jr, Morbidelli, Jedicke, Petit,
  Levison, Michel, \& Metcalfe}]{bottke2002debiased}
Bottke~Jr, W.~F., Morbidelli, A., Jedicke, R., {et~al.} 2002, Icarus, 156, 399

\bibitem[{Brown {et~al.}(2013)Brown, Marchenko, Moser, Weryk, \&
  Cooke}]{brown2013meteorites}
Brown, P., Marchenko, V., Moser, D.~E., Weryk, R., \& Cooke, W. 2013,
  Meteoritics \& Planetary Science, 48, 270

\bibitem[{Brown {et~al.}(2010)Brown, Wong, Weryk, \&
  Wiegert}]{brown2010meteoroid}
Brown, P., Wong, D., Weryk, R., \& Wiegert, P. 2010, Icarus, 207, 66

\bibitem[{Brown {et~al.}(2000)Brown, Hildebrand, Zolensky, Grady, Clayton,
  Mayeda, Tagliaferri, Spalding, MacRae, Hoffman, {et~al.}}]{brown2000fall}
Brown, P.~G., Hildebrand, A.~R., Zolensky, M.~E., {et~al.} 2000, Science, 290,
  320

\bibitem[{Campbell-Brown \& Hildebrand(2004)}]{campbell2004new}
Campbell-Brown, M. \& Hildebrand, A. 2004, Earth, Moon, and Planets, 95, 489

\bibitem[{Carusi {et~al.}(1985)Carusi, Kres{\'a}k, Perozzi, \&
  Valsecchi}]{carusi1985long}
Carusi, A., Kres{\'a}k, L., Perozzi, E., \& Valsecchi, G. 1985, Bristol: Hilger

\bibitem[{{Cassini}(1685)}]{cassini_1685}
{Cassini}, G.~D. 1685, {D{\'e}couverte de la lumi{\`e}re celeste qui paroist
  dans le zodiaque}

\bibitem[{{Ceplecha}(1987)}]{Ceplecha_1987}
{Ceplecha}, Z. 1987, Bulletin of the Astronomical Institutes of Czechoslovakia,
  38, 222

\bibitem[{Chen \& Jewitt(1994)}]{chen1994rate}
Chen, J. \& Jewitt, D. 1994, Icarus, 108, 265

\bibitem[{Colas {et~al.}(2020)Colas, Zanda, Bouley, Jeanne, Malgoyre, Birlan,
  Blanpain, Gattacceca, Jorda, Lecubin, {et~al.}}]{colas2020fripon}
Colas, F., Zanda, B., Bouley, S., {et~al.} 2020, Astronomy \& Astrophysics,
  644, A53

\bibitem[{Consolmagno {et~al.}(2008)Consolmagno, Britt, \&
  Macke}]{consolmagno2008significance}
Consolmagno, G., Britt, D., \& Macke, R. 2008, Geochemistry, 68, 1

\bibitem[{Devillepoix {et~al.}(2019)Devillepoix, Bland, Sansom, Towner, Cupák,
  Howie, Hartig, Jansen-Sturgeon, \& Cox}]{devillepoix2019observation}
Devillepoix, H.~A., Bland, P.~A., Sansom, E.~K., {et~al.} 2019, Monthly Notices
  of the Royal Astronomical Society, 483, 5166

\bibitem[{Devillepoix {et~al.}(2021)Devillepoix, Jenniskens, Bland, Sansom,
  Towner, Shober, Cup{\'a}k, Howie, Hartig, Anderson,
  {et~al.}}]{devillepoix2021taurid}
Devillepoix, H.~A., Jenniskens, P., Bland, P.~A., {et~al.} 2021, The Planetary
  Science Journal, 2, 223

\bibitem[{Devillepoix {et~al.}(2020)Devillepoix, Cupák, Bland, Sansom, Towner,
  Howie, Hartig, Jansen-Sturgeon, Shober, Anderson, Benedix, Busan, Sayers,
  Jenniskens, Albers, Herd, Carlson, Hill, Brown, Krzeminski, Osinski,
  Aoudjehane, Shisseh, Benkhaldoun, Jabiri, Guennoun, Barka, Darhmaoui, Daly,
  Collins, McMullan, Suttle, Shaw, Young, Alexander, Mardon, Ireland, Bonning,
  Baeza, Alrefay, Horner, Swindle, Hergenrother, Fries, Tomkins, Langendam,
  Rushmer, O'Neill, Janches, \& Hormaechea}]{devillepoix2020global}
Devillepoix, H. A.~R., Cupák, M., Bland, P.~A., {et~al.} 2020, A Global
  Fireball Observatory

\bibitem[{{Devillepoix} {et~al.}(2022){Devillepoix}, {Sansom}, {Shober},
  {Anderson}, {Towner}, {Lagain}, {Cup{\'a}k}, {Bland}, {Howie},
  {Jansen-Sturgeon}, {Hartig}, {Sokolowski}, {Benedix}, \&
  {Forman}}]{Devillepoix_Madura_Cave}
{Devillepoix}, H. A.~R., {Sansom}, E.~K., {Shober}, P., {et~al.} 2022,
  Meteoritics \& Planetary Science, 57, 1328

\bibitem[{Di~Sisto {et~al.}(2009)Di~Sisto, Fern{\'a}ndez, \&
  Brunini}]{disisto2009population}
Di~Sisto, R.~P., Fern{\'a}ndez, J.~A., \& Brunini, A. 2009, Icarus, 203, 140

\bibitem[{Dones {et~al.}(2015)Dones, Brasser, Kaib, \&
  Rickman}]{dones2015origin}
Dones, L., Brasser, R., Kaib, N., \& Rickman, H. 2015, Space Science Reviews,
  197, 191

\bibitem[{Drummond(1981)}]{drummond1981test}
Drummond, J.~D. 1981, Icarus, 45, 545

\bibitem[{Duncan {et~al.}(2004)Duncan, Levison, \& Dones}]{duncan2004dynamical}
Duncan, M., Levison, H., \& Dones, L. 2004, Comets II, 193, 204

\bibitem[{Duncan {et~al.}(1988)Duncan, Quinn, \& Tremaine}]{duncan1988origin}
Duncan, M., Quinn, T., \& Tremaine, S. 1988, The Astrophysical Journal, 328,
  L69

\bibitem[{Duncan \& Levison(1997)}]{duncan1997disk}
Duncan, M.~J. \& Levison, H.~F. 1997, Science, 276, 1670

\bibitem[{Egal {et~al.}(2021)Egal, Wiegert, Brown, Spurn{\`y}, Borovi{\v{c}}ka,
  \& Valsecchi}]{egal2021dynamical}
Egal, A., Wiegert, P., Brown, P., {et~al.} 2021, Monthly Notices of the Royal
  Astronomical Society, 507, 2568

\bibitem[{Fern{\'a}ndez(1980)}]{fernandez1980existence}
Fern{\'a}ndez, J.~A. 1980, Monthly Notices of the Royal Astronomical Society,
  192, 481

\bibitem[{Fern{\'a}ndez {et~al.}(2002)Fern{\'a}ndez, Gallardo, \&
  Brunini}]{fernandez2002there}
Fern{\'a}ndez, J.~A., Gallardo, T., \& Brunini, A. 2002, Icarus, 159, 358

\bibitem[{Fern{\'a}ndez \& Morbidelli(2006)}]{fernandez2006population}
Fern{\'a}ndez, J.~A. \& Morbidelli, A. 2006, Icarus, 185, 211

\bibitem[{Fern{\'a}ndez \& Sosa(2015)}]{fernandez2015jupiter}
Fern{\'a}ndez, J.~A. \& Sosa, A. 2015, Planetary and Space Science, 118, 14

\bibitem[{Fern{\'a}ndez {et~al.}(2014)Fern{\'a}ndez, Sosa, Gallardo, \&
  Guti{\'e}rrez}]{fernandez2014assessing}
Fern{\'a}ndez, J.~A., Sosa, A., Gallardo, T., \& Guti{\'e}rrez, J.~N. 2014,
  Icarus, 238, 1

\bibitem[{Fern{\'a}ndez(2009)}]{fernandez2009s}
Fern{\'a}ndez, Y.~R. 2009, Planetary and Space Science, 57, 1218

\bibitem[{Fern{\'a}ndez {et~al.}(2005)Fern{\'a}ndez, Jewitt, \&
  Sheppard}]{fernandez2005albedos}
Fern{\'a}ndez, Y.~R., Jewitt, D.~C., \& Sheppard, S.~S. 2005, The Astronomical
  Journal, 130, 308

\bibitem[{Fuse {et~al.}(2007)Fuse, Yamamoto, Kinoshita, Furusawa, \&
  Watanabe}]{fuse2007observations}
Fuse, T., Yamamoto, N., Kinoshita, D., Furusawa, H., \& Watanabe, J.-i. 2007,
  Publications of the Astronomical Society of Japan, 59, 381

\bibitem[{Gallardo {et~al.}(2012)Gallardo, Hugo, \& Pais}]{gallardo2012survey}
Gallardo, T., Hugo, G., \& Pais, P. 2012, Icarus, 220, 392

\bibitem[{Genge {et~al.}(2020)Genge, Van~Ginneken, \&
  Suttle}]{genge2020micrometeorites}
Genge, M.~J., Van~Ginneken, M., \& Suttle, M.~D. 2020, Planetary and Space
  Science, 187, 104900

\bibitem[{Gomes {et~al.}(2005)Gomes, Gallardo, Fern{\'a}ndez, \&
  Brunini}]{gomes2005origin}
Gomes, R.~S., Gallardo, T., Fern{\'a}ndez, J.~A., \& Brunini, A. 2005,
  Celestial Mechanics and Dynamical Astronomy, 91, 109

\bibitem[{Granvik \& Brown(2018)}]{granvik2018identification}
Granvik, M. \& Brown, P. 2018, Icarus, 311, 271

\bibitem[{Granvik {et~al.}(2016)Granvik, Morbidelli, Jedicke, Bolin, Bottke,
  Beshore, Vokrouhlick{\`y}, Delb{\`o}, \& Michel}]{granvik2016super}
Granvik, M., Morbidelli, A., Jedicke, R., {et~al.} 2016, Nature, 530, 303

\bibitem[{Granvik {et~al.}(2018)Granvik, Morbidelli, Jedicke, Bolin, Bottke,
  Beshore, Vokrouhlick{\'y}, Nesvorn{\'y}, \& Michel}]{granvik2018debiased}
Granvik, M., Morbidelli, A., Jedicke, R., {et~al.} 2018, Icarus, 312, 181

\bibitem[{Graykowski \& Jewitt(2019)}]{graykowski2019fragmented}
Graykowski, A. \& Jewitt, D. 2019, The Astronomical Journal, 158, 112

\bibitem[{Gritsevich \& Stulov(2006)}]{gritsevich2006extra}
Gritsevich, M. \& Stulov, V. 2006, Solar System Research, 40, 477

\bibitem[{{Gritsevich}(2007)}]{gritsevich2007og}
{Gritsevich}, M.~I. 2007, Solar System Research, 41, 509

\bibitem[{{Gural}(2012)}]{Gural_2012M&PS}
{Gural}, P.~S. 2012, Meteoritics \& Planetary Science, 47, 1405

\bibitem[{Gustafson(1994)}]{gustafson1994physics}
Gustafson, B.~A. 1994, Annual Review of Earth and Planetary Sciences, 22, 553

\bibitem[{Halliday {et~al.}(1978)Halliday, Blackwell, \&
  Griffin}]{halliday1978innisfree}
Halliday, I., Blackwell, A.~T., \& Griffin, A.~A. 1978, Journal of the Royal
  Astronomical Society of Canada, vol. 72, Feb. 1978, p. 15-39., 72, 15

\bibitem[{Halliday {et~al.}(1996)Halliday, Griffin, \&
  Blackwell}]{halliday1996detailed}
Halliday, I., Griffin, A.~A., \& Blackwell, A.~T. 1996, Meteoritics \&
  Planetary Science, 31, 185

\bibitem[{Hauser {et~al.}(1984)Hauser, Gillett, Low, Gautier, Beichman,
  Neugebauer, Aumann, Baud, Boggess, Emerson, {et~al.}}]{hauser1984iras}
Hauser, M., Gillett, F., Low, F., {et~al.} 1984, Astrophysical Journal, Part
  2-Letters to the Editor (ISSN 0004-637X), vol. 278, March 1, 1984, p.
  L15-L18., 278, L15

\bibitem[{Hoffleit(1988)}]{hoffleit1988yale}
Hoffleit, D. 1988, Vistas in astronomy, 32, 117

\bibitem[{Howie {et~al.}(2017{\natexlab{a}})Howie, Paxman, Bland, Towner,
  Cupák, Sansom, \& Devillepoix}]{howie2017build}
Howie, R.~M., Paxman, J., Bland, P.~A., {et~al.} 2017{\natexlab{a}},
  Experimental Astronomy, 43, 237

\bibitem[{Howie {et~al.}(2017{\natexlab{b}})Howie, Paxman, Bland, Towner,
  Sansom, \& Devillepoix}]{howie2017submillisecond}
Howie, R.~M., Paxman, J., Bland, P.~A., {et~al.} 2017{\natexlab{b}},
  Meteoritics \& Planetary Science, 52, 1669

\bibitem[{Hsieh \& Haghighipour(2016)}]{hsieh2016potential}
Hsieh, H.~H. \& Haghighipour, N. 2016, Icarus, 277, 19

\bibitem[{Hsieh \& Jewitt(2006)}]{hsieh2006population}
Hsieh, H.~H. \& Jewitt, D. 2006, Science, 312, 561

\bibitem[{Hsieh {et~al.}(2020)Hsieh, Novakovi{\'c}, Walsh, \&
  Sch{\"o}rghofer}]{hsieh2020potential}
Hsieh, H.~H., Novakovi{\'c}, B., Walsh, K.~J., \& Sch{\"o}rghofer, N. 2020, The
  Astronomical Journal, 159, 179

\bibitem[{Hughes {et~al.}(2022)Hughes, Sankar, Davis, Palotai, \&
  Free}]{hughes2022analysis}
Hughes, A., Sankar, R., Davis, K.~E., Palotai, C., \& Free, D.~L. 2022,
  Meteoritics \& Planetary Science, 57, 575

\bibitem[{Jansen-Sturgeon {et~al.}(2019)Jansen-Sturgeon, Sansom, \&
  Bland}]{jansen2019comparing}
Jansen-Sturgeon, T., Sansom, E.~K., \& Bland, P.~A. 2019, Meteoritics \&
  Planetary Science, 54, 2149

\bibitem[{Jansen-Sturgeon {et~al.}(2020)Jansen-Sturgeon, Sansom, Devillepoix,
  Bland, Towner, Howie, \& Hartig}]{jansen2020dynamic}
Jansen-Sturgeon, T., Sansom, E.~K., Devillepoix, H.~A., {et~al.} 2020, The
  Astronomical Journal, 160, 190

\bibitem[{Jeanne(2020)}]{jeanne2020methode}
Jeanne, S. 2020, PhD thesis, Universit{\'e} Paris sciences et lettres

\bibitem[{Jeanne {et~al.}(2019)Jeanne, Colas, Zanda, Birlan, Vaubaillon,
  Bouley, Vernazza, Jorda, Gattacceca, Rault, {et~al.}}]{jeanne2019calibration}
Jeanne, S., Colas, F., Zanda, B., {et~al.} 2019, Astronomy \& Astrophysics,
  627, A78

\bibitem[{Jenniskens(2004)}]{jenniskens20042003}
Jenniskens, P. 2004, The Astronomical Journal, 127, 3018

\bibitem[{Jenniskens(2006)}]{jenniskens2006meteor}
Jenniskens, P. 2006, Meteor showers and their parent comets (Cambridge
  University Press)

\bibitem[{Jenniskens {et~al.}(2016)Jenniskens, N{\'e}non, Albers, Gural,
  Haberman, Holman, Morales, Grigsby, Samuels, \&
  Johannink}]{jenniskens2016established}
Jenniskens, P., N{\'e}non, Q., Albers, J., {et~al.} 2016, Icarus, 266, 331

\bibitem[{Jenniskens \& Vaubaillon(2008)}]{jenniskens2008minor}
Jenniskens, P. \& Vaubaillon, J. 2008, The Astronomical Journal, 136, 725

\bibitem[{Jenniskens \& Vaubaillon(2010)}]{jenniskens2010minor}
Jenniskens, P. \& Vaubaillon, J. 2010, The Astronomical Journal, 139, 1822

\bibitem[{Jewitt {et~al.}(2016)Jewitt, Mutchler, Weaver, Hui, Agarwal,
  Ishiguro, Kleyna, Li, Meech, Micheli, {et~al.}}]{jewitt2016fragmentation}
Jewitt, D., Mutchler, M., Weaver, H., {et~al.} 2016, The Astrophysical Journal
  Letters, 829, L8

\bibitem[{Jiang \& Ji(2021)}]{jiang2021thermophysical}
Jiang, H. \& Ji, J. 2021, The Astronomical Journal, 162, 40

\bibitem[{Jones {et~al.}(2006)Jones, Williams, \& Porubcan}]{jones2006kappa}
Jones, D., Williams, I., \& Porubcan, V. 2006, Monthly Notices of the Royal
  Astronomical Society, 371, 684

\bibitem[{Jopek(1993)}]{jopek1993remarks}
Jopek, T.~J. 1993, Icarus, 106, 603

\bibitem[{Jopek {et~al.}(2008)Jopek, Rudawska, \&
  Bartczak}]{jopek2008meteoroid}
Jopek, T.~J., Rudawska, R., \& Bartczak, P. 2008, Advances in Meteoroid and
  Meteor Science, 73

\bibitem[{Jorgensen {et~al.}(2021)Jorgensen, Benn, Connerney, Denver,
  Jorgensen, Andersen, \& Bolton}]{jorgensen2021distribution_juno}
Jorgensen, J., Benn, M., Connerney, J., {et~al.} 2021, Journal of Geophysical
  Research: Planets, 126, e2020JE006509

\bibitem[{Kasuga {et~al.}(2010)Kasuga, Balam, \& Wiegert}]{kasuga2010comet}
Kasuga, T., Balam, D.~D., \& Wiegert, P.~A. 2010, The Astronomical Journal,
  140, 1806

\bibitem[{Kehoe {et~al.}(2015)Kehoe, Kehoe, Colwell, \&
  Dermott}]{kehoe2015signatures}
Kehoe, A.~E., Kehoe, T., Colwell, J., \& Dermott, S. 2015, The Astrophysical
  Journal, 811, 66

\bibitem[{Kelsall {et~al.}(1998)Kelsall, Weiland, Franz, Reach, Arendt, Dwek,
  Freudenreich, Hauser, Moseley, Odegard, Silverberg, \& Wright}]{kelsall_cobe}
Kelsall, T., Weiland, J.~L., Franz, B.~A., {et~al.} 1998, The Astrophysical
  Journal, 508, 44

\bibitem[{{King} {et~al.}(2022){King}, {Daly}, {Rowe}, {Joy}, {Greenwood},
  {Devillepoix}, {Suttle}, {Chan}, {Russell}, {Bates}, {Bryson}, {Clay},
  {Vida}, {Lee}, {O'Brien}, {Hallis}, {Stephen}, {Tart{\`e}se}, {Sansom},
  {Towner}, {Cupak}, {Shober}, {Bland}, {Findlay}, {Franchi}, {Verchovsky},
  {Abernethy}, {Grady}, {Floyd}, {Van Ginneken}, {Bridges}, {Hicks}, {Jones},
  {Mitchell}, {Genge}, {Jenkins}, {Martin}, {Sephton}, {Watson}, {Salge},
  {Shirley}, {Curtis}, {Warren}, {Bowles}, {Stuart}, {Di Nicola}, {Gy{\"o}re},
  {Boyce}, {Shaw}, {Elliott}, {Steele}, {Povinec}, {Laubenstein}, {Sanderson},
  {Cresswell}, {Jull}, {S{\'y}kora}, {Sridhar}, {Harrison}, {Willcocks},
  {Harrison}, {Hallatt}, {Wozniakiewicz}, {Burchell}, {Alesbrook}, {Dignam},
  {Almeida}, {Smith}, {Clark}, {Humphreys-Williams}, {Schofield}, {Cornwell},
  {Spathis}, {Morgan}, {Perkins}, {Kacerek}, {Campbell-Burns}, {Colas},
  {Zanda}, {Vernazza}, {Bouley}, {Jeanne}, {Hankey}, {Collins}, {Young},
  {Shaw}, {Horak}, {Jones}, {James}, {Bosley}, {Shuttleworth}, {Dickinson},
  {McMullan}, {Robson}, {Smedley}, {Stanley}, {Bassom}, {McIntyre}, {Suttle},
  {Fleet}, {Bastiaens}, {Ih{\'a}sz}, {McMullan}, {Boazman}, {Dickeson},
  {Grindrod}, {Pickersgill}, {Weir}, {Suttle}, {Farrelly}, {Spencer}, {Naqvi},
  {Mayne}, {Skilton}, {Kirk}, {Mounsey}, {Mounsey}, {Mounsey}, {Godfrey},
  {Bond}, {Bond}, {Wilcock}, {Wilcock}, \& {Wilcock}}]{King_Winchcombe2022SciA}
{King}, A.~J., {Daly}, L., {Rowe}, J., {et~al.} 2022, Science Advances, 8,
  eabq3925

\bibitem[{Knight {et~al.}(2016)Knight, Fitzsimmons, Kelley, \&
  Snodgrass}]{knight2016comet}
Knight, M.~M., Fitzsimmons, A., Kelley, M.~S., \& Snodgrass, C. 2016, The
  Astrophysical Journal Letters, 823, L6

\bibitem[{Kozai(1979)}]{kozai1979secular}
Kozai, Y. 1979, in Symposium-International Astronomical Union, Vol.~81,
  Cambridge University Press, 231--237

\bibitem[{{Kresak}(1979)}]{kresak1979}
{Kresak}, L. 1979, in Asteroids, ed. T.~{Gehrels} \& M.~S. {Matthews}, 289--309

\bibitem[{{Levison}(1996)}]{levison1996taxonomy}
{Levison}, H.~F. 1996, in Astronomical Society of the Pacific Conference
  Series, Vol. 107, Completing the Inventory of the Solar System, ed.
  T.~{Rettig} \& J.~M. {Hahn}, 173--191

\bibitem[{Levison \& Duncan(1994)}]{levison1994long}
Levison, H.~F. \& Duncan, M.~J. 1994, Icarus, 108, 18

\bibitem[{Levison \& Duncan(1997)}]{levison1997kuiper}
Levison, H.~F. \& Duncan, M.~J. 1997, Icarus, 127, 13

\bibitem[{Levison {et~al.}(2006)Levison, Terrell, Wiegert, Dones, \&
  Duncan}]{levison2006origin}
Levison, H.~F., Terrell, D., Wiegert, P.~A., Dones, L., \& Duncan, M.~J. 2006,
  Icarus, 182, 161

\bibitem[{Li \& Jewitt(2015)}]{li2015disappearance}
Li, J. \& Jewitt, D. 2015, The Astronomical Journal, 149, 133

\bibitem[{Lindgren(1992)}]{lindgren1992dynamical}
Lindgren, M. 1992, in Asteroids, Comets, Meteors 1991

\bibitem[{Lyytinen \& Gritsevich(2016)}]{lyytinen2016implications}
Lyytinen, E. \& Gritsevich, M. 2016, Planetary and Space Science, 120, 35

\bibitem[{Madiedo {et~al.}(2014)Madiedo, Trigo-Rodr{\'\i}guez, Zamorano,
  Ana-Hern{\'a}ndez, Izquierdo, Ortiz, Castro-Tirado, de~Miguel, Oca{\~n}a,
  Pastor, {et~al.}}]{madiedo2014trajectory}
Madiedo, J.~M., Trigo-Rodr{\'\i}guez, J.~M., Zamorano, J., {et~al.} 2014,
  Astronomy \& Astrophysics, 569, A104

\bibitem[{{Martino} {et~al.}(2019){Martino}, {Tancredi}, {Monteiro}, {Lazzaro},
  \& {Rodrigues}}]{martino_s_2019}
{Martino}, S., {Tancredi}, G., {Monteiro}, F., {Lazzaro}, D., \& {Rodrigues},
  T. 2019, Planetary and Space Science, 166, 135

\bibitem[{Matlovi{\v{c}} {et~al.}(2017)Matlovi{\v{c}}, T{\'o}th, Rudawska, \&
  Korno{\v{s}}}]{matlovivc2017spectra}
Matlovi{\v{c}}, P., T{\'o}th, J., Rudawska, R., \& Korno{\v{s}}, L. 2017,
  Planetary and Space Science, 143, 104

\bibitem[{Matsumoto {et~al.}(2018)Matsumoto, Tsumura, Matsuoka, \&
  Pyo}]{matsumoto_pioneer10}
Matsumoto, T., Tsumura, K., Matsuoka, Y., \& Pyo, J. 2018, The Astronomical
  Journal, 156, 86

\bibitem[{Meech {et~al.}(2004)Meech, Hainaut, \& Marsden}]{meech2004comet}
Meech, K., Hainaut, O., \& Marsden, B. 2004, Icarus, 170, 463

\bibitem[{Mommert {et~al.}(2015)Mommert, Harris, Mueller, Hora, Trilling,
  Bottke, Thomas, Delbo, Emery, Fazio, {et~al.}}]{mommert2015exploreneos}
Mommert, M., Harris, A., Mueller, M., {et~al.} 2015, The Astronomical Journal,
  150, 106

\bibitem[{{Nesvorn{\'y}} {et~al.}(2023){Nesvorn{\'y}}, {Deienno}, {Bottke},
  {Jedicke}, {Naidu}, {Chesley}, {Chodas}, {Granvik}, {Vokrouhlick{\'y}},
  {Bro{\v{z}}}, {Morbidelli}, {Christensen}, {Shelly}, \&
  {Bolin}}]{2023AJ....166...55N}
{Nesvorn{\'y}}, D., {Deienno}, R., {Bottke}, W.~F., {et~al.} 2023, The
  Astronomical Journal, 166, 55

\bibitem[{Nesvorn{\`y} {et~al.}(2011)Nesvorn{\`y}, Janches, Vokrouhlick{\`y},
  Pokorn{\`y}, Bottke, \& Jenniskens}]{nesvorny2011dynamical}
Nesvorn{\`y}, D., Janches, D., Vokrouhlick{\`y}, D., {et~al.} 2011, The
  Astrophysical Journal, 743, 129

\bibitem[{Nesvorn{\'y} {et~al.}(2010)Nesvorn{\'y}, Jenniskens, Levison, Bottke,
  Vokrouhlick{\'y}, \& Gounelle}]{nesvorny2010cometary}
Nesvorn{\'y}, D., Jenniskens, P., Levison, H.~F., {et~al.} 2010, The
  Astrophysical Journal, 713, 816

\bibitem[{Nesvorn{\'y} {et~al.}(2017)Nesvorn{\'y}, Vokrouhlick{\'y}, Dones,
  Levison, Kaib, \& Morbidelli}]{nesvorny2017origin}
Nesvorn{\'y}, D., Vokrouhlick{\'y}, D., Dones, L., {et~al.} 2017, The
  Astrophysical Journal, 845, 27

\bibitem[{Opik(1971)}]{opik1971comet}
Opik, E. 1971, Irish Astronomical Journal, vol. 10 (1/2), p. 35, 10, 35

\bibitem[{Ott {et~al.}(2017)Ott, Drolshagen, Koschny, G{\"u}ttler, Tubiana,
  Frattin, Agarwal, Sierks, Bertini, Barbieri, {et~al.}}]{ott2017dust}
Ott, T., Drolshagen, E., Koschny, D., {et~al.} 2017, Monthly Notices of the
  Royal Astronomical Society, 469, S276

\bibitem[{{Pecina} \& {Ceplecha}(1983)}]{pecina1983BAICz..34..102P}
{Pecina}, P. \& {Ceplecha}, Z. 1983, Bulletin of the Astronomical Institutes of
  Czechoslovakia, 34, 102

\bibitem[{Pe{\~n}a-Asensio {et~al.}(2022)Pe{\~n}a-Asensio,
  Trigo-Rodr{\'\i}guez, \& Rimola}]{pena2022orbital}
Pe{\~n}a-Asensio, E., Trigo-Rodr{\'\i}guez, J.~M., \& Rimola, A. 2022, The
  Astronomical Journal, 164, 76

\bibitem[{Plane(2012)}]{plane2012cosmic}
Plane, J.~M. 2012, Chemical Society Reviews, 41, 6507

\bibitem[{Popova {et~al.}(2011)Popova, Borovi{\v{c}}ka, Hartmann, Spurn{\'y},
  Gnos, Nemtchinov, \& TRIGO-RODR{\'I}GUEZ}]{popova2011very}
Popova, O., Borovi{\v{c}}ka, J., Hartmann, W.~K., {et~al.} 2011, Meteoritics \&
  Planetary Science, 46, 1525

\bibitem[{{Rein} \& {Liu}(2012)}]{rebound2012A&A...537A.128R}
{Rein}, H. \& {Liu}, S.~F. 2012, Astronomy \& Astrophysics, 537, A128

\bibitem[{Rein \& Spiegel(2015)}]{rein2015ias15}
Rein, H. \& Spiegel, D.~S. 2015, Monthly Notices of the Royal Astronomical
  Society, 446, 1424

\bibitem[{Rein \& Tamayo(2015)}]{rein2015whfast}
Rein, H. \& Tamayo, D. 2015, Monthly Notices of the Royal Astronomical Society,
  452, 376

\bibitem[{Rickman {et~al.}(1990)Rickman, Fernandez, \&
  Gustafson}]{rickman1990formation}
Rickman, H., Fernandez, J., \& Gustafson, B. 1990, Astronomy and Astrophysics
  (ISSN 0004-6361), vol. 237, no. 2, Oct. 1990, p. 524-535. Research supported
  by NFR and Alexander von Humboldt-Stiftung., 237, 524

\bibitem[{Rigley \& Wyatt(2022)}]{rigley2022comet}
Rigley, J.~K. \& Wyatt, M.~C. 2022, Monthly Notices of the Royal Astronomical
  Society, 510, 834

\bibitem[{Robitaille {et~al.}(2013)Robitaille, Tollerud, Greenfield,
  Droettboom, Bray, Aldcroft, Davis, Ginsburg, Price-Whelan, Kerzendorf,
  {et~al.}}]{robitaille2013astropy}
Robitaille, T.~P., Tollerud, E.~J., Greenfield, P., {et~al.} 2013, Astronomy \&
  Astrophysics, 558, A33

\bibitem[{Sansom {et~al.}(2015)Sansom, Bland, Paxman, \&
  Towner}]{sansom2015novel}
Sansom, E.~K., Bland, P., Paxman, J., \& Towner, M. 2015, Meteoritics \&
  Planetary Science, 50, 1423

\bibitem[{Sansom {et~al.}(2019)Sansom, Gritsevich, Devillepoix,
  Jansen-Sturgeon, Shober, Bland, Towner, Cupák, Howie, \&
  Hartig}]{sansom2019determining}
Sansom, E.~K., Gritsevich, M., Devillepoix, H.~A., {et~al.} 2019, The
  Astrophysical Journal, 885, 115

\bibitem[{Sekanina \& Chodas(2005)}]{sekanina2005origin}
Sekanina, Z. \& Chodas, P.~W. 2005, The Astrophysical Journal Supplement
  Series, 161, 551

\bibitem[{Shober {et~al.}(2019)Shober, Jansen-Sturgeon, Sansom, Devillepoix,
  Bland, Cup{\'a}k, Towner, Howie, \& Hartig}]{shober2019identification}
Shober, P., Jansen-Sturgeon, T., Sansom, E., {et~al.} 2019, The Astronomical
  Journal, 158, 183

\bibitem[{Shober {et~al.}(2022)Shober, Devillepoix, Sansom, Towner, Cup{\'a}k,
  Anderson, Benedix, Forman, Bland, Howie, {et~al.}}]{shober2022arpu}
Shober, P.~M., Devillepoix, H.~A., Sansom, E.~K., {et~al.} 2022, Meteoritics \&
  Planetary Science, 57, 1146

\bibitem[{Shober {et~al.}(2020{\natexlab{a}})Shober, Jansen-Sturgeon, Bland,
  Devillepoix, Sansom, Towner, Cup{\'a}k, Howie, \& Hartig}]{shober2020using}
Shober, P.~M., Jansen-Sturgeon, T., Bland, P., {et~al.} 2020{\natexlab{a}},
  Monthly Notices of the Royal Astronomical Society, 498, 5240

\bibitem[{Shober {et~al.}(2020{\natexlab{b}})Shober, Jansen-Sturgeon, Sansom,
  Devillepoix, Towner, Bland, Cupák, Howie, \& Hartig}]{shober2020did}
Shober, P.~M., Jansen-Sturgeon, T., Sansom, E.~K., {et~al.} 2020{\natexlab{b}},
  The Astronomical Journal, 159, 191

\bibitem[{Shober {et~al.}(2021)Shober, Sansom, Bland, Devillepoix, Towner,
  Cup{\'a}k, Howie, Hartig, \& Anderson}]{shober2021main}
Shober, P.~M., Sansom, E.~K., Bland, P.~A., {et~al.} 2021, The Planetary
  Science Journal, 2, 98

\bibitem[{Shober \& Vaubaillon(2024)}]{shober2024generalizable}
Shober, P.~M. \& Vaubaillon, J. 2024, A generalizable method for estimating
  meteor shower false positives

\bibitem[{Shober {et~al.}(2023)Shober, Vaubaillon, Anghel, Malgoyre,
  Devillepoix, Sansom, Vida, \& Colas}]{shober2023comparison}
Shober, P.~M., Vaubaillon, J., Anghel, S., {et~al.} 2023, in Proceedings of the
  Asteroids Comets Meteors Conference 2023, Lunar and Planetary Institute,
  Houston, Texas, lPI Contribution No. 2851, Abstract \#2402

\bibitem[{Southworth \& Hawkins(1963)}]{southworth1963statistics}
Southworth, R. \& Hawkins, G. 1963, Smithsonian Contributions to Astrophysics,
  7, 261

\bibitem[{Spurn{\'y} {et~al.}(2017)Spurn{\'y}, Borovi{\v{c}}ka, Baumgarten,
  Haack, Heinlein, \& S{\o}rensen}]{spurny2017ejby}
Spurn{\'y}, P., Borovi{\v{c}}ka, J., Baumgarten, G., {et~al.} 2017, Planetary
  and Space Science, 143, 192

\bibitem[{Spurny {et~al.}(2013)Spurny, Borovicka, Haack, Singer, Keuer, \&
  Jobse}]{spurny2013trajectory}
Spurny, P., Borovicka, J., Haack, H., {et~al.} 2013, in Meteoroids 2013
  conference, Poznan, Poland

\bibitem[{Spurn{\`y} {et~al.}(2017)Spurn{\`y}, Borovi{\v{c}}ka, Mucke, \&
  Svore{\v{n}}}]{spurny2017discovery}
Spurn{\`y}, P., Borovi{\v{c}}ka, J., Mucke, H., \& Svore{\v{n}}, J. 2017,
  Astronomy \& Astrophysics, 605, A68

\bibitem[{Spurn{\`y} {et~al.}(2020)Spurn{\`y}, Borovi{\v{c}}ka, \&
  Shrben{\`y}}]{spurny2020vzvdar}
Spurn{\`y}, P., Borovi{\v{c}}ka, J., \& Shrben{\`y}, L. 2020, Meteoritics \&
  Planetary Science, 55, 376

\bibitem[{Spurn{\`y} {et~al.}(2003)Spurn{\`y}, Oberst, \&
  Heinlein}]{spurny2003photographic}
Spurn{\`y}, P., Oberst, J., \& Heinlein, D. 2003, Nature, 423, 151

\bibitem[{Sykes(1988)}]{sykes1988iras}
Sykes, M.~V. 1988, Astrophysical Journal, Part 2-Letters (ISSN 0004-637X), vol.
  334, Nov. 1, 1988, p. L55-L58. Research supported by NASA IRAS General
  Investigator Program and USAF., 334, L55

\bibitem[{Sykes \& Walker(1992)}]{sykes1992cometary}
Sykes, M.~V. \& Walker, R.~G. 1992, Icarus, 95, 180

\bibitem[{Szalay {et~al.}(2021)Szalay, Pokorný, Malaspina, Pusack, Bale,
  Battams, Gasque, Goetz, Krüger, McComas, Schwadron, \&
  Strub}]{szalay_parker_solar_probe}
Szalay, J.~R., Pokorný, P., Malaspina, D., {et~al.} 2021

\bibitem[{Takir \& Emery(2012)}]{takir2012outer}
Takir, D. \& Emery, J.~P. 2012, Icarus, 219, 641

\bibitem[{Tancredi(1995)}]{tancredi1995dynamical}
Tancredi, G. 1995, Astronomy and Astrophysics, 299, 288

\bibitem[{Tancredi(1998)}]{tancredi1998chaotic}
Tancredi, G. 1998, Celestial Mechanics and Dynamical Astronomy, 70, 181

\bibitem[{Tancredi(2014)}]{tancredi2014criterion}
Tancredi, G. 2014, Icarus, 234, 66

\bibitem[{Tancredi \& Rickman(1992)}]{tancredi1992evolution}
Tancredi, G. \& Rickman, H. 1992, in Symposium-International Astronomical
  Union, Vol. 152, Cambridge University Press, 269--274

\bibitem[{Taylor(2005)}]{taylor2005topcat}
Taylor, M.~B. 2005, in Astronomical Data Analysis Software and Systems XIV,
  Vol. 347, 29

\bibitem[{Toliou \& Granvik(2023)}]{toliou2023resonant}
Toliou, A. \& Granvik, M. 2023, Monthly Notices of the Royal Astronomical
  Society, 521, 4819

\bibitem[{Toliou {et~al.}(2021)Toliou, Granvik, \&
  Tsirvoulis}]{toliou2021minimum}
Toliou, A., Granvik, M., \& Tsirvoulis, G. 2021, Monthly Notices of the Royal
  Astronomical Society, 506, 3301

\bibitem[{Trigo-Rodriguez {et~al.}(2019)Trigo-Rodriguez, Madiedo, Blanch,
  Chioare, Tilve, Llorca, Herrero-P{\'e}rez, Gonz{\'a}lez, Jover~Benjumea,
  Pujols, {et~al.}}]{trigo2019jupiter}
Trigo-Rodriguez, J., Madiedo, J., Blanch, E., {et~al.} 2019, in 50th Annual
  Lunar and Planetary Science Conference No. 2132, 2368

\bibitem[{Trigo-Rodr{\'\i}guez {et~al.}(2009)Trigo-Rodr{\'\i}guez, Madiedo,
  Williams, Castro-Tirado, Llorca, Vitek, \& Jelinek}]{trigo2009observations}
Trigo-Rodr{\'\i}guez, J.~M., Madiedo, J.~M., Williams, I.~P., {et~al.} 2009,
  Monthly Notices of the Royal Astronomical Society, 394, 569

\bibitem[{Valsecchi {et~al.}(1999)Valsecchi, Jopek, \&
  Froeschl{\'e}}]{valsecchi1999meteoroid}
Valsecchi, G., Jopek, T., \& Froeschl{\'e}, C. 1999, Monthly Notices of the
  Royal Astronomical Society, 304, 743

\bibitem[{Vida {et~al.}(2020)Vida, Gural, Brown, Campbell-Brown, \&
  Wiegert}]{vida2020estimating}
Vida, D., Gural, P.~S., Brown, P.~G., Campbell-Brown, M., \& Wiegert, P. 2020,
  Monthly Notices of the Royal Astronomical Society, 491, 2688

\bibitem[{{Weryk} \& {Brown}(2012)}]{Weryk_Brown_2012P&SS}
{Weryk}, R.~J. \& {Brown}, P.~G. 2012, Planetary and Space Science, 62, 132

\bibitem[{Wiegert \& Brown(2005)}]{wiegert2005quadrantid}
Wiegert, P. \& Brown, P. 2005, Icarus, 179, 139

\bibitem[{Wiegert {et~al.}(2020)Wiegert, Brown, Pokorn{\`y}, Ye, Gregg,
  Lenartowicz, Krzeminski, \& Clark}]{wiegert2020supercatastrophic}
Wiegert, P., Brown, P., Pokorn{\`y}, P., {et~al.} 2020, The Astronomical
  Journal, 159, 143

\bibitem[{Wiegert {et~al.}(2013)Wiegert, Brown, Weryk, \&
  Wong}]{wiegert2013return}
Wiegert, P.~A., Brown, P.~G., Weryk, R.~J., \& Wong, D.~K. 2013, The
  Astronomical Journal, 145, 70

\bibitem[{Ye(2018)}]{ye2018meteor}
Ye, Q.-Z. 2018, Planetary and Space Science, 164, 7

\bibitem[{Ye {et~al.}(2016)Ye, Brown, \& Pokorn{\`y}}]{ye2016dormant}
Ye, Q.-Z., Brown, P.~G., \& Pokorn{\`y}, P. 2016, Monthly Notices of the Royal
  Astronomical Society, 462, 3511

\end{thebibliography}

\end{document}